%% file: arxiv-version.tex
\def\verTR{1}    

\def\submit{0}   
\def\AlgVer{1}   
\def\RedAna{2}   
\def\LatexVer{2} 
\ifnum\LatexVer=1
  \def\inclFIG{0} 
\else
  \def\inclFIG{1} 
\fi

\ifnum\LatexVer=1
\documentstyle[11pt,fullpage]{article}
\input psfig
\else
\documentclass[11pt]{article}
\usepackage{amsfonts,fullpage,epsfig,color,psfrag}
\fi

\title{Finding Cycles and Trees in Sublinear Time}

\author{Artur Czumaj 
\and Oded Goldreich
\and Dana Ron
\and C.~Seshadhri\footnote{Employee of Sandia National Laboratories. Sandia National Laboratories is a multi-program laboratory managed and operated by Sandia Corporation, a wholly owned subsidiary of Lockheed Martin Corporation, for the U.S. Department of Energy's National Nuclear Security Administration under contract DE-AC04-94AL85000.}
\and Asaf Shapira
\and Christian Sohler 
}

\newcommand{\onote}[1]{\begin{quote}{\sf Oded's Note:} {\sl{#1}} \end{quote}}

\newcommand{\eqref}[1]{{Eq.~(\ref{#1})}}
\newcommand{\thmref}[1]{{Theorem~\ref{#1}}}
\newcommand{\secref}[1]{{Section~\ref{#1}}}
  \newcommand{\SAref}[1]{{Section~\ref{#1}}}
  \newcommand{\myfoot}[1]{\footnote{#1}}

\newcommand{\mybibitem}[1]{\bibitem[#1]{#1}}
\newcommand{\triag}[1]{\bigtriangledown_{\stackrel{}{\{#1\}}}}
\def\FullBox{\hbox{\vrule width 8pt height 8pt depth 0pt}}
\newcommand{\QED}{\(\;\;\;\FullBox\)}
\newcommand{\eps}{\epsilon}

\newcommand{\Prob}{{\rm Pr}}
\newcommand{\tildeO}{{\widetilde{O}}}
\newcommand{\tildeT}{{\widetilde{\Theta}}}
\newcommand{\xpath}{{\mbox{\rm--}\cdots\mbox{\rm--}}}
\newcommand{\xedge}{{\mbox{\rm---}}}
\newcommand{\ceil}[1]{\lceil{#1}\rceil}

\newcommand{\ang}[1]{\langle{#1}\rangle}
\newcommand{\mst}{\;\mbox{\rm s.t. }}
\newcommand{\FS}{\mbox{\tt FS}}
\newcommand{\pS}{{R'}}

\ifnum\LatexVer=1

\newcommand{\emph}[1]{{\em{#1}}\/}
\newcommand{\textbf}[1]{{\bf{#1}}}
\newcommand{\cH}{{\cal{H}}}
\newcommand{\cP}{{\cal{P}}}
\else

\newcommand{\cH}{\mathcal{H}}
\newcommand{\cP}{\mathcal{P}}
\fi

\newcommand{\minor}{\mbox{\tt minor}}
\newcommand{\free}{\mbox{\tt free}}
\newcommand{\cut}{\mbox{\tt cut}}
\newenvironment{proof}{\noindent{\bf Proof:~~}}{\QED}
\newcommand{\BPF}{\begin{proof}} \newcommand{\EPF}{\end{proof}}
\newenvironment{proofof}[1]{\noindent{\bf Proof of {#1}.~}}{\endproof}
\newcommand{\BPFOF}{\begin{proofof}} \newcommand {\EPFOF}{\end{proofof}}
\newenvironment{proofout}{\noindent{\bf Proof outline:~~}}{\QED}
\newcommand{\BPFOUT}{\begin{proofout}} \newcommand{\EPFOUT}{\end{proofout}}

\newcommand{\QT}{{\tt{QT}}}

\def\etal{{\it et~al.}}
\newcommand{\eqdef}{\stackrel{\rm def}{=}}
\newcommand{\bitset}{\{0,1\}}

\newcommand{\xth}{{\rm th}}

\newcommand{\e}{\epsilon}
\newcommand{\f}{\zeta}

\newtheorem{Ithm}{Theorem}[section]     
\newcommand{\BIT}{\begin{Ithm}}   \newcommand{\EIT}{\end{Ithm}}
\newtheorem{lem}{Lemma}[section]  
\newcommand{\BL}{\begin{lem}}   \newcommand{\EL}{\end{lem}}
\newtheorem{thm}[lem]{Theorem}      
\newcommand{\BT}{\begin{thm}}   \newcommand{\ET}{\end{thm}}
\newtheorem{dfn}[lem]{Definition}      %
\newcommand{\BD}{\begin{dfn}}   \newcommand{\ED}{\end{dfn}}
\newtheorem{const}[lem]{Construction}      %
\newcommand{\BCT}{\begin{const}} \newcommand{\ECT}{\end{const}}
\newtheorem{alg}[lem]{Algorithm}      %
\newcommand{\BA}{\begin{alg}} \newcommand{\EA}{\end{alg}}
\newtheorem{prop}[lem]{Proposition}
\newcommand{\BP}{\begin{prop}}   \newcommand{\EP}{\end{prop}}
\newtheorem{clm}[lem]{Claim}            %
\newcommand{\BCM}{\begin{clm}}   \newcommand{\ECM}{\end{clm}}
\newtheorem{corr}[lem]{Corollary}      %
\newcommand{\BCR}{\begin{corr}} \newcommand{\ECR}{\end{corr}}
\newtheorem{fact}[lem]{Fact}            %
\newcommand{\BF}{\begin{fact}}   \newcommand{\EF}{\end{fact}}
\newtheorem{rem}[lem]{Remark}            %
\newcommand{\BR}{\begin{rem}}   \newcommand{\ER}{\end{rem}}
\newcommand{\BE}{\begin{enumerate}}
\newcommand{\EE}{\end{enumerate}}
\newcommand{\BI}{\begin{itemize}}
\newcommand{\EI}{\end{itemize}}
\newcommand{\BDes}{\begin{description}}
\newcommand{\EDes}{\end{description}}
\def\blackslug
{\hbox{\hskip 1pt\vrule width 8pt height 8pt depth 1.5pt\hskip 1pt}}
\def\qed{\quad\blackslug\lower 8.5pt\null\par}

\newcommand{\cS}{{\cal S}}
\newcommand{\cG}{{\cal G}}
\newcommand{\lab}{{\Lambda}}
\newcommand{\tteq}{{\tt eq}}
\newcommand{\ttneq}{{\tt neq}}

\renewcommand{\th}{{\rm th}}

\newcommand{\dist}{{\rm dist}}

\newcommand{\N}{{\mathbb N}}

\newcommand{\prob}{{\rm Pr}}
\newcommand{\poly}{{\rm poly}}

\renewcommand{\P}{{\cal P}}

\newcommand{\kh}{\hat{k}}
\newcommand{\find}{{\tt find}}
\newcommand{\bound}[2]{\partial_{#1}(#2)}
\newcommand{\nbound}[2]{\overline{\partial}_{#1}(#2)}
\ifnum\AlgVer=1
  
\else
  
\fi

\newcommand{\davg}{{d_{\rm avg}}}

\newcommand{\BEQ}{\begin{equation}} \newcommand{\EEQ}{\end{equation}}
\newcommand{\BEQN}{\begin{eqnarray}}\newcommand{\EEQN}{\end{eqnarray}}

  \newcommand{\tsf}{\sf}
  \newcommand{\dsf}{\sf}

\begin{document}


\begin{titlepage}
\maketitle
\thispagestyle{empty}

\begin{abstract}
We present sublinear-time (randomized) algorithms
for finding simple cycles of length at least $k\geq3$
and tree-minors in bounded-degree graphs.
The complexity of these algorithms is related to the distance
of the graph from being $C_k$-minor free
(resp., free from having the corresponding tree-minor).
In particular,
if the graph is $\Omega(1)$-far from being cycle-free
(i.e., a constant fraction of the edges must be deleted
to make the graph cycle-free),
then the algorithm finds a cycle of polylogarithmic length
in time $\tildeO(\sqrt{N})$,
where $N$ denotes the number of vertices.
This time complexity is optimal up to polylogarithmic factors.

The foregoing results are the outcome of our study of the complexity
of {\em one-sided error}\/ property testing algorithms
in the bounded-degree graphs model.
For example, we show that cycle-freeness of $N$-vertex graphs
can be tested with one-sided error
within time complexity $\tildeO(\poly(1/\e)\cdot\sqrt{N})$,
where $\e$ denotes the proximity parameter.
This matches the known $\Omega(\sqrt{N})$ query lower bound
for one-sided error cycle-freeness testing,
and contrasts with the fact that any minor-free property admits
a {\em two-sided error}\/ tester of query complexity
that only depends on $\e$.
We show that the same upper bound holds for testing whether
the input graph has a simple cycle of length at least~$k$,
for any $k\geq3$.
On the other hand, for any fixed tree $T$,
we show that $T$-minor freeness has a one-sided error tester
of query complexity that only depends on the proximity parameter $\e$.

Our algorithm for finding cycles in bounded-degree graphs
extends to general graphs, where distances are measured with
respect to the actual number of edges. Such an extension is not
possible with respect to finding tree-minors in $o(\sqrt{N})$ complexity.
\end{abstract}



\vfill

\medskip\noindent{~}
\pagenumbering{Roman}
\end{titlepage}

\tableofcontents \newpage
\pagenumbering{arabic}

\section{Introduction}\label{intro.sec}
%
%

Consider the algorithmic problem of finding a (simple) cycle in a
bounded degree graph (assuming one exists), where the aim is to find
such a cycle in (randomized) sublinear time. In general, finding a cycle
in sublinear time may not be possible, since the graph may contain only
cycles of length $\Omega(n)$. This may also be the case if one needs to
remove a constant {\em number\/} of the edges of the graph in order to make it
cycle-free. But suppose one needs to remove a constant {\em fraction\/} of the
graph's edges in order to make it cycle free. Can we then devise
a sublinear time algorithm? One of our results in this
paper is an affirmative answer to this question.
Furthermore, the running time of that algorithm is (essentially) optimal.

\subsection{Our main results}
\label{intro:finding.sec}
As we have mentioned above, we consider graphs of bounded degree $d$
with $N$ vertices.
We say that a graph is $\epsilon$-far from being cycle-free
if one has to remove at least $\epsilon d N$ edges from $G$
in order to make it cycle free.%
\footnote{In some sources, being $\e$-far from a property
means that and $\e$ fraction of the function's values should
be changed so to obtain a function that has the property.
In our case, such a definition would translate to an omission
of $\e dN/2$ edges, since each edge appears twice
(i.e., once in each of its endpoints). Nevertheless,
for sake of simplicity, we chose to measure distance in terms of $dN$
(rather than in terms of $dN/2$).\label{edge-count:fn}}
In all our results,
vertex manipulation operations are counted at unit cost.
We can now formally state our first result.

\BT{\em(finding cycles):}
\label{intro:find-cycle.thm}
There exists a randomized algorithm that, on input an $N$-vertex graph $G$
of 
degree bound $d$ that is $\e$-far from being cycle-free,
finds a simple cycle in $G$
in expected time
$\tildeO(\poly(d/\e)\cdot\sqrt{N})$.
Furthermore, the cycle found has length at most $\poly(\e^{-1}d\log N)$.
\ET
Using the connection to one-sided error property testing
(detailed in \secref{intro:PT-connect.sec}),
we infer that the algorithm of \thmref{intro:find-cycle.thm}
is optimal; that is, no randomized $o(\sqrt{N})$-time algorithm
can find cycles in (bounded-degree) graphs
that are $\Omega(1)$-far from being cycle-free.
Furthermore, one cannot expect to find simple cycles
of length $o(\log N)$, since such may not exist
(even if the graph is far from being cycle-free).
%
The result of \thmref{intro:find-cycle.thm} can be extended to
finding a simple cycle of length at least~$k$, for any fixed $k>3$
(where the case $k=3$ is covered by \thmref{intro:find-cycle.thm}).

\BT{\em(finding cycles of length at least $k$):}
\label{intro:find-Ck.thm}
For every constant $k>3$,
there exists a randomized algorithm that, on input an $N$-vertex
graph $G$ of degree bound $d$ that is $\e$-far from having
no simple cycles of length at least $k$, finds such a cycle
in expected time $\tildeO(\poly(d^k/\e)\cdot\sqrt{N})$.
Furthermore, the cycle found
has length at most $ \poly(d^k \cdot\e^{-1}\log N)$.
\ET
Again, the algorithm obtained is optimal in the sense discussed above.

We note that our results can be stated in terms of finding
graph minors.
A graph $G$ has an {\tsf $H$-minor}
 if $H$ can be obtained from $G$ through a series of
 vertex removals, edge removals, and edge contractions.
 A graph $G$ is {\tsf $H$-minor free}, if it contains no $H$-minor.
Note that cycles of length at least $k$ in $G$
 correspond to $C_k$-minors of $G$,
 where $C_k$ denotes the $k$-vertex cycle.

We next turn from finding cycles to finding tree-structures
in graphs; that is, finding tree-minors.
Consider the following interesting special case.
For any constant $k$, we want to find
a tree with at least $k$ leaves.
One of our results is a randomized algorithm that finds such trees
in expected time that is polynomially related to $k$
and to the distance of the input graph from a graph
having no such trees.
This problem corresponds to finding minors that are $k$-vertex stars.
More generally, we prove the following result.

\BT{\em(finding tree minors):}
\label{intro:find-tree.thm}
For any fixed tree $T$ with $k$ vertices,
there exists a randomized algorithm that,
on input an $N$-vertex graph $G$ of constant degree bound $d$ that
is $\e$-far from being $T$-minor free, finds a $T$-minor
in expected time $\poly(d^D)$, where $D=k(16d/\e)^{4k+2}$.
\ET
We highlight the fact that finding tree minors can be done within
complexity that does not depend on the size of the graph
(but rather depends only on ($d,k$ and) $\e$),
whereas finding cycles requires $\Omega(\sqrt{N})$ time
(also for constant $\e>0$).
In fact, we show that \thmref{intro:find-tree.thm} extends to
any cycle-free graph (forest) $H$, and on the other hand
we prove that {\em for any $H$ that contains a cycle
finding $H$-minors requires $\Omega(\sqrt{N})$ queries}\/
(see \thmref{lower-bound.thm}).%
Thus, we obtain the following characterization:

\BCR{\em(finding graph minors, a dichotomy):}
\label{intro:characterization.cor}
Finding $H$-minors in a constant degree graph that
is $\e$-far from being $H$-minor free can be done
in complexity that only depends on $\e$
if and only if $H$ is cycle-free.
\ECR
%

\subsection{The property testing connection}\label{intro:PT-connect.sec}
Loosely speaking, property testing refers to sublinear time
probabilistic algorithms for deciding whether a given object has
a predetermined property or is far from any object having this property
(see the surveys~\cite{F,R2,R3}).
Such algorithms, called testers, obtain local views of the object
by making suitable queries; that is, the object is seen as a function
and the tester gets oracle access to this function
(and thus may be expected to work in time that
is sublinear in the size of the object).

Randomization is essential to natural testers (i.e., testers of natural
properties that have sublinear query-complexity)~\cite{GS}.
The same holds also for error probability, at least on some instances,
but the question is whether a (small) error probability must
appear on all instances.
In particular, {\em should we allow}\/ (small)
{\em error probability both on instances that have the property and on
instances that are far from having it?}\footnote{Recall that,
in any case, the basic paradigm of property testing allows arbitrary error
in case the instance neither has the property nor is far from having it.}

Indeed, testers come in two basic flavors
referring to the foregoing question:
{\tsf two-sided error} testers allow (small) error
probability both on instances that have the property
and on instances that are far from having it,
whereas {\tsf one-sided error} testers only allow (small) error
probability on instances that are far from having the property.
That is, in one-sided error testers, any instance that has the property
is accepted with probability~1.

An important observation regarding one-sided error testers is that
{\em whenever such a tester rejects some instance,
it always has a certificate that this
instance does not have the property, where this certificate
is the partial view of the instance as obtained by the tester}.
Indeed, in the case of one-sided error, rejecting an instance
based on a specific partial view means that there exists
no instance that has the property and is consistent
with this partial view.
Furthermore, in some cases (as those addressed in the current work),
this partial view contains some natural structures
(e.g., a cycle or a tree of interest).

Consider, for example, the case of testing cycle-freeness
(with one-sided error). In this case, whenever the tester rejects,
its partial view must contain a cycle. Thus, {\em any one-sided tester
of cycle-freeness may be used for finding cycles in graphs that
are far from being cycle-free}. A similar observation applies
to finding $T$-minors, for any fixed tree $T$.

We mention that in most of the property testing literature,
one-sided error is viewed as a secondary feature
that some testers have and others may lack.
The foregoing connection demonstrates the fundamental
advantage of one-sided error testers over standard
(two-sided error) testers.
(Other advantages are discussed in \SAref{fur-refl:one-sided}.)

Lower bounds on the complexity of one-sided error testers
that significantly exceeds the performance guarantees of
known two-sided error testers have been observed,
starting with~\cite[Sec.~10.1.6]{GGR}.
However, so far, no study has been devoted to providing
a {\em one-sided error tester of optimal complexity,
in the case where this complexity significantly exceeds
that of the corresponding two-sided error tester}. %

To the best of our knowledge, the text that seems closest
to addressing this issue is the discussion in~\cite[Sec.~2]{AS:sat}
that refers to the complexity of testing $K_{t,t}$-freeness
in the {\em adjacency matrix model}\/ (introduced in~\cite{GGR}).
Specifically,~\cite[Clm.~2.2]{AS:sat}
asserts a two-sided tester of $K_{t,t}$-freeness
having query complexity $O(1/\e)$,
whereas~\cite[Clm.~2.3]{AS:sat} (combined with~\cite[Thm.~2]{GT})
asserts that one-sided error testing of $K_{t,t}$-freeness
requires $\Omega(\e^{-t/4})$ queries.
As noted at the end of~\cite[Sec.~2]{AS:sat},
this is tight up to a polynomial function
(i.e., there exists two-sided tester of $K_{t,t}$-freeness
having query complexity $\e^{-O(t)}=\poly(\e^{-t/4})$).
It is telling that~\cite[Sec.~2]{AS:sat} leaves
the complexity of one-sided error testing undetermined
(at the ``polynomial slackness'' level).
Indeed, like other prior works that address the complexity
of one-sided error testers, their interest is in demonstrating the gap
between the complexities of two-sided and one-sided error testing,
and not in determining the latter.

In contrast, our work is aimed at providing
one-sided error testers of (almost) optimal complexity,
in cases in which this complexity significantly exceed
the complexity of the corresponding two-sided error tester.
For example, recall that Goldreich and Ron
provided a two-sided error tester for cycle-freeness
of $\poly(1/\e)$ query complexity~\cite[Thm.~4.2]{GR1},
where $\e$ denotes the desired proximity parameter
(i.e., the tester distinguishes cycle-free graphs from graphs
that are $\e$-far from being cycle-free).
In contrast,~\cite[Prop.~4.3]{GR1} asserts that cycle-freeness
has no one-sided error tester that makes $o(\sqrt{N})$ queries
(even for $\e=1/3$),
where $N$ denotes the number of vertices in the input graph.
In that context, \thmref{intro:find-cycle.thm} is equivalent to

\BT{\em(one-sided error tester for cycle-freeness):}
\label{intro:c3.thm}
Cycle-freeness of constant degree $N$-vertex graphs can be tested
with one-sided error
within time complexity $\tildeO(\poly(d/\e)\cdot\sqrt{N})$.
Furthermore, whenever the tester rejects, it outputs
a simple cycle of length $\poly(\e^{-1}d\log N)$.
\ET
Indeed, by the foregoing discussion,
whenever the tester asserted in \thmref{intro:c3.thm} rejects,
it is the case that it explored a subgraph that is not cycle-free.
Moreover, the furthermore clause of \thmref{intro:c3.thm}
asserts that in this case the explored subgraph actually contains
a simple cycle of length $\poly(\e^{-1}d\log N)$.
Thus, \thmref{intro:c3.thm} implies \thmref{intro:find-cycle.thm}.
Similarly, {\em Theorem~\ref{intro:c3.thm} extends to
testing $C_k$-minor freeness, for any $k>3$,
which in turn is equivalent to \thmref{intro:find-Ck.thm}}.
And, similarly, \thmref{intro:find-tree.thm} is equivalent
to the existence of a {\em tester for $T$-minor freeness of
query complexity that only depends on the proximity parameter,
for any tree $T$}.

\subsection{Techniques}\label{intro:techniques}
As stated at the end of Section~\ref{intro:finding.sec},
all our results are obtained via the study of the complexity
of one-sided error testers for the corresponding properties.

Our testers for $C_k$-minor freeness are all obtained by local reductions.
Specifically, our cycle-freeness (i.e., $C_3$-minor freeness) tester
is obtained by a {\em randomized}\/ reduction to testing bipartiteness,
whereas our $C_k$-minor freeness tester is obtained by a
{\em deterministic}\/ reduction to testing cycle-freeness.

\subsubsection{Testing cycle-freeness}
We mention that
the two-sided error cycle-freeness tester of~\cite{GR1}
does not even try to find a simple cycle.
It just estimates the number of edges
in the graph and rejects if this estimate exceed the number of edges
that correspond to any forest that spans the set of connected
components of the graph.\myfoot{Note that any cycle-free graph
is a forest, and if the number of trees in this forest is $t$,
then the difference between the number of vertices and
the number of edges in the graph equals $t$.
The two-sided error tester of~\cite{GR1} estimates the number
of edges and the number of connected components in the graph,
and conducts the adequate computation. The number of connected
components is estimated by the number of connected components
that have $O(1/\e)$ vertices, whereas the latter number
is approximated by exploring the neighborhood of a few
randomly selected vertices.}
We also mention that as observed by Bollob\'as and
Thomason~\cite[Thm.~5]{BT97},
a ``girth versus edge-density'' lower bound
implies that any graph $G=([N],E)$
that is $\e$-far from being cycle-free (and hence
contains $N+\Omega(\eps N)$ edges)
must have a simple cycle of length 
$O(\log N + 1/\eps)$.
The problem, however, is finding such a cycle in sublinear time.

Our one-sided error tester of cycle-freeness finds a cycle
in the original graph by randomly reducing this problem
to the problem of finding an odd-length cycle in
an auxiliary graph. Specifically, the input graph $G=([N],E)$
is randomly transformed into an auxiliary graph such that
each edge $e\in E$ is replaced, with probability~$1/2$
by a 2-vertex path (with an auxiliary vertex),
and remains intact otherwise.
Thus, with probability~$1/2$,
each cycle in $G$ is transformed into an odd-length cycle.
Furthermore, we show that
{\em if $G$ is $\e$-far from being cycle-free, then}\/
(w.h.p.) {\em the resulting graph is $\Omega(\e)$-far from
being bipartite}.

A crucial feature of the foregoing randomized reduction
is that it is local in the sense that each operation
on the transformed graph can be implemented by a constant number
of operations on the original graph. Thus, we can emulate the
execution of a bipartite tester (i.e., the one of~\cite{GR2})
on the transformed graph.
This allows us to establish \thmref{intro:c3.thm}.

\subsubsection{Testing $C_k$-minor freeness, for any $k>3$}
Recall that the set of $C_k$-minor-free graphs coincides with
the set of graphs that have no simple cycle of length at least~$k$.
Theorem~\ref{intro:find-Ck.thm} is proved by a (local) reduction
of testing $C_k$-minor-freeness to testing cycle-freeness.
For example, in the case of $k=4$ we replace each triangle
by a 3-vertex star; that is, we omit the original edges
of this triangle, and introduce an auxiliary vertex that
is connected to the three corresponding vertices.
We then prove that if the original graph
is $C_4$-minor-free then the resulting graph is cycle-free,
whereas {\em if the original graph is $\e$-far from
being $C_4$-minor-free then the resulting graph
is $\Omega(\e)$-far from being cycle-free}.

For larger values of $k$,
a more sophisticated local replacement is used;
that is, replacing all small cycles by auxiliary
vertices will not do.
To illustrate the difficulty of dealing with $k>4$,
note that, unlike in the case $k=4$, a $C_k$-minor free
graph may contain cycles of length smaller than $k$
that share some common edges, and so the simple
replacement will not yield a cycle-free graph.%

\subsubsection{Testing $H$-minor freeness, for any cycle-free $H$}

The main challenge for this problem is testing $T$-minor freeness,
where $T$ is an arbitrary tree.
The simple case in which $T$ is a $k$-vertex star, for some $k\geq2$,
provides a good illustration to the underlying main idea.
In this case we may select a random vertex and start a
Breadth First Search (BFS)
at this vertex, stopping whenever
either we encounter a layer with at least $k$ vertices
or we explored more than $4k/\e$ layers
(or we explored the entire connected component).
In the first case, we found a desired minor
and can safely reject, whereas in the second case
we found a set of at least $4k/\e$ vertices that is separated from
the rest of the graph by less than $dk$ edges.
Thus, if the graph $G=([N],E)$ contains
at least $(1-\e/4)\cdot N$ start vertices
that do not lead the algorithm to reject,
then $G$ can be decomposed to connected components
that are each $T$-minor free by omitting at most $\e d N/2$ edges
(i.e., the edges that are incident at the $\e N/4$ exceptional
vertices and the edges of the aforementioned small cuts).

The case of a general tree $T$ is much more complex,
but the governing principle remains a tight relation between
having few start vertices that contain a $T$-minor at their
vicinity and the ability to decompose the graph to connected
components with few edges between them. This relation is captured
by the following result, which may be of independent interest.

\BT{\em(``local expansion'' and tree minors):}
\label{intro:minor-cuts.thm}
For every $d$ and $k$ there exists an $r=r(d,k)$
such that if the $r$-neighborhood of a vertex $s$
in a graph of degree bound $d$ does not contain
a $T$-minor of some tree $T$ with at most $k$ vertices,
then this neighborhood contains a set $S$ that is separated from
the rest of the graph by less than $\e d|S|/4$ edges.
\ET
In other words,
if all ``sub-neighborhoods'' of the $r$-neighborhood of $s$
are ``expanding'' (i.e., are not separated from the rest by small cuts),
then this $r$-neighborhood contains a $T$-minor of every tree $T$
with at most $k$ vertices.
(We mention that the problem of finding small trees in locally
expanding graphs has been studied before (cf., e.g.~\cite{FP}).
However, our Theorem~\ref{intro:minor-cuts.thm} seems incomparable,
since we seek specific {\em tree minors}\/ rather than specific trees,
whereas our expansion condition is very weak.)

Finally, we reduce finding $H$-minors, where $H$ is an arbitrary
cycle-free graph (i.e., a forest), to finding disjoint tree minors.
Again, the reduction is local,
and in this case it is almost straightforward,
where the subtlety is related to the fact that
we refer to one-sided error.
Specifically, if $H$ consists of the connected components $H_1,\dots,H_m$,
then it does not necessarily hold that $G$ is $H$-minor free
if and only if $G$ is $H_i$-minor free for all $i\in[m]$.
Still, this is ``almost true'' and so a small modification
of the straightforward reduction will do.

\subsection{Another perspective: Finding arbitrary forbidden minors}
Our results may be viewed as
progress in resolving an open problem,
posed by Benjamini, Schramm, and Shapira~\cite{BSS},
that refers to one-sided error testing of $H$-minor-freeness,
for any finite graph $H$ (or even a finite family of such graphs).
%
Specifically, Benjamini \etal~\cite{BSS} proved that, for any $H$,
the property of being $H$-minor-free can be tested
within query complexity that only depends on the proximity parameter,%
\myfoot{The query complexity obtained in~\cite{BSS} is
triple-exponential in~$1/\e$. The complexity was 
improved to exponential in~$1/\e$~\cite{HKNO}.}
{\em when allowing two-sided error}.
They conjectured that for any non-forest $H$,
there exists an $H$-minor-freeness tester
with query complexity $O(\sqrt{N})$.
%
%
%
Viewed from that perspective,
our results prove the aforementioned conjecture
in the special case of $H=C_k$, for every $k\geq3$.

%
We note that finding cycles seems the ``hard'' part of finding minors;
that is, cycles are the source of the $\Omega(\sqrt{N})$ query lower bound.
Specifically,
recall that~\cite[Prop.~4.3]{GR1} establishes an $\Omega(\sqrt{N})$
query lower bound for any algorithm that finds $C_3$-minors
(or, in other words, a one-sided property tester for cycle-freeness).
In~\cite{BSS} it was suggested that this lower bound can
be deduced by adapting the lower bound argument from~\cite{GR1}.
We present a proof of this fact, thus establishing
{\em an $\Omega(\sqrt{N})$ query lower bound for any algorithm
that finds minors that contain cycles}.
Recall that this stands in contrast to \thmref{intro:find-tree.thm}
(which asserts that finding cycle-free minors can be done in a number
of queries that is independent of the size of the graph).

\paragraph{A wider perspective on finding forbidden minors.}
The first result dealing
with graph minors is the well known
Kuratowski-Wagner theorem~\cite{K30,W37} that states that
any non-planar graph contains a $K_5$ or $K_{3,3}$ minor.
Consider a property $\P$ such that if $G \in \P$,
then, for any minor $H$ of $G$, it holds that $H \in \cP$.
Such a property is {\em minor-closed}.
It was conjectured by Wagner
that for \emph{any} minor-closed property $\cP$,
there is a finite set of graphs $\cH_\cP$ such that
$G \in \cP$ if and only if $G$ is $H$-minor free, for all $H \in \cH_\cP$.
Robertson and Seymour had a long series of papers,
which culminated in the proof of this conjecture~\cite{RS:20},
called the Graph-Minor Theorem.
{From} an algorithmic perspective, one of the milestones in this
series was a polynomial time algorithm that checked $H$-minor freeness,
for any (constant-size) graph $H$~\cite{RS:13}.
(We will use a recent improvement
on that by Kawarabayashi, Kobayashi, and Reed~\cite{KaKoRe},
which gives a quadratic time algorithm for this problem.)

It is natural to consider a sublinear variant of
the above algorithmic question; that is, given a graph $G$
that is far from being minor-free, and we find an $H$-minor by looking
at a sublinear portion of the graph?
An affirmative answer would, in particular, imply that such a graph
contains sublinear sized $H$-minors, which is an interesting
combinatorial conjecture. Needless to say, this paper provides
an affirmative answer in the special case that $H$ is a cycle.


\subsection{Further reflections regarding one-sided error}
\label{fur-refl:one-sided}
The relative power
of two-sided versus one-sided error
randomized decision procedures has been the focus
of considerable study in many settings,
including in the context of property testing.
Indeed, in any setting, one-sided error procedures
offer the advantage of never rejecting yes-instances.
However, as we already saw in \secref{intro:PT-connect.sec},
this advantage has a special appeal in the context of property testing,
since it yields algorithms for very efficiently finding some desired
structures (whenever the graph is far from being ``free of them'').
Additional benefits of one-sided error testers are discussed next.

Firstly, we note that property testing is asymmetric in nature:
It refers to distinguishing objects that {\em perfectly}\/ satisfy
a predetermined property from objects that are {\em far}\/ from
satisfying this property.
Indeed, property testing is a relaxation of the original decision task
(which refers to distinguishing objects that satisfy the property
{from} objects that do not satisfy it),
where the relaxation is applied to one type of instances
but not to the other. In this context, it is natural
to apply the probabilistic relaxation also
to one type of instances (i.e., the far-away instances)
but not to the other.

Secondly, we note that one of the main applications of property testers
is their potential use as a preliminary ``fast but crude'' decision step,
which when coupled with an exact (but slow) decision procedure
yields a procedure that is always correct and often very fast.
That is, we envision using a property tester as a ``sieve''
that rejects ``on the spot'' (i.e., ``fast'') very bad instances
(i.e., those that are far from satisfying this property),
while passing the rest of the instances for further examination.
In such a context, we can afford passing very bad instances
for further examination (since all this means is a waste of time),
but we cannot afford not passing a good instance.

Lastly, we consider the relationship between property testing
and local structures in the tested property. Intuitively,
the existence of a property tester means that a global structure
(i.e., distance of the object to the property)
is reflected in (or co-related with) a local structure
(i.e., the part of the object being probed by the tester).
In the general case (of two-sided error),
this co-relation is statistical,
whereas in the case of one-sided error this correlation
is actually a (``robust'') characterization.

The last aspect is particularly clear in the current study.
Firstly, the notion of local structure is most appealing in the
bounded-degree model, where it refers to graph neighborhoods.
Secondly, the different types of local structures underlying
the two-sided and one-sided error testers is most striking in the
case of cycle-freeness. The two-sided error tester of~\cite{GR1}
relies on the fact that distance from cycle-freeness
in connected graphs is reflected by the difference
between the number of edges and the number of vertices,
whereas these numbers can be estimated (with two-sided error)
by sampling the graph's vertices.
Note that such {\em estimates}\/ cannot yield a characterization
(let alone a robust one) of the cycle-free graphs.
In contrast, our one-sided error tester relies on the fact
that distance from cycle-freeness is reflected in the density
of short simple cycles in the graph, whereas such cycles
can be found by an appropriate randomized exploration of the graph.
Indeed, this yields a (robust) characterization of the set of
cycle-free graphs (i.e., a graph is cycle-free if and only if it contains no
simple cycle, and the farther the graph is from being cycle-free
the shorter and more abundant these cycles are).


\subsection{The general (unbounded-degree) graph model}
\label{intro:general-model}
Although our upper bounds (e.g., \thmref{intro:find-cycle.thm})
state the dependence of the complexities on the degree bound, $d$,
so far we thought of $d$ as being a constant
(or at least as being extremely small in comparison to $N$).
Indeed, an upper bound as stated in \thmref{intro:find-cycle.thm}
(i.e., an arbitrary polynomial dependence on $d$)
is not meaningful, when $d=N$ (or even $d=\sqrt N$).
Nevertheless, it is possible to obtain a better result
than stated in \thmref{intro:find-cycle.thm}~--
specifically, eliminate the dependence on $d$.
That is, there exists a randomized algorithm that,
on input an $N$-vertex graph $G$ of degree bound $d$
that is $\e$-far from being cycle-free,
finds a simple cycle (of length $\poly(\e^{-1}\log N)$) in $G$
in expected time $\tildeO(\poly(1/\e)\cdot\sqrt{N})$.

The foregoing algorithm can be extended to the general graphs model
(i.e., the model in~\cite{PR}),
where distances are measured with respect to the actual number of edges
(see \SAref{general-model}).%
\myfoot{Algorithms in this model use the same type of incidence
queries as in the main (bounded-degree) model we consider.
The difference is that a graph $G=([N],E)$ is said to
be $\e$-far from $H$-minor-freeness if $2\e|E|$ edges
(rather than $\e  d N$ edges) must be removed from $G$
in order to obtain an $H$-minor-free subgraph.
The point is that the number of edges is related
to the average degree of $G$ rather than to its
degree (upper) bound, which may be significantly smaller.
Thus, distances under this model are possibly larger,
and thus the testing requirement is possibly harder.}
%
This follows by an alternative presentation of the basic
randomized reduction, which may be viewed as reducing
cycle-freeness to a generalization of 2-colorability.
In this generalization, edges of the graph are labeled
by either $\tt eq$ or $\tt neq$, and a legal 2-coloring
(of the vertices) is one in which every two vertices
that are connected by an edge labeled $\tt eq$ (resp. $\tt neq$)
are assigned the same color (resp., opposite colors).
We observe that the (one-sided error) Bipartite testers
of~\cite{GR2,KKR} extend to this generalization of 2-colorability.

We mention that analogous extensions do not work
for testing $C_k$-minor freeness, for $k>3$,
nor for testing tree-minor-freeness.
In fact, in the general graph model,
it is not possible to find tree-minors
(or even test tree-minor freeness with two-sided error)
by using $o(\sqrt{N})$ queries.

\subsection{Organization}
Section~\ref{prelim.sec} contains a formal statement
of the relevant definitions and terminology.
The testers of $C_k$-minor freeness are presented
in Sections~\ref{C3:section}--\ref{Ck:section}.
Our first result (i.e., the one-sided error tester of cycle-freeness)
is presented in Section~\ref{C3:section}.
The reduction of testing $C_k$-minor freeness to testing cycle-freeness
is presented in Section~\ref{Ck:section}, but Section~\ref{C4:section}
provides an adequate warm-up by treating the case of $k=4$.

In Section~\ref{lower-bound.sec}, we prove the lower bound claimed
in~\cite{BSS} regarding the query complexity of one-sided error
testing $H$-minor freeness, when $H$ contains a cycle.
In contrast, in Section~\ref{tree.section} we consider the
case that $H$ is cycle-free, and present the improved
testers for $H$-minor freeness in this case
(i.e., when $H$ is a forest).

Finally, in Section~\ref{general-model}
we consider the unbounded-degree model,
discussed in \secref{intro:general-model},
and in Section~\ref{open:section} we compile a list
of open problems that are scattered throughout the paper.

\section{Preliminaries}\label{prelim.sec}
This work refers mainly to the bounded-degree model
(introduced in~\cite{GR1}).
The only exception is Section~\ref{general-model},
where we consider the unbounded-degree model,
also discussed in \secref{intro:general-model}.
%
The bounded-degree model refers to a fixed degree bound, denoted $d$,
where a tester is given oracle access to an
$N$-vertex graph $G=([N],E)$ of maximum degree $d$.
Specifically, for any $v \in [N]$ and $1 \leq i \leq d$,
the tester can ask for the $i^{\rm th}$ neighbor of
vertex $v$. If $v$ has less than $i$ neighbors, then the
answer returned is $0$ (and no assumption is made on the
order of the neighbors of any vertex).

\ifnum\submit=0
\BD{\em(testers in the bounded-degree model):}
\else
\BD[testers in the bounded-degree model]
\fi
\label{testing.def}
Let $d\in\N$ be fixed and $\Pi$ be a property of graphs
with maximum degree at most $d$.
We denote the restriction of $\Pi$ to $N$-vertex graphs by $\Pi_N$.
A randomized oracle machine $T$ is called a {\dsf tester for $\Pi$}
if the following two conditions hold:
\BE
\item For every $N\in\N$ and $\e\in[0,1]$,
on input $(N,\e)$ and when given oracle access to any $G\in\Pi_N$
the machine $T$ accepts with probability at least~$2/3$;
that is, $\prob[T^G(N,\e)=1]\geq2/3$.
\item For every $N\in\N$ and $\e\in[0,1]$,
and every $N$-vertex graph $G$ that is $\e$-far from $\Pi_N$,
it holds that $\prob[T^G(N,\e)=1]\leq1/3$,
where {\dsf $G=([N],E)$ is $\e$-far from $\Pi_N$}
if for every $G'=([N],E')\in\Pi_N$ it holds that
the symmetric difference of $E$ and $E'$
contains more than $\e\cdot d N$ elements.
\EE
In case the first condition holds with probability~1,
we say that $T$ has {\dsf one-sided error}.
Otherwise, we say that $T$ has {\dsf two-sided error}.
\ED
Throughout our study,
the degree bound $d$ is a constant,
and sometimes O/Omega-notions hide a dependence on $d$.
The query and time complexities of testers are stated as functions
of the graph size, $N$, and the proximity parameter, $\e$.
When discussing time complexity,
basic vertex-manipulation operations are counted at unit cost.
We may assume without loss of generality that $d \geq 3$,
where in order to obtain a result for $d=2$, we can simply
run the algorithm with $d=3$ and a proximity parameter of $2\eps/3$
(and for $d=1$ all problems become trivial).


\paragraph{Notation.}
For a graph $G=([N],E)$, we denote the set of neighbors of $v\in [N]$
(in $G$) by $\Gamma_G(v)$;
that is, $\Gamma_G(v)=\{u\!\in\![N]:\{u,v\}\!\in\!E\}$.

\paragraph{Terminology.}
By a {\tsf cycle} in a graph $G=([N],E)$
we mean a sequence of vertices $(v_1,\ldots,v_t,v_{t+1})$
such that $v_1=v_{t+1}$ and for every $i\in[t]$
it holds that $\{v_i,v_{i+1}\}\in E$;
that is, $(u,v,w,v,u)$ (or even $(u,v,u)$) is considered a cycle.
A {\tsf simple cycle} is a cycle as above
in which $t \geq 3$ and $|\{v_i:i\in[t]\}| = t$.

\paragraph{A useful bound.}
For any positive integer $a$ and fraction $0 < \alpha < 1/2$
we have:
\BEQ\label{ent-bound.eq}
{a \choose \alpha a} \;<\; 2^{H_2(\alpha)\cdot a}
\EEQ
where $H_2(\alpha) = \alpha\log(1/\alpha) + (1-\alpha)\log(1/(1-\alpha))$
is the binary entropy function.
\section{Testing Cycle-Freeness}
\label{C3:section}
As stated in the introduction, we reduce testing cycle-freeness
to testing bipartiteness. Recall that we consider bounded-degree
graphs, where the degree bound $d$ is assumed to be a constant
(for the general case, see \SAref{general-model}).
The reduction is randomized
and local so that operations in the resulting graph are easily
implemented via operations in the original graph. Wishing to
avoid a general definition of (randomized) local reductions,
we explicitly present the tester obtained by it.

For a fixed graph $G=([N],E)$ and function $\tau:E\to\{1,2\}$,
we denote by $G_\tau$ the graph obtained from $G$
by replacing each edge $e\in E$ such that $\tau(e)=2$
by a 2-edge path (with an auxiliary intermediate vertex).
Each edge $e\in E$ such that $\tau(e)=1$
remains an edge in $G_\tau$.
That is, the graph $G_\tau=(V_\tau,E_\tau)$ is defined as follows:
\BEQN
V_\tau &\eqdef& [N]\cup\{a_e:e\in E \wedge \tau(e)=2\} \\
E_\tau &\eqdef& \{e:e\in E \wedge \tau(e)=1\}
             \cup\{\{u,a_e\},\{a_e,v\}:e=\{u,v\}\in E \wedge \tau(e)=2\}
\EEQN
Note that $|V_\tau|\leq(d+1)\cdot N$
and that $G_\tau$ preserves the degree bound, $d$.
We first establish the next lemma concerning features of
the transformation from $G$ to $G_\tau$, and later turn to
discuss the tester in detail.

\BL{\em(analysis of the randomized transformation):}
\label{cycle-free.ana}
\BE
\item
If $G$ is cycle-free, then,
for every choice of $\tau:E\to\{1,2\}$,
the graph $G_\tau$ is bipartite.
\item
If $G$ is not cycle-free, then, with probability
at least $1/2$ over the random choice of $\tau:E\to\{1,2\}$,
the graph $G_\tau$ is not bipartite.
\item
There exist universal constants $c_1 > 1$ and $c_2,c_3>0$
such that, for every $\eps \geq c_1/(d N)$,
if $G$ is $\e$-far from being cycle free, then, with probability
at least $1-\exp(-c_2\e dN)$ over the random choice of $\tau:E\to\{1,2\}$,
the graph $G_\tau$ is $c_3\cdot\e/2d$-far from being bipartite.
\EE
\EL

\medskip
\BPF
The first item follows from the fact that
if $G$ is cycle-free, then, for every $\tau:E\to\{1,2\}$,
the graph $G_\tau$ is also cycle-free, and thus bipartite.
The second item follows by observing that any cycle in $G$
is transformed with probability~$1/2$ to an odd-length cycle in $G_\tau$.
Turning to the last item, we consider an arbitrary graph $G$
that is not cycle-free.
Denoting by $\Delta$ the actual number of edges (not its fraction)
that should be omitted from $G$ in order to obtain a cycle-free graph,
we shall show the following.
For $\Delta$ that is at least some constant (i.e., $\Delta\geq c_1$),
with probability  $1-\exp(-\Omega(\Delta))$,
the number of edges that should be omitted from $G_\tau$ in order to
obtain a bipartite graph is $\Omega(\Delta)$.
(Note that the second item in the lemma holds
for any $\Delta\geq 1$, which may be below this constant.)

We start by considering the case that the graph $G$ is connected.
We later address the case in which $G$ contains more than one connected
component.
We may assume without loss of generality
that $G$ has no vertices of degree~1,
since removing such vertices maintains the value of $\Delta$
(i.e., the absolute distance from being cycle-free)
as well as (the distribution of) the number of edges
that have to be removed to make $G_\tau$ bipartite.
We also observe that except in the case that $G$ is a simple cycle,
which is covered by the second item in the lemma,
we may assume that there are no vertices of degree~2.
This is true since we can contract paths that only contain intermediate vertices
of degree~2 to a single edge, while again preserving $\Delta$
as well as (the distribution of) the number of edges that
have to be removed to make $G_\tau$ bipartite.
The latter assertion follows from the fact
that the distribution of the parity of the path-lengths in $G_\tau$
is maintained (i.e., both the original path and the contracted path
in $G_\tau$ have odd/even length with probability $1/2$).
We also mention that the contracted graph $G$ may contain self-loops
and parallel edges, but the rest of the argument holds in this case too.
We stress that the contracted graph is merely a mental experiment
for proving the current lemma.

In light of the foregoing,
we consider a connected graph $G=([N],E)$,
which may have self-loops and parallel edges,
in which each vertex has degree at least~3.
It follows that $\Delta=|E|-(N-1)>N/2$.
We shall prove that,
with high probability over the choice of $\tau$,
for some constant $c_3>0$,
more than $c_3\cdot\Delta=c_3\e dN$ edges must be omitted from
the graph $G_\tau$ in order to obtain a bipartite graph.
Since the number of vertices in $G_\tau$ is upper bounded by $(d+1)N$
(and its degree bound is $d$),
we get that $G_\tau$ is at least $(c_3\eps/2d)$-far from bipartite,
since $\frac{c_3\e dN}{d(d+1)N}>\frac{c_3\eps}{2d}$.

For each $E'\subset E$ of size $c_3\Delta$,
let $G'_\tau$ denotes the subgraph of $G_\tau$
obtained by applying the foregoing randomized reduction to
the graph $G'=([N],E\setminus E')$ rather than to $G=([N],E)$.
We consider the probability that $G'_\tau$ is bipartite.
Note that $G_\tau$ is at (absolute) distance at most $c_3\Delta$ from
being bipartite if and only if there exists a set $E'$ of size $c_3\Delta$
such that $G'_\tau$ is bipartite.
%
\ifnum\RedAna=1 
We prove below that, {\em for a uniformly distributed $\tau$,
the probability that $G'_\tau$ is bipartite
is at most $2^{-(\Delta-|E'|)}$}.
Using this claim, we have
\begin{eqnarray*}
p &\eqdef& \Prob_\tau[\mbox{\rm $\exists E'\subset E$
such that $|E'|=c_3\Delta$ and $G'_\tau$ is bipartite}] \\
&\leq& \sum_{E'\subset E:\,|E'|=c_3\Delta}
       \Prob_\tau[\mbox{\rm $G'_\tau$ is bipartite}] \\
&\leq& {{|E|}\choose{c_3\Delta}}\cdot2^{-(\Delta-c_3\Delta)}
\end{eqnarray*}
Substituting $|E|$ by $(N-1)+\Delta$ and using $\Delta\geq N/2$
(and $c_3<1/2$), we get
\else 
Thus, the probability that $G_\tau$ is at distance at most $c_3\Delta$
{from} being bipartite is given by
\begin{eqnarray*}
p &\eqdef& \Prob_\tau[\mbox{\rm $\exists E'\subset E$
such that $|E'|=c_3\Delta$ and $G'_\tau$ is bipartite}] \\
&\leq& \sum_{E'\subset E:\,|E'|=c_3\Delta}
       \Prob_\tau[\mbox{\rm $G'_\tau$ is bipartite}] \\
&\leq& {{|E|}\choose{c_3\Delta}}\cdot 2^{N-1}\cdot2^{-(|E|-c_3\Delta)}
\end{eqnarray*}
where the second inequality is due to considering
all possible 2-partitions of $[N]$,
and noting that for each edge $e$ in $E\setminus E'$
and each 2-partition $\pi$, with probability~$1/2$ over
the choice of $\tau(e)\in\{1,2\}$, the partition $\pi$
is inconsistent with the value of $\tau(e)$.
(In such a case we say that $e$ violates the 2-partition $\pi$.)
Specifically, if $\pi(u)=\pi(v)$ and $\tau(\{u,v\})=1$,
then the edge $\{u,v\}$ violates the 2-partition $\pi$,
and ditto if $\pi(u)\neq\pi(v)$ and $\tau(\{u,v\})=2$.
Note that the hypothesis that $G$ is (connected and is)
at (absolute) distance $\Delta$ from being cycle-free
implies that $|E|=(N-1)+\Delta$.
Now,
substituting $|E|$ by $(N-1)+\Delta$,  using $\Delta\geq N/2$
and Equation~(\ref{ent-bound.eq}) we get
\fi
\begin{eqnarray*}
p &\leq& {{N-1+\Delta}\choose{c_3\Delta}}\cdot 2^{-(\Delta-c_3\Delta)} \\
&<& {{3\Delta}\choose{c_3\Delta}}\cdot 2^{-(1-c_3)\Delta} \\
&<& 2^{H_2(3 c_3)\cdot 3\Delta - (1-c_3)\Delta}
\end{eqnarray*}
which vanishes exponentially in $\Delta$ provided
that $c_3>0$ is a sufficiently small constant.
\ifnum\RedAna=1 
Thus it remains to prove that (for a uniformly distributed $\tau$)
the probability that $G'_\tau$ is bipartite
is at most $2^{-(\Delta-|E'|)}$.
This follows by viewing $G'_\tau$ as a result of
applying the randomized reduction to the graph $G'=([N],E\setminus E')$,
noting that at least $\Delta'=\Delta-|E'|$ edges must be
omitted from $G'$ in order to obtain a cycle-free graph,
and proving the following.

\medskip\noindent{\sf Claim:}
Suppose that $k$ edges must be omitted from the graph $G'$
in order to make it cycle-free.
Then, the probability that $G'_\tau$ is bipartite
is at most $2^{-k}$.

\medskip\noindent{\sf Proof:}
Consider a spanning forest $F$ of $G'$,
and note that $G'$ must contain $k$ non-tree edges
(i.e., edges not in the forest).
Now, consider selecting a random $\tau$ in two stages
such that in the first stage the value of $\tau$ is determined
on the tree edges, and in the second stage $\tau$ is determined
for the non-tree edges. Then, for any possible setting
of the first stage, there exists a unique setting of the
non-tree edges that makes the resulting graph bipartite.
This is the case since the $\tau$-values of tree edges
(fixed in Stage~1) determine a bi-partition of each connected component,
and for each non-tree edge $e$ there is a unique value of $\tau(e)$
that fits this bi-partition. The probability that this unique
setting of all non-tree edges is selected in Stage~2 equals $2^{-k}$.
\fi

It remains to address the case in which $G$ is not connected.
Let $C_1,\ldots, C_t$ be the connected components of $G$ where
$t>1$. For each $1 \leq i \leq t$, let $N_i$ be the
number of vertices in $C_i$, and let $\Delta_i$ be the number
of edges that should be removed from $C_i$ in order to make
it cycle-free. Thus, $\sum_{i=1}^t N_i = N$ and
$\sum_{i=1}^t \Delta_i = \Delta$.
Let $\tau_i$ be the restriction of $\tau$ to the edges
in $C_i$, let $C_{i,\tau}$ be the graph obtained by
applying the transformation defined by $\tau_i$ to $C_i$,
and let $\Delta_{i,\tau}$ be the number of edges that
should be removed from $C_i$ to make it bipartite.

By applying the argument detailed above to each $C_i$
separately, we get that
$\Pr[\Delta_{i,\tau} < c_3 \Delta_i] \leq 2^{-c_4\Delta_i}$
(for constants $0 < c_3,c_4 <1$).
We would like to infer that
\BEQ\label{sumDeltas.eq}
\Pr\left[\sum_{i=1}^t \Delta_{i,\tau} < c'_3 \sum_{i=1}^t \Delta_i\right]
   \leq \exp\left(-\Omega\left(\sum_{i=1}^t \Delta_i\right)\right)
= \exp(-\Omega(\Delta))\;
\EEQ
for some constant $c_3'$.
To this end, for each $C_i$ we define $m_i = c_3 \Delta_i$
independent $0/1$ random variables, $X_{i,1},\ldots,X_{i,m_i}$,
such that $\Pr[X_{i,j}=1] =  2^{-c_4/c_3}$.
Observe that $\Pr[\sum_{j=1}^{m_i} X_{i,j} \leq m_i]=1$
and $\Pr[\sum_{j=1}^{m_i} X_{i,j} = 0] = 2^{-c_4\Delta_i}$.
This implies that
$\Pr[\sum_{j=1}^{m_i} X_{i,j} < \beta]
   \geq \Pr[\Delta_{i,\tau} < \beta]$
for every threshold $\beta$,
which means that the random variables $\sum_{j=1}^{m_i} X_{i,j}$
and $\Delta_{i,\tau}$ can be coupled (i.e., defined over
the same sample space) such that the value of the first
is always upper bounded by the value of the second.
Hence, in order to prove Equation~(\ref{sumDeltas.eq}),
it suffices to bound the probability that
$\sum_{i=1}^t\sum_{j=1}^{m_i} X_{i,j}
  < c'_3  \sum_{i=1}^t \Delta_i = c_3' \Delta$.
But since these ($\sum_{i=1}^t m_i = c_3\Delta$)
 random variables are independent,
we can apply a multiplicative Chernoff bound,
which gives us that the probability that
$\sum_{i=1}^t\sum_{j=1}^{m_i} X_{i,j} <  c_3' \Delta$
for $c_3' = c_3 \cdot  2^{-c_4/c_3 - 1}$ (i.e., half the
expected value of the sum), is $\exp(-\Omega(\Delta))$.
\EPF

\paragraph{The Tester For Cycle-Freeness.}
The tester emulates the execution of the bipartiteness testing
algorithm~\cite{GR2} on $G_\tau$ by performing queries to $G$.
We next state the main theorem proved in~\cite{GR2}.

\BT{\em\cite{GR2}} \label{gr-bip:thm}
There exist an algorithm {\sf Test-Bipartite}
for testing bipartiteness of bounded-degree graphs
whose query complexity and running time are
 $\poly((\log \tilde{N})/\tilde{\eps})\cdot \sqrt{\tilde{N}}$
where $\tilde{N}$ denotes the number of vertices in the graph
and $\tilde{\eps}$ is the given proximity parameter.
The algorithm uniformly selects random vertices and performs random
walks from them.
Whenever the algorithm rejects a graph it outputs a {\em certificate\/}
to the non-bipartiteness of the graph in form of an odd-length cycle
of length $\poly(\tilde{\eps}^{-1}\log \tilde{N})$.
\ET
As stated in Theorem~\ref{gr-bip:thm},
algorithm  {\sf Test-Bipartite} performs two types of operations:
(1)~selecting a vertex uniformly at random,
and (2)~taking random walks by querying vertices on their neighbors.%
\footnote{Actually, {\sf Test-Bipartite} requires
also a rough estimate of the number of vertices in the graph,
since such an estimate is used to determine a couple of parameters
(i.e., the number of random walks performed and their length).
It is clear that our reduction provides such an estimate,
since $|V_\tau|=\Theta(N)$.}
Thus the execution of the cycle-freeness tester boils down
to emulating these operations, as described next.

\BA{\em(the cycle-freeness tester):}
\label{cycle-free.alg}
Given input graph $G=([N],E)$,
the tester selects uniformly at random a function $\tau:E\to\{1,2\}$
and invokes {\sf Test-Bipartite} on the graph $G_\tau$ with
the proximity parameter set to $c_3\eps/2d$
{\em(where $c_3$ is the constant
from the last item in Lemma~\ref{cycle-free.ana})},
emulating its operations as follows.%
\footnote{Actually, the function $\tau:E\to\{1,2\}$
is selected on-the-fly; that is, whenever the tester
needs the value of $\tau$ on some edge in $E$,
it retrieves it from its memory in case it was determined already
and selects it at random (and stores it for future use) otherwise.}
\BE
\item
If {\sf Test-Bipartite} wishes to select a random vertex in $G_\tau$,
then the tester first selects uniformly a vertex $v\in[N]$.
It then outputs $v$ with probability $1/(d+1)$,
and if $v$ is not output, then it selects each neighbor of $v$
with probability $1/(2(d+1))$
and outputs $a_{\{u,v\}}$ if $\tau(\{u,v\})=2$,
where $u$ denotes the selected neighbor.

Indeed, the foregoing process outputs a vertex in $G_\tau$
with probability at least $1/(d+1)$,
and in case no vertex is output, the procedure
is repeated {\em(up to $O(\log N)$ times)}.
\item
If {\sf Test-Bipartite} queries for the $i^\xth$ neighbor
of vertex $v\in[N]\subseteq V_\tau$,
then the tester queries for the $i^\xth$ neighbor of $v$ in $G$,
and answers accordingly.
Specifically, if the answer to this query was $u$
{\em(i.e., $u$ is the $i^\xth$ neighbor of $v$ in $G$)},
then $u$ is given to {\sf Test-Bipartite} if $\tau(\{u,v\})=1$
and otherwise $a_{\{u,v\}}$ is given.
{\em(If the answer was~0,
indicating that $v$ has less than $i$ neighbors,
then~0 is returned as answer to {\sf Test-Bipartite}.)}

Finally, if {\sf Test-Bipartite} queries for the $i^\xth$ neighbor of
a vertex $a_{\{u,v\}}$ such that $u<v$,
then the tester answer with $u$ if $i=1$, with $v$ if $i=2$,
and with~0 if $i>2$.
\EE
When {\sf Test-Bipartite} halts, the current tester halts with the same verdict.
\EA
Furthermore, if {\sf Test-Bipartite} provides
an odd-length cycle in $G_\tau$, then we can easily
obtain a corresponding cycle in $G$
(by contracting the 2-vertex paths that appear on it
into single edges).

Note that in each iteration of the process detailed in Step~1,
each vertex of $G_\tau$ (regardless if it is an original vertex
of $G$ or an auxiliary vertex) is output with probability
exactly $\frac1N\cdot\frac1{d+1}$ (and with probability
$1-|V_\tau|\cdot \frac1N\cdot\frac1{d+1}$ no vertex is output),
Thus, conditioned on a vertex being selected in Step~1
(which happens
with very high probability since the process is  repeated
sufficiently many times),
Step~1 implements a uniform random selection
of vertices in $G_\tau$.


\paragraph{Conclusion.}
\sloppy
Combining Lemma~\ref{cycle-free.ana} with
Theorem~\ref{gr-bip:thm}
we conclude that Algorithm~\ref{cycle-free.alg}
is a one-sided error tester for cycle-freeness.
Its complexity is $\tildeO(\poly(d/\eps)\cdot\sqrt{N})$
and  if it rejects the graph $G$ then it outputs a cycle of length
$\poly(\eps^{-1} d \log N)$.
This establishes Theorem~\ref{intro:c3.thm}.

\section{Testing $C_4$-Minor-Freeness}
\label{C4:section}
As a warm-up towards testing $C_k$-minor-freeness, for any $k\geq3$,
we present the treatment of the special case of $k=4$.
We actually reduce the task of {\em testing $C_4$-minor-freeness}\/
to the task of {\em testing $C_3$-minor-freeness}.
Loosely speaking, the reduction replaces each triangle $\{u,v,w\}$
in the input graph by an auxiliary vertex (denoted $\triag{u,v,w}$)
that is connected to the corresponding three vertices.
The reduction is summarized in the following construction.

\ifnum\submit=0
 \BCT{\em(the reduction):}
\else
\BCT[the reduction]
\fi
\label{reducing:C4-C3.const}
Given a graph $G=([N],E)$ {\em(of max degree $d$)},
we {\em(locally)} construct the auxiliary graph $G'=([N]\cup T,E')$
such that $T$ contains the vertex $\triag{u,v,w}$
{\em(referred to as a ``triangle'' vertex)}
if and only if $\{u,v\},\{v,w\},\{w,u\}\in E$ and
\begin{equation}
\label{C4:edges.eq}
E' = \left(E \setminus
           \left(\bigcup_{u,v,w:\triag{u,v,w}\in T}\{u,v\}\right)\right)
     \cup \left\{\{u,\triag{u,v,w}\}:\triag{u,v,w}\in T\right\}.
\end{equation}
Specifically, the set of neighbors of $v\in[N]$ in $G'$,
denoted $\Gamma_{G'}(v)$,
consists of the following elements of $[N]\cup T$.
\BE
\item
Neighbors of $v$ in $G$ that do not reside in $G$ on a triangle
together with $v$; that is, $u\in\Gamma_G(v)$ is in $\Gamma_{G'}(v)$
if and only if $\Gamma_G(u)\cap\Gamma_G(v)=\emptyset$.
\item
Each triangle that contains $v$ in $G$;
that is, $\triag{u,v,w}$ is in $\Gamma_{G'}(v)$
if and only if $u,w\in\Gamma_G(v)$ and $\{w,u\}\in E$.
\EE
The set of neighbors of $\triag{u,v,w}\in T$ equals $\{u,v,w\}$.
Noting that $d+{d\choose2}\leq d^2$,
we view $G'$ as a graph of maximal degree $d^2$.
\ECT
\ifnum\inclFIG=1
For an illustration of Construction~\ref{reducing:C4-C3.const} see
Figure~\ref{C4b.fig}.
\ifnum\submit=0
\begin{figure}[htb]
\begin{center}
\input{C4c.pstex_t}  
\end{center}
\caption{\small An illustration for Construction~\protect\ref{reducing:C4-C3.const}.
On the left, $G$ is $C_4$-minor free, and indeed $G'$ is cycle-free; while on the
right, $G$ is not $C_4$-minor free, and $G'$ contains cycles (but
no cycles of length $3$ (triangles).)}
\label{C4b.fig}
\end{figure}
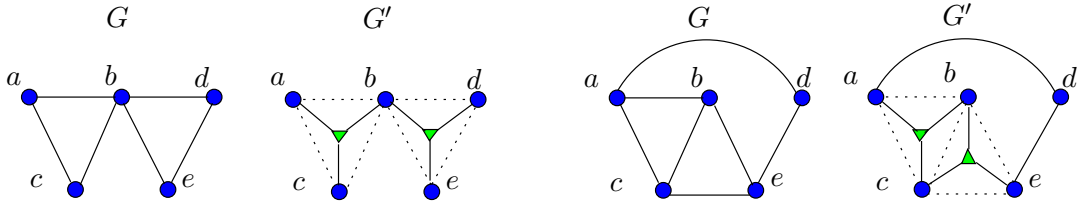
\fi
\fi
Note that given any $v\in [N]$, we can easily determine its
neighbors in $G'$ by checking the foregoing conditions.
Similarly, for every $u,v,w$, we can easily determine
whether $\triag{u,v,w}$ is in $G'$.
Lastly, note that we can select a vertex of $G'$ uniformly
by using the following procedure.

\BE
\item Select uniformly $v\in[N]$.
\item Select one of the following two instructions at random
with equal probability.
\BE
\item(Generating a vertex of $G$):

Output $v$ with probability $d^{-2}$.
\item(Generating a triangle):

Query all neighbors of $v$ to obtain $\Gamma_G(v)$.
and select uniformly $u,w\in\Gamma_G(v)$ such that $u\neq w$.
If $\{u,w\}\in E$, then output $\triag{u,v,w}$
with probability $p_v=d^{-2}\cdot {|\Gamma_G(v)|\choose 2}/3$.
\EE
In all the other cases, there is no output.
\EE
Thus, this process outputs each vertex of $G$
with probability $N^{-1}\cdot0.5\cdot d^{-2} = d^{-2}/2N$,
and outputs each $\triag{u,v,w}\in T$ with probability
$\sum_{x\in\{u,v,w\}}N^{-1}\cdot0.5\cdot {|\Gamma_G(x)| \choose 2}\cdot p_x
 = d^{-2}/2N$.
Since there are at least $N$ vertices in $G'$, the probability
that the process does not output {\em any\/} vertex in $G'$
is at most $(1-d^{-2})$. If we repeat the process
$\Theta(\log N)$ times (recall that $d$ is assumed to be a
constant), then the probability that we get no output is
$1/\poly(N)$. Since the total size of the sample needed is $o(N)$,
by a union bound, the probability that
 this occurs at any step of the algorithm, is negligible, and
this can be accounted for in the one-sided error probability
by letting the algorithm accept in case sampling fails.

\BA{\em(the $C_4$-minor-freeness tester):}
\label{reducing:C4-C3.alg}
Given input graph $G=([N],E)$, the tester emulates
the execution of Algorithm~\ref{cycle-free.alg}
on the graph $G'=([N]\cup T,E')$ as defined in
Construction~\ref{reducing:C4-C3.const}.
In the emulation, vertices of $G'$ are selected at random and
their neighbors are 
explored on the fly, as detailed above.
\EA
The analysis of Algorithm~\ref{reducing:C4-C3.alg}
reduces to an analysis of Construction~\ref{reducing:C4-C3.const}.

\BCM
\label{reducing:C4-C3.pos-clm}
If $G$ is $C_4$-minor-free, then $G'$ is cycle-free.
\ECM

\BPF
We first give a high-level idea of the proof and then give a
detailed argument.
By the hypothesis, the only simple cycles in $G$ are triangles,
and they are replaced in  $G'$ by stars centered at
auxiliary vertices. Specifically, the triangle $\{u,v,w\}$
(i.e., the edges $\{u,v\},\{v,w\},\{w,u\}$)
is replaced by a star-tree centered at $\triag{u,v,w}$
and having the leaves $u,v,w$.
Note that this replacement can form no simple cycles in $G'$,
because the simple paths in $G'$ correspond to simple paths in $G$
(where the sub-path $v\xedge\triag{u,v,w}\xedge w$
corresponds to the edge $v\xedge w$).

The corresponding detailed argument proceeds as follows.
Assume, contrary to the claim, that
there exists a simple cycle $\psi' = v_1\xedge v_2\xpath v_t\xedge v_{t+1}=v_1$ in $G'$.
%
Consider replacing each length-2 subpath
$u \xedge  \triag{u,w,x} \xedge w$
in $\psi'$
by the edge (in $G$) between $u$ and $w$ (where this edge exists because
$u$ and $w$ belong to a common triangle and $u\neq w$). Since, by construction of $G'$,
there are no edges in $G'$ between {\em triangle\/} vertices, this way
we obtain a cycle in $G$, which we denote by $\psi$.
We next show that $\psi$ is a simple cycle of length greater than $3$,
and we reach a contradiction to the hypothesis that $G$ is $C_4$-minor-free.

We first verify that the length of $\psi$ is greater than $2$.
This is true because otherwise,
the cycle $\psi'$ is either of the form $u \xedge \triag{u,w,x} \xedge w \xedge u$,
or it is of the form $u \xedge  \triag{u,w,x_1} \xedge w \xedge  \triag{u,w,x_2} \xedge u$.
In the first case $\psi'$ contains an edge $\{w,u\}$ of a triangle in $G$, which
is not possible by construction of $G'$. In the second case, since
$\psi'$ is simple (so that $x_1 \neq x_2$), there is a simple 4-cycle
$u \xedge x_1 \xedge w \xedge x_2 \xedge u$ in $G$ (contradicting the
hypothesis that $G$ is $C_4$-minor-free). It follows that $\psi$
is a simple cycle and it remains to verify that its length is greater than $3$.

Suppose that the length of $\psi$ is $3$, that is, $\psi =u \xedge w \xedge v \xedge u$
is a triangle in $G$. It follows that none of the edges
$\{u,w\},\{w,v\},\{v,u\}$ belong to $G'$ and therefore,
$\psi' = u \xedge \triag{u,w,x_1} \xedge w \xedge  \triag{w,v,x_2}  \xedge v
      \xedge \triag{v,u,x_3} \xedge u$, where
the triangles are distinct and hence at least one of them
does not equal $\triag{u,w,v}$.
But this implies that 
 there exists a simple 4-cycle in $G$
(contradicting the hypothesis that $G$ is $C_4$-minor-free).
\EPF

\BCM
\label{reducing:C4-C3.neg-clm}
If $G$ is $\eps$-far from being $C_4$-minor-free,
then $G'$ is $\Omega(\eps)$-far from being cycle-free,
where the Omega-notation hides a polynomial in $1/d$.
\ECM

\BPF
Suppose that $G'$ is $\delta$-close to being cycle-free,
where the distance refers to the degree bound of $G'$,
which is $d^2$ as well as the number of vertices in $G'$
which is $N+|T|$.
Let $R'$ be a set of at most $\delta\cdot d^2\cdot (N+|T|)$ edges
such that removing $R'$ from $G'$ yields a cycle-free graph,
$([N]\cup T,E'\setminus R')$.
Let $R\subseteq E$ be a set of edges that consists of
(1) all edges of $E$ that are in $R'$,
and (2) each edge $\{u,v\}\in E$
such that $\{u,\triag{u,v,w}\}$ is in $R'$.
Hence, $|R| \leq 2|R'| < \delta\cdot d^4N$,
where we use $|T| \leq {d\choose2}\cdot N$.
We next prove that removing $R$ from $G$
yields a graph that is $C_4$-minor-free,
and it follows that $G$ is $2d^3\delta$-close to being $C_4$-minor-free.

\sloppy
Assume, contrary to the claim, that for some $t\geq4$
there exists a {\em simple}\/ cycle $v_1\xedge v_2\xpath v_t\xedge v_1$
in the resulting graph
(i.e., in the graph $([N],E\setminus R)$).
We consider the corresponding (not necessarily simple) cycle in
the graph $([N]\cup T,E'\setminus R')$:
\BDes
\item[{\sf Case 1:}]
If the edge $\{v_i,v_{i+1}\}\in E\setminus R$
is not a part of any triangle in $G$,
then $\{v_i,v_{i+1}\}\in E'\setminus R'$,
because $\{v_i,v_{i+1}\}$ is an edge of $G'$ and it cannot be in $R'$
(since this would imply that $\{v_i,v_{i+1}\}\in R$).
In this case, we just use the edge $\{v_i,v_{i+1}\}$ on the cycle in
the graph $([N]\cup T,E'\setminus R')$.
\item[{\sf Case 2:}]
If the edge $\{v_i,v_{i+1}\}\in E\setminus R$
is part of a triangle $v_i,v_{i+1},w$ (in $G$),
then $\{v_i,\triag{v_i,v_{i+1},w}\}\in E'\setminus R'$
and $\{v_{i+1},\triag{v_i,v_{i+1},w}\}\in E'\setminus R'$,
because both pairs are edges of $G'$ and cannot be in $R'$
(since this would imply that $\{v_i,v_{i+1}\}\in R$).
In this case, we replace the edge $\{v_i,v_{i+1}\}\in E\setminus R$
by the length-two-path $v_i\xedge\triag{v_i,v_{i+1},w}\xedge v_{i+1}$
(in 
the graph $([N]\cup T,E'\setminus R')$).
\EDes
Observe that the ``triangle'' vertices used in Case~(2)
need not be distinct, but they can collide only when
they refer to three consecutive vertices on the original $t$-cycle
(i.e., if $\triag{v_i,v_{i+1},w_1}=\triag{v_j,v_{j+1},w_2}$, for $i<j$,
then $v_j=v_{i+1}$ must hold, and $w_1=v_{j+1}=v_{i+2}$ follows).
Such collisions can be eliminated
at the cost of omitting a single ``non-triangle'' vertex
(i.e., the path
$v_i\xedge\triag{v_i,v_{i+1},v_{i+2}}\xedge v_{i+1}\xedge
    \triag{v_i,v_{i+1},v_{i+2}}\xedge v_{i+2}$
is replaced by the
path $v_i\xedge\triag{v_i,v_{i+1},v_{i+2}}\xedge v_{i+2}$).
Thus, we derive a simple cycle of length at least $t\geq4$ in 
the graph $([N]\cup T,E'\setminus R')$
(since we have a ``triangle'' vertex per each
omitted ``non-triangle'' vertex).
This contradicts the hypothesis that 
$([N]\cup T,E'\setminus R')$
is cycle-free, and so the claim follows.
\EPF

\paragraph{Conclusion.}
Combining Claims~\ref{reducing:C4-C3.pos-clm}
and~\ref{reducing:C4-C3.neg-clm}
with Theorem~\ref{intro:c3.thm} and the fact that the number of
vertices in $G'$ is linear in $N$ (for constant $d$),
we conclude that {\em there exists a one-sided error tester
of complexity $\tildeO(\poly(1/\eps)\cdot\sqrt{N})$
for $C_4$-minor-freeness.}

\section{Testing $C_k$-Minor-Freeness, for any $k\geq4$}
\label{Ck:section}
In this section we show that, for any $k\geq4$,
the task of {\em testing $C_k$-minor-freeness}\/
reduces to the task of {\em testing $C_3$-minor-freeness}.
The reduction extends the ideas underlying the reduction
of {\em testing $C_4$-minor-freeness}\/
to {\em testing $C_3$-minor-freeness}\/
(as presented in Section~\ref{C4:section}).

\ifnum\submit=0
The basic idea of the reduction is replacing simple cycles
that have length smaller than $k$ by stars. Actually, we
replace certain subgraphs that contain such cycles by stars.
We start by defining the class of (induced) subgraphs that
we intend to replace by stars.
These subgraphs (or rather their vertex sets) will be called {\em spots\/}.
Below, the term {\tsf 2-connectivity} means {\em 2-vertex connectivity};
that is, a graph is called 2-connected if every two vertices
in the graph can be connected by two vertex-disjoint paths.
\BD{\em(spots):}
\label{spot.def}
A set $S\subseteq[N]$ is called a {\dsf $k$-spot} of the
graph $G=([N],E)$ if the following three conditions hold:
\BE
\item The subgraph induced by $S$, denoted $G_S$, contains no
simple cycle of length at least $k$; that is, $G_S$ is $C_k$-minor-free.
\item The subgraph induced by $S$ is 2-connected and $|S|\geq3$.
\item For every $u, v\in S$ such that $u \neq v$,
      either $u$ and $v$ are not connected by any path
      that is external to $G_S$
      or the length of every such external path
      is at least $\ell(k)\eqdef2k$.
Here, by a path {\dsf external to $G_S$}
we mean a path that does not use any edge
that is incident to a vertex in $S$
with the exception of the endpoints $u$ and $v$
{\em (i.e., all intermediate vertices
of the path belong to $[N]\setminus S$)}.
\EE
\ED
For example, every 4-spot of $G$ induces a triangle in $G$,
whereas the set of possible subgraphs induced by 5-spots of $G$
consists of the following graphs: the $4$-cycle (i.e., $C_4$),
the $4$-cycle augmented by a chord, the $4$-clique (i.e., $K_4$),
and the graphs $K_{2,n}$ and $K'_{2,n}$ for every $n\geq3$,
where $K'_{2,n}$ is the graph $K_{2,n}$ augmented by a
single edge that connects the two vertices on the small side.%
\footnote{Recall that $K_{m,n}$ denotes the complete bipartite graph
with $m$ vertices on one side and $n$ vertices on the other side;
that is, $K_{m,n}=([m+n],\{\{i,m+j\}:i\!\in\![m],j\!\in\![n]\})$.}
(Indeed, in Section~\ref{C4:section} we essentially used a relaxed
notion of a 4-spot in which the third condition was not required.)
\else

\noindent
We next restate Definition~\ref{spot.def}.

\ifnum\submit=0
\medskip\noindent{\bf Definition~\ref{spot.def}}~
\else
\medskip\noindent{\bf Definition~\ref{spot.def}.}~
\fi
{\it
A set $S\subseteq[N]$ is called a {\dsf $k$-spot} of the
graph $G=([N],E)$ if the following three conditions hold:
\BE
\item The subgraph induced by $S$, denoted $G_S$, contains no
simple cycle of length at least  $k$; that is, $G_S$ is $C_k$-minor-free.
\item The subgraph induced by $S$ is 2-vertex-connected and $|S|\geq 3$.
\item For every $u, v\in S$ where $u \neq v$,
      either $u$ and $v$ are not connected by any path
      that is external to $G_S$
      or the length of every such external path
      is at least $\ell(k)\eqdef2k$.
Here, by a path {\dsf external to $G_S$}
we mean a path that does not use any edge
that is incident to a vertex in $S$ with the
exception of the endpoints $u$ and $v$
{\em(i.e., all intermediate vertices
of the path belong to $[N]\setminus S$)}.
\EE
}

\fi

\subsection{Some basic facts regarding spots}
Since $k$ is fixed throughout the rest of our discussion,
we may omit it from the notations and refer to $k$-spots as spots.
A few basic properties of spots are listed below.

\BCM
\label{diam.clm}
If $S$ is a $k$-spot of $G$,
then the diameter of $G_S$ is smaller than $k/2$.
\ECM
It follows from the claim that for every $k$-spot $S$
where $k \geq 4$,
\BEQ
|S|\;<\;\sum_{i=0}^{k/2} d^i \;<\; 2d^{k/2} \;<\; d^{k-1}
\label{k-spot-ub.eq}
\EEQ
(since $d\geq3$).\footnote{We mention
that there may exists spots of size $d^{(k-1)/2}$.
Consider, for example, a graph that consists of two copies
of a depth $(d-1)$-ary tree of depth $(k-1)/2$ such that each
vertex in one tree is connected to its mirror vertex in the
second tree. To see that this graph is $C_k$-minor-free,
consider the correspondence between cycles on this graphs
and traversals of parts of the original tree, and note that
simple cycles correspond to traversals in which each edge
is used at most twice. Since such traversals have length at most
twice the depth of the tree, the claim follows.}
\medskip

\BPF
Assume, contrary to the claim that the diameter of
$G_S$ is at least $k/2$
and consider $u,v\in S$ such that the distance
between $u$ and $v$ in $G_S$ is at least $k/2$.
Since $G_S$ is 2-connected, there exists a simple cycle
in $G_S$ that passes through both $u$ and $v$, and it follows
that this cycle has length at least $k$, which contradicts
the hypothesis that $G_S$ is $C_k$-minor-free.
\EPF

\medskip
Note that, for any spot $S$ and every three
distinct vertices $u,v,w\in S$, the subgraph $G_S$ contains
a simple path that goes from $u$ to $v$ via $w$.
This holds by the very fact that $G_S$ is 2-connected
(i.e., the second condition in Definition~\ref{spot.def}).
By Claim~\ref{diam.clm} the length of this path
is less than $d^{k-1}$. As we shall show next, a much better bound
follows by using the fact that $G_S$ is $C_k$-minor-free
(i.e., the first condition in Definition~\ref{spot.def}),

\BCM
\label{spot:u-w-v.clm}
For every $k$-spot $S$ and distinct vertices $u,v,w\in S$,
the subgraph $G_S$ contains a simple path
of length at most $2k-1$ that goes from $u$ to $v$ via $w$.
\ECM

\ifnum\inclFIG=1
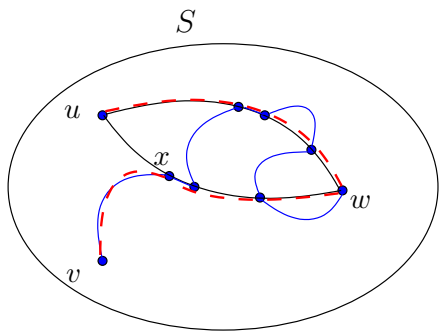
\begin{figure}[htb]
\begin{center}
\input{spot-paths.pstex_t}
\end{center}
\caption{\small An illustration for the proof of Claim~\protect\ref{spot:u-w-v.clm}.
The jotted line is the path between $u$ and $v$ that passes through $w$.}
\label{spot-paths.fig}
\end{figure}
\fi
\BPF
We just take a closer look at the standard proof
that the fact that a graph is 2-connected implies
the existence of a $u\xpath w\xpath v$ path
(for every three vertices $u,v,w$ in the graph).
\ifnum\inclFIG=1
For an illustration of the argument
that follows, see Figure~\ref{spot-paths.fig}.
\fi
The proof starts by considering two different
vertex-disjoint $u\xpath w$ paths,
and an arbitrary path between $v$ and $w$.
In the current case (i.e., by $C_k$-minor-freeness),
we may assume that the total length of the first two paths
is smaller than $k$.
Similarly, without loss of generality,
the length of the third path is smaller than $k$.
Proceeding as in the standard proof, we ask whether
the third path (i.e., the $v\xpath w$ path)
intersects both the $u\xpath w$ paths.
If the answer is negative, then we are done
(as we obtain the desired simple path by concatenating
the path $v\xpath w$ to the $w\xpath u$ path that does not intersect it).

Otherwise, let $x$ be the ``closest to $v$'' vertex on the
path $v\xpath w$ that appear on either of the $u\xpath w$ paths;
that is, $x$ is on one of the $u\xpath w$ paths
and the sub-path $v\xpath x$ (of the path $v\xpath w$)
contains no vertex from either the $u\xpath w$ paths.
Note that $x=v$ is possible (but $x=w$ is not),
and assume, w.l.o.g.,
that $x$ resides on the first $u\xpath w$ path.
Then, consider the path obtained by combining
the following three path segments:
(1)~the segment $v\xpath x$ of the path $v\xpath w$,
(2)~the segment $x\xpath w$ of the first $u\xpath w$ path,
and (3)~the second $u\xpath w$ path.
Note that the total length of this path is at most $2(k-1)$
(i.e., the total length of the three paths),
and that the three segment do not intersect
(since the $v\xpath x$ segment does not intersect
the $x\xpath w$ segment nor the $u\xpath w$ path
by the choice of $x$).
\EPF

\ifnum\inclFIG=1
\begin{figure}[htb]
\begin{center}
\input{2spot.pstex_t}
\end{center}
\caption{\small An illustration for the proof of Claim~\ref{intersection.clm}.}
\label{2spot.fig}
\end{figure}
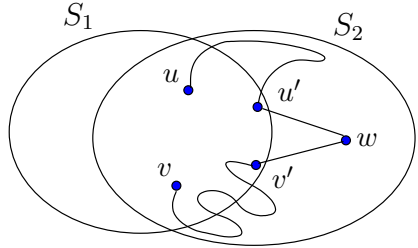
\fi

\BCM
\label{intersection.clm}
If $S_1\neq S_2$ are $k$-spots of $G$, then $|S_1\cap S_2|\leq1$.
\ECM

\BPF
Assume, contrary to the claim that $|S_1\cap S_2|\leq1$
for two $k$-spots $S_1\neq S_2$.
Consider (w.l.o.g.)
$u,v\in S_1\cap S_2$ such that $u \neq v$ and $w\in S_2\setminus S_1$
(as in Figure~\ref{2spot.fig}).
By Claim~\ref{spot:u-w-v.clm},
the subgraph $G_{S_2}$ contains a simple path
of length at most $2k-1$ that goes from $u$ to $v$ via $w$.
Let $u'$ (resp., $v'$) be the last (resp., first) vertex
of $S_1$ that appears on this path before reaching $w$
(resp., after leaving $w$).
Then, we get a simple path (in $G$)
from $u'\in S_1$ to $v'\in S_1\setminus\{u'\}$ such that this
path contains only intermediate vertices of $S_2\setminus S_1$.
Recalling that this path has length at most $2k-1$,
we reach a contradiction to the hypothesis that $S_1$ is a $k$-spot
(specifically to the third condition of Definition~\ref{spot.def}).
\EPF

As a corollary of Claim~\ref{intersection.clm} we get:
\BCR\label{intersection.cor}
Every vertex $v$ may belong to at most $|\Gamma(v)|/2$ spots
and hence the number of spots in a graph $G$
is upper-bounded by the number of edges in $G$.
\ECR
\BPF
The second part of the corollary follows directly from the first,
and so we only need to establish the first part.
Since by the definition of a spot, it must contain at least 3 vertices,
every spot $S$ that contains a vertex $v$
must also contain at least two of $v$'s neighbors.
However, by Claim~\ref{intersection.clm},
spots that contain $v$ may not share any other vertex.
\EPF

\BCM
\label{cycle-spot.clm}
Each simple cycle in any $C_k$-minor-free graph $G$
is a subset of some $k$-spot of $G$.
\ECM

\BPF
Consider the following iterative process of constructing a spot $S$
that contains the aforementioned cycle. Initially, we set $S$
to equal the set of vertices that reside on this cycle.
Clearly, this set $S$ satisfies the first two conditions
of the definition of a spot (i.e., Definition~\ref{spot.def}),
which is an invariant that we
shall maintain throughout the iterative process.
The process ends once all three conditions are satisfied.
Since the size of the spot increases in each iteration,
the process must eventually end.
Thus, at the start of each iteration of the process we have a
set $S$ that satisfies the first two conditions in Definition~\ref{spot.def}
but does not satisfy the third condition.
That is, there
exists a simple path external to $S$ that connects
two of its vertices $u,v\in S$.
Adding this path to $S$ we obtain a new set that satisfies
Condition~1 (since $G$ is $C_k$-minor-free).
To see that the new set satisfies Condition~2,
we need to show that there exist two disjoint paths between
each pair of vertices that are not both in $S$.
\ifnum\inclFIG=1
For an illustration of the argument that follows,
 see Figure~\ref{cycle-spot.fig}.
\fi

\ifnum\inclFIG=1
\begin{figure}[htb]
\begin{center}
\input{cycle-spot.pstex_t}
\end{center}
\caption{\small An illustration for the proof of Claim~\protect\ref{cycle-spot.clm}.}
\label{cycle-spot.fig}
\end{figure}
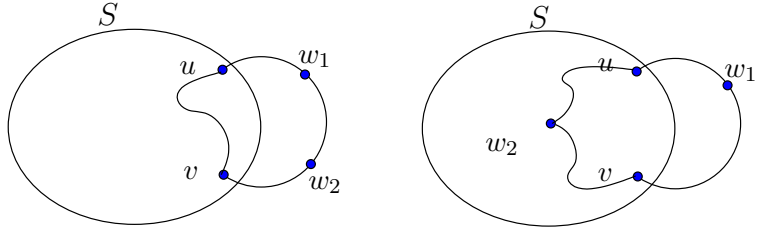
\fi

In the case that $w_1$ and $w_2$ are both new vertices
(which reside on the aforementioned $S$-external path),
we connect them by the direct path that resides outside of $S$
as well as by a simple path that (without loss of generality)
connects $w_1$ to $u$
(via the external path), connects $u$ and $v$ via $S$,
and connects $v$ and $w_2$ (via the external path).
In the case that $w_1$ is new but $w_2\in S$,
we use the external path to connect $w_1$ to $u$ and $v$, respectively,
and use the fact that there are vertex disjoint paths in $G_S$
that connect $u$ and $v$ to $w_2$.
Thus the new set satisfies the first two conditions in
Definition~\ref{spot.def}, as desired.
\EPF

\subsection{The actual reduction}
Using these facts, we are ready to present our reduction.

\ifnum\submit=0
 \BCT{\em(the reduction):}
\else
\BCT[the reduction]
\fi
\label{reducing:Ck-C3.const}
Given a graph $G=([N],E)$ {\em(of max degree $d$)},
we {\em(locally)} construct
the auxiliary graph $G'=([N]\cup\{\ang{S}:S\!\in\!\cS\},E')$
such that $\cS$ is the set of all spots of $G$
and
\begin{equation}
\label{Ck:edges.eq}
E' = \Big(E \setminus
           \Big(\bigcup_{S\in\cS}\{\{u,v\}:u,v\!\in\!S\}\Big)\Big)
     \cup \Big\{\{v,\ang{S}\}:S\in\cS,v\!\in\!S\Big\}.
\end{equation}
Specifically, the set of neighbors of $v\in[N]$ in $G'$,
denoted $\Gamma_{G'}(v)$,
consists of the following elements of $[N]\cup\{\ang{S}:S\!\in\!\cS\}$.
\BE
\item
Neighbors of $v$ in $G$ that do not reside in any spot
together with $v$; that is, $u\in\Gamma_G(v)$ is in $\Gamma_{G'}(v)$
if and only if $\{u,v\}$ is not a subset of any $S\in\cS$.
\item
Each spot that contains $v$ in $G$;
that is, $\ang{S}$ is in $\Gamma_{G'}(v)$
if and only if $S\in\cS$ and $v\in S$.
\EE
For any $S\in\cS$,
the set of neighbors of $\ang{S}$ in $G'$ equals $S$.
Recalling that by Equation~(\ref{k-spot-ub.eq})
 each $S\in\cS$ has size at most $d^{k-1}$,
we view $G'$ as a graph of maximal degree $d^{k-1}$.
\ECT
Observe that the set of spots that contain a vertex $v\in[N]$
is determined by the $(k+\ell(k))$-neighborhood of $v$ in $G$,
where the $t$-neighborhood of $v$ contains all vertices that are
at distance at most $t$ from $v$.
Thus, we can determine the set of neighbors of each vertex in $G'$.
We note that the process of determining the spots that
contain a vertex may fail if a cycle of length at least $k$ is encountered.
In such a case the algorithm can clearly reject.
Lastly, note that we can select a vertex of $G'$ uniformly
by using the following procedure.

\BE
\item Select uniformly $v\in[N]$.
\item Select one of the following two instructions at random
with equal probability.
\BE
\item(Generating a vertex of $G$):

Output $v$ with probability $1/d$.
\item(Generating a spot):

Select uniformly a spot $S$ that contain $v$ (i.e., $S\in\cS_v$),
and output $\ang{S}$ with probability
$p_v(S)=\frac{|{\cS}_v|}{d|S|}$,
where ${\cS}_v\eqdef\{S\in{\cS}:v\in S\}$.
(Recall that by Corollary~\ref{intersection.clm},
$|{\cS}_v| \leq d/2$, so that $p_v(S) < 1$).
\EE
In all the other cases, there is no output.
\EE
Thus, this process output each vertex of $G$
with probability $N^{-1}\cdot0.5\cdot d^{-1} = 1/(2d N)$,
and outputs each spot $\ang{S}\in\cS$ with probability
$\sum_{v\in S} N^{-1}\cdot0.5\cdot|\cS_v|^{-1}\cdot p_v(S) = 1/(2d N)$.

\BA{\em(the $C_k$-minor-freeness tester):}
\label{reducing:Ck-C3.alg}
Given input graph $G=([N],E)$, the tester emulates
the execution of Algorithm~\ref{cycle-free.alg}
on the graph $G'$ as defined in Construction~\ref{reducing:Ck-C3.const}.
In the emulation, vertices of $G'$ are selected at random and
their neighbors are being explored on the fly, as detailed above.
\EA
The analysis of Algorithm~\ref{reducing:Ck-C3.alg}
reduces to an analysis of
Construction~\ref{reducing:Ck-C3.const}.

\ifnum\submit=0
\BCM{\em(yes-instances):}
\else
\BCM[yes-instances]
\fi
\label{reducing:Ck-C3:yes.clm}
If $G$ is $C_k$-minor-free, then $G'$ is cycle-free.
\ECM

\BPF
Suppose, contrary to the claim, that $v_1\xedge v_2\xpath v_t\xedge v_1$
is a simple cycle in $G'$. We consider two cases.
\BDes
\item[{\sf Case 1:} {\it All $v_i$'s are vertices of $G$.}]
In this case, the edges $\{v_i,v_{i+1}\}$ in $G'$ must
be edges of $G$ (since the only edges in $G'$ that are not
edges in $G$ are incident to spot-vertices).
On the other hand $t<k$ must hold, because $G$ is $C_k$-minor-free.
But this yields a contradiction, because, by Claim~\ref{cycle-spot.clm},
the set $\{v_i:i\in[t]\}$ must be a subset of some spot of $S$,
which means that none of the edges $\{v_i,v_{i+1}\}$ may exist in $G'$.
\item[{\sf Case 2:} {\it Some $v_i$ represents a spot of $G$.}]
Let $v_i=\ang{S}$, for some $S\in{\cS}$.
By the definition of the neighborhood relations in
$G'$ we have that $v_{i+1},v_{i-1}\in S$.
Now, consider a minimal sub-path of $v_{i+1},\dots,v_t,v_1,\dots,v_{i-1}$
that starts in a vertex of $S$, denoted $u$, and ends in a vertex of $S$,
denoted $v$. That is, we consider a sub-path that starts and ends in
vertices of $S$, but has no intermediate vertices in $S$.
This sub-path (in $G'$) cannot consist of a single edge
(because the edge $\{u,v\}\subset S$ cannot appear in $G'$),
it cannot contain the vertex $\ang{S}$
(because $\ang{S}$ already appears as $v_i$),
and it cannot be a path of length 2 that goes through another spot
(because, by Claim~\ref{intersection.clm},
no other spot may contain both $u$ and $v$).
Since this path may not contain intermediate vertices in $S$,
and since spot-vertices cannot be adjacent in $G'$,
it follows that this path must contain a vertex $w\in[N]\setminus S$.
That is, we get a path in $G'$ that goes from $u$ to $v$ via $w$,
without passing through any vertex in $S$.

We now obtain a corresponding path in $G$;
that is, a path in $G$ that goes from $u$ to $v$ via $w$,
without passing through any vertex in $S$.
This is done by replacing any length-2 subpath $u'\xedge\ang{S'}\xedge v'$
(in $G'$) by a sub-path $u'\xpath v'$ (in $G$)
that does not pass through $S$,
where the latter path exists by the fact that $u',v'\in S'$
are connected by vertex-disjoint paths (internal to $S'$)
such that their intersection with $S$ contains at most a single
vertex (see Claim~\ref{intersection.clm}).
It follows that $G$ itself contains a path between $u$ and $v$
that passes through $w$ and does not pass through $S$,
where $u,v\in S$ but $w\not\in S$.
Thus, $G$ itself contains a simple (non-edge) path between $u$ and $v$
that does not pass through $S$ (i.e., an external path).
By the third condition in Definition~\ref{spot.def},
the length of this external path is at least $\ell(k)>k$,
but this contradicts the hypothesis that $G$ is $C_k$-minor-free
(because $u$ and $v$ are connected in $G_S$ and $\ell(k)\geq k$,
yielding a simple cycle of length at least $k$).
\EDes
The claim follows.
\EPF

\ifnum\submit=0
 \BCM{\em(no-instances):}
\else
\BCM[no-instances]
\fi
\label{reducing:Ck-C3:no.clm}
If $G$ is $\eps$-far from being $C_k$-minor-free,
then $G'$ is $\Omega(\eps)$-far from being cycle-free,
where the Omega-notation hides a $d^{-k}$ factor.
\ECM

\BPF
Suppose that $G'$ is $\delta$-close to being cycle-free,
where the distance refers to the degree bound of $G'$,
which is $d^{k-1}$. Recall that by Corollary~\ref{intersection.cor}
 $|\cS|\leq|E|\leq d N/2$.
Let $R'$ be a set of
at most $\delta\cdot d^{k-1}(N+|{\cS}|) < \delta\cdot d^{k}N$ edges
such that removing $R'$ from $G'$ yields a cycle-free graph.
Let $R\subseteq E$ be a set of edges that consists of
(1) all edges of $E$ that are in $R'$,
and (2) each edge $\{v,w\}\in E$
such that $\{v,\ang{S}\}$ is in $R'$.
Hence, $|R| \leq d|R'| < \delta\cdot d^{k+1}N$.
We next prove that removing $R$ from $G$
yields a graph that is $C_k$-minor-free,
and it follows that $G$ is $\delta\cdot d^k$-close
to being $C_k$-minor-free.

Suppose, contrary to the claim, that for $t\geq k$
there exists a simple cycle $v_1\xedge v_2\xpath v_t\xedge v_1$
in the resulting graph (i.e., in the graph $([N],E\setminus R)$).
We first show that there exists a corresponding (not necessarily simple)
cycle in $E'\setminus R'$. Specifically, for
each $\{v_i,v_{i+1}\}\in E\setminus R$, we consider two cases.
\BDes
\item[{\sf Case 1:} {\it This edge is not a subset of any spot in $G$}.]
In this case, $\{v_i,v_{i+1}\}\in E'\setminus R'$,
because this edge is in $E'$ and cannot be in $R'$
(or else it would have been in $R$).
So we just use this edge in the cycle (in $E'\setminus R'$).
\item[{\sf Case 2:} {\it This edge is a subset of a spot $S$ in $G$}.]
In this case, $\{v_i,\ang{S}\},\{v_{i+1},\ang{S}\}\in E'\setminus R'$,
because both these edges are in $E'$ and cannot be in $R'$
(or else $\{v_i,v_{i+1}\}$ would have been in $R$).
In this case, we replace the edge $\{v_i,v_{i+1}\}\in E\setminus R$
by the length-two-path $v_i\xedge\ang{S}\xedge v_{i+1}$.
\EDes
Thus, we obtain a cycle
in $([N]\cup\{\ang{S}:S\!\in\cS\},E'\setminus R')$
that contains the vertices $v_1,\dots,v_t\in[N]$
as well as (possibly) some elements in $\{\ang{S}:S\!\in\cS\}$.
Since the latter elements may appear in multiple copies,
the foregoing cycle is not necessarily simple.
Note that a simple cycle
in $([N]\cup\{\ang{S}:S\!\in\cS\},E'\setminus R')$
yields a contradiction to the hypothesis
that this graph is cycle-free, and thus establishes our claim
that the graph $([N],E\setminus R)$ is $C_k$-minor-free.
We obtain a simple cycle, in two steps, as follows.

First, we replace every maximal sub-path of the form
$v_i\xedge\ang{S}\xedge v_{i+1}\xedge\ang{S}\xpath\ang{S}\xedge v_{j}$,
where $j\neq  i$ (or else $S$ contains a $t$-cycle for $t\geq k$),
by a length-two path $v_i\xedge\ang{S}\xedge v_{j}$.
If the resulting cycle contains distinct spot (representative)
vertices, then we are done (since we obtain a simple cycle).
Otherwise, we obtain a cycle of the form
$$u_1\xpath u_{t_1}\xedge \ang{S_1}
  \xedge u_{t_1+1} \xpath u_{t_1+t_2}\xedge \ang{S_2}
  \xedge u_{t_1+t_2+1} \xpath u_{t_1+t_2+t_3}\xedge \ang{S_3}
  \cdots \ang{S_m} \xedge u_1$$
where the $u_i$'s are all distinct
and adjacent $S_i$'s are distinct
(but non-adjacent $S_i$'s may be identical).
Next, we consider a sub-path of the foregoing cycle
such that the endpoints of this sub-path are two copies
of the same spot $S$ and no other spot appears more than once
on this sub-path. This sub-path cannot have length two
(because adjacent $S_i$'s are distinct),
which means that it is actually a simple cycle,
and we are done.
\EPF

\paragraph{Conclusion.}
Combining Claims~\ref{reducing:Ck-C3:yes.clm}
and~\ref{reducing:Ck-C3:no.clm}
with Theorem~\ref{intro:c3.thm} and the fact that the number of
vertices in $G'$ is linear in $N$ (for constant $d$ and $k$),
we conclude that {\em Algorithm~\ref{reducing:Ck-C3.alg}
is a one-sided error tester for $C_k$-minor-freeness,
and its complexity is $\tildeO(\poly(d^k/\eps)\cdot\sqrt{N})$.}
This establishes Theorem~\ref{intro:find-Ck.thm}.

\section{Proof of the Lower Bound}
\label{lower-bound.sec}
Recall that Goldreich and Ron proved a $\Omega(\sqrt{N})$ query
lower bound on the complexity of one-sided error testers
for cycle-freeness~\cite[Prop.~4.3]{GR1}.
As stated in the introduction,
Benjamini, Schramm, and Shapira~\cite{BSS} suggested
that this lower bound can be extended to testing $H$-minor freeness
for any $H$ that is not a forest.
This is indeed the case, as proved next.

\BT{\em(lower bound for one-sided error testing of $H$-minor freeness,
for any $H$ that contains cycles):}
\label{lower-bound.thm}
For any fixed $H$ that contains a simple cycle,
the query complexity of {\em one-sided error}
testing of $H$-minor freeness is $\Omega(\sqrt{N})$.
\ET
Indeed, as can been seen in the case that $H$ is a single edge,
the lower bound does not hold in case $H$ contains no simple cycles.
A general study of testing $H$-minor freeness for any cycle-free $H$
is initiated in Section~\ref{tree.section}.
\medskip

\BPF
Following the proof of~\cite[Prop.~4.3]{GR1},
we show that for sufficiently large $N$, with high probability,
the random $N$-vertex graphs considered in~\cite[Sec.~7]{GR1}
are far from being $H$-minor free.
Once this is done, the theorem follows,
since it was shown in~\cite[Sec.~7]{GR1} that
a probabilistic algorithm that makes $o(\sqrt{N})$
queries is unlikely to find a cycle in such a random graph
(and this algorithm must accept whenever it fails to see a cycle,
because otherwise it will reject some $H$-minor free graph
with positive probability).
Furthermore, it suffices to show that,
for any fixed $k$ and sufficiently large $N$, with high probability,
such a random graph is far from being $K_k$-minor free,
because containing a minor of $K_k$ implies containing
a minor of any $k$-vertex graph $H$.

The random graphs considered in~\cite[Sec.~7]{GR1}
are graphs uniformly chosen in the family $\cG_N$
(which is denoted $\cG^N_1$ in~\cite{GR1}).
Each ($N$-vertex) graph in $\cG_N$ consists
of the union of a simple $N$-vertex (Hamiltonian) cycle
and a perfect matching of these $N$ vertices.
(Indeed, each graph in $\cG_N$ is 3-regular.)
Furthermore, the cycle is fixed to be $(1,2,\dots,N,1)$
and so a random graph in $\cG_N$ corresponds to a random choice
of a perfect matching.
Our aim is to prove that, with high probability,
such a random graph is $\e/3$-far from being $K_k$-minor free,
where $\e=1/ck^2$ for a sufficiently large constant $c$
(to be determined below).
We first show that any graph having a specific property
(which is stated in the conditions of the following claim)
is far from being $K_k$-minor free.

\BCM \label{special.clm}
Suppose that the vertices of the $N$-vertex graph $G$
can be partitioned into $\widehat{N} = 2\e N$ equal-sized sets,
denoted $S_1, S_2, \ldots, S_{\widehat{N}}$,
such that the following conditions hold:
\BE
\item The subgraph induced by each $S_i$ is connected.
\item For every two disjoint collections of sets, $C$ and $C'$,
such that $|C|=|C'|\geq\widehat{N}/6k$,
there are at least $\e N + 1$ edges between vertices
in $U=\bigcup_{i\in C}S_i$ and vertices in $U'=\bigcup_{i\in C'}S_i$.
\EE
Then, removing any set of $\e N$ edges from $G$,
yields a graph that contains an $K_k$-minor
{\em(i.e., $G$ is $\e/3$-far from being $K_k$-minor free)}.
\ECM

\BPF
We prove the claim by contradiction.
Suppose that there is a subset $E'$
of at most $\e N$ edges whose removal from $G$ results
in a graph, denoted $G'$, that is $K_k$-minor free.
First, note that at most $\e N$ of the sets $S_i$ can become disconnected.
Thus, (1)~at least $\widehat{N} - \e N= \widehat{N}/2$
of the $S_i$'s induced connected subgraphs in $G'$.
Furthermore, (2)~for every $U$ and $U'$ as defined in the claim,
there exists at least one edge in $G'$ between $U$ and $U'$.

Starting with (1), assume, w.l.o.g.,
that for $i=1,...,\widetilde{N}\eqdef\widehat{N}/2$,
the subgraph of $G'$ induced by $S_i$ is connected.
We partition these sets into $k$ equal-sized parts;
that is, for $1 \leq i \leq k$,
let $T_i = \bigcup_{r = ((i-1)\widetilde{N}/k)+1}^{i\widetilde{N}/k} S_r$
and $G'_i$ be the subgraph of $G'$ induced by $T_i$.

We first show that for each $G'_i$ has a connected
component that contains at least $\widetilde{N}/3k$ sets $S_r$
(which are contained in $T_i$).
Let $W_1,...,W_t$ denote the connected components of $G'_i$,
and suppose towards the contradiction that each of them
contains less than $\widehat{N}/3k$ sets $S_r$.
Then, there exists $I\subset[t]$ such that
both $W=\bigcup_{i\in I}W_i$ and $W'=\bigcup_{i\in[t]\setminus I}W_i$
contain at least $\widetilde{N}/3k=\widehat{N}/6k$ sets $S_r$.
But, then a contradiction is reached,
because by~(2) there must be an edge in $G'$
between some vertex of $W$ and some vertex of $W'$.

Hence, each $G'_i$ contain a connected component,
denoted $C_i$, that has at least $\widetilde{N}/3k$ sets $S_r$.
Applying~(2) again, we infer that there must be an edge in $G'$
going between any two of the $C_i$'s.
By contracting each $C_i$ to a node, we obtain a $K_k$-minor in $G'$,
which contradicts the hypothesis that $G'$ is $K_k$-minor free.
\EPF

\medskip
It remains to show that, with high probability,
a graph $G$ drawn from $\cG_N$ satisfies the two conditions
stated in Claim~\ref{special.clm}.
Recalling that graph $G$ in $\cG_N$ consists of a Hamiltonian cycle
augmented by a matching, we obtain the desired
sets $S_1, \ldots, S_{\widehat{N}}$,
by partitioning the Hamiltonian cycle into $\widehat{N} = 2\e N$
contiguous segments, each of length $1/2\e$.
Clearly, these $S_i$'s satisfy the first condition
(i.e., the subgraph of $G$ induced by each $S_i$ is connected).
We shall show that the second condition holds too,
by considering all relevant sets $U$ and $U'$,
showing that, with probability $1- \exp(-\Omega(N/k^2))$
(over the choice of a random matching),
there are $\Omega(N/k^2)>\e N+1$ edges going between $U$ and $U'$,
and applying a union bound.

Specifically, we fix two arbitrary disjoint collections
of sets, $C$ and $C'$, such that $|C|=|C'|=\widehat{N}/6k$,
and consider the sets $U=\bigcup_{i\in C}S_i$
and $U'=\bigcup_{i\in C'}S_i$.
Since $|U|=|U'|=(\widehat{N}/6k)\cdot(1/2\e)=N/6k$,
we expect the number of edges between $U$ and $U'$ to be $N/(36k^2)$.
Intuitively, with very high probability
(i.e., with probability $1- \exp(-\Omega(N/k^2))$),
the number of edges is a constant fraction of its expected size;
in fact, this is the case as shown in Claim~\ref{many-edges.clm} (below).

Applying a union bound over all possible choices of $C$ and $C'$
(which underlie the choice of $U$ and $U'$),
we infer that the second condition stated in Claim~\ref{special.clm}
is satisfied with probability at least
$1-{{2\e N}\choose{2\e N/(6k)}}^2\cdot\exp(-\Omega(N/k^2))$,
which is lower bounded
by $1 - 2^{4\e N}\cdot \exp(-\Omega(N/k^2))=1-\exp(-\Omega(N/k^2))$,
since $\e=1/ck^2$ for a sufficiently large $c$
(which is determined at this point to make assertion hold).
Thus, modulo Claim~\ref{many-edges.clm} (below),
the theorem follows.
%

\BCM\label{many-edges.clm}
For some universal constant $c'>0$,
the following holds for all $N$ and $t$.
Consider selecting a matching between $N$ vertices uniformly at random,
and let $T_1$ and $T_2$ be two disjoint sets of $N/t$ vertices each.
Then, with probability at least $1-\exp(-c'\cdot N/t^2)$,
over the choice of the matching edges,
there exist $c' N/t^2$ edges going between these sets.
\ECM
\BPF
A random matching can be selected in $N/2$ steps, where at each step
we pick an arbitrary yet-unmatched vertex $v$, and select, uniformly at
random, another yet-unmatched vertex $u$ to match $v$ to.
In particular, we can start by matching the vertices in $T_1$,
and once they are all matched we continue in an arbitrary order
with the remaining unmatched vertices.

Observe that the number of steps it takes to match all vertices
in $T_1$ is at least $N'\eqdef N/(2t)$,
and we shall lower bound the number of edges obtained
between $T_1$ and $T_2$ (only) in the first $N'$ steps.
Let $X_1,\ldots,X_{N'}$ be $0/1$ random variables,
where $X_i =1$ if and only if the selected edge in the $i^\th$
step has as its second endpoint a vertex in $T_2$.
By the definition of the matching process, $\Pr[X_1=1] = N'/(N-1)$.
More generally, $\Pr[X_i=1]$
equals the fraction of yet-unmatched vertices in $T_2$
at the start of the $i^\th$ step over $N-2(i-1)-1$.
Since we consider only the first  $N'$ steps,
which means that at most half of the vertices in $T_2$ can be matched,
we have that $\Pr[X_i=1] \geq \frac{N'}{N-2i+1} > \frac{1}{2t}$
for every $i\in\{1,2,...,N'\}$.
Hence, the expected value of  $\sum_{i=1}^{N'} X_i$,
which is a lower bound on the expected number of edges
between $T_1$ and $T_2$, is at least $N'/2t=N/(4 t^2)$.

We would like to show that
\BEQ\label{sumXi.eq}
\Pr\left[\sum_{i=1}^{N'} X_i < \frac{N}{8 t^2}\right]
  < \exp(-\Omega(N/t^2))\;.
\EEQ
Since the $X_i$'s are not independent random variables
(as the probability that $X_i = 1$ depends on $X_1,\ldots,X_{i-1}$),
we cannot simply apply a multiplicative Chernoff bound in order
to obtain Equation~(\ref{sumXi.eq}).
However, we shall define a related sequence of independent random variables
that will give us the bound in the equation.

For every $x_1,...,x_{N'}\in\bitset$ and every $i\in[N']$,
let $f(x_1\cdots x_{i-1})$ denote the probability that $X_i=1$
conditioned on $X_j=x_j$ for every $j\in[i-1]$ (i.e.,
$f(x_1\cdots x_{i-1})=\Pr[X_i=1|X_1\cdots X_{i-1}=x_1\cdots x_{i-1}]$).
Recall that $f(x_1\cdots x_{i-1})\geq1/2t$.
Define random variables $Y_1,...,Y_{N'}$
such that $Y_i$ depends on $X_1,...,X_{i-1}$
and $\Pr[Y_i=1]=\frac{1/2t}{f(X_1\cdots X_{i-1})}$.
Lastly, define $Z_1,...,Z_{N'}$ such that $Z_i=1$
if and only if $X_i=Y_i=1$.
Observe that for every $x_1,...,x_{i-1}\in\bitset$
and $z_1,...,z_{i-1}\in\bitset$ it holds that
\begin{eqnarray*}
\lefteqn{\Pr[Z_i=1|X_1\cdots X_{i-1}=x_1\cdots x_{i-1}
      \wedge Z_1\cdots Z_{i-1}=z_1\cdots z_{i-1}]} \\
&=& \Pr[X_i=1|X_1\cdots X_{i-1}=x_1\cdots x_{i-1}]
    \cdot\frac{1/2t}{f(x_1\cdots x_{i-1})} \\
&=& \frac{1}{2t}
\end{eqnarray*}
Hence, the $Z_i$'s are independent random variables
and $\sum_{i=1}^{N'}Z_i\leq \sum_{i=1}^{N'} X_i$
always holds (since $Z_i =1$ occurs only when $X_i = 1$).
Now the claim (or rather Equation~(\ref{sumXi.eq}))
follows by applying a multiplicative Chernoff bound
to the sum of the $Z_i$'s.
\EPF

\medskip\noindent
As stated above,
the proof of Claim~\ref{many-edges.clm}
completes the proof of the theorem.
\EPF

\section{Testing Tree-Minor Freeness}
\label{tree.section}
As noted in Section~\ref{lower-bound.sec},
the $\Omega(\sqrt{N})$ lower bound of Theorem~\ref{lower-bound.thm}
does not hold in the case that the forbidden minor is a tree.
This is easiest to see in the case that the forbidden
minor is a single edge.
We show that, for any cycle-free graph $H$,
the set of $H$-minor free graphs can be tested with one-sided
error with query complexity that is independent of the input graph's
size (and that only depends on the proximity parameter and on $H$).

To begin, we provide a reduction of the case where $H$ is a forest
to the case where $H$ is a tree.
Actually, this reduction works for any $H$ (regardless of cycle-freeness)
allowing to focus on the connected components of $H$.
Next, we turn to two special cases (which are easy to handle):
the case that $H$ is a $k$-path
and the case that $H$ is a $k$-star.
Since these cases correspond to the two possible extremes,
it is tempting to hope that all cases can be treated easily.
We warn, however,
that the extreme cases have simple characterizations,
which are not available in non-extreme cases.
Nevertheless, the case of stars provides some intuition
towards the more complicated treatment of general trees.
\ifnum\verTR=1    
Further intuition can be obtained from the case of depth-two trees,
treated in \secref{tree-depth2.sec},
where we also obtain better complexity than in the general case.
\fi

\subsection{A reduction of unconnected $H$ to connected $H$}
Let $H$ be a graph with connected components $H_1,\dots,H_m$.
Then, essentially (but not exactly),
a graph $G$ is $H$-minor free if and only if
for some $i\in[m]$ the graph $G$ is $H_i$-minor free;
in other words, $G$ has an $H$-minor if and only if for every $i\in[m]$
the graph $G$ contains an $H_i$-minor.
The alternative formulation reveals the small inaccuracy:
it may be that the $H_i$-minors contained in $G$ are not disjoint
(and in such a case $G$ does not necessarily have an $H$-minor).
Still, for our purposes (of studying one-sided error testers
of sublinear query complexity),
this problem can be overcome (as done next).

Indeed, we focus on one-sided error testers
of sublinear query complexity.
Given such testers for $H_i$-minor freeness,
we present the following one-sided error tester for $H$-minor freeness.

\BA{\em(the $H$-minor-freeness tester for cycle-free $H$):}
\label{reducing:forest-tree.alg}
On input $G=([N],E)$ and proximity parameter $\e$,
set $G_0=G$ and proceed in $m$ iterations, as follows.
For $i=1$ to $m$,
\BE
\item Invoke the $H_i$-minor tester on input $G_{i-1}$,
using error parameter $1/3m$ and proximity parameter $\e/2$.
\item If the answer is positive then accept.
\item Otherwise, omit from $G_{i-1}$ all vertices that were visited
by the tester, obtaining a residual graph $G_i$.
\EE
If all iterations rejected, then reject.
\EA
If Algorithm~\ref{reducing:forest-tree.alg} rejects,
then (by the one-sided error feature of the tests)
the $m$ exploration contain corresponding (disjoint) $H_i$-minors,
and so $G$ contains an $H$-minor.
Thus, Algorithm~\ref{reducing:forest-tree.alg} satisfies
the one-sided error condition.
On the other hand, if $G$ is $\e$-far from being $H$-minor free,
then, for every $i\in[m]$, the graph $G$ must be $\e$-far from
being $H_i$-minor free (because otherwise $G$ is $\e$-close
to an $H_i$-minor free graph, which in turn is $H$-minor free).
Furthermore, for every $i\in[m]$, the graph $G_{i-1}$
is $\e/2$-far from being $H_i$-minor free,
because $G_{i-1}$ is obtained from $G$ by omitting $o(N)$ edges
(since all testers have sublinear query complexity).
Thus, in each iteration $i$, with probability at least $1-(1/3m)$,
the corresponding tester rejects.
It follows that Algorithm~\ref{reducing:forest-tree.alg}
rejects $G$ with probability at least $2/3$ (as required).
We thus get the following result.

\BP
Let $H$ have connected components $H_1,\dots,H_m$,
and suppose that $H_i$-minor freeness can be tested
by a one-sided error tester of query complexity $q_i(N,\e)$.
Suppose that $q_i(N,\e)$ is monotonically non-decreasing with $N$.
Then, $H$-minor freeness can be tested by a one-sided error tester
of query complexity $q(N,\e) = O(\log m)\cdot\sum_{i=1}^n q_i(N,\e/2)$.
\EP
(The $O(\log m)$ factor is due to error reduction that is
employed on each of the testers.)

\paragraph{Detour.}
For sake of elegance, it would be nice to prove a similar reduction
also for the case of two-sided error testers.
Naturally, for testing $H$-minor freeness with two-sided error,
we may just run all $H_i$-minor freeness tests
(with error probability parameter set to $1/3m$)
and accept if and only if at least one of these tests accepted
(i.e., reject if and only if all these tests rejected).
Clearly, if $G$ is $\e$-far from being $H$-minor free,
then, for every $i$, the graph $G$ must be $\e$-far from
being $H_i$-minor free (see above), and so in this case,
with probability at least $2/3$,
all tests will reject, and so will we.
But what is missing is proving that if $G$ is $H$-minor free,
then the above tester accepts with high probability.
(Indeed, it is not necessarily the case that if $G$ is $H$-minor free
then for some $i$ it holds that $G$ is $H_i$-minor free).

\subsection{Testing that the graph contains no simple $k$-length path}
Here we consider the special case where $H=P_k$,
where $P_k$ denotes the $k$-length path.
Note that a graph $G$ is $P_k$-minor free
if and only if $G$ contains no simple path of length $k$.
Thus, we just search for such a path at random.
Specifically, we select uniformly a start vertex
and take a random $k$-step walk, rejecting if and only if
the walk corresponds to a simple path.
Clearly, we never reject a $P_k$-minor free graph.

\BCM
If $G$ is $\e$-far from being a $P_k$-minor free graph,
then we reject with probability at least $\e/2d^k$.
\ECM
Thus, $P_k$-minor freeness can be tested by a one-sided error tester
of query complexity $q\eqdef O(d^k k/\e)$
and time complexity $O(q)$.
%
We note that it may be possible to reduce the query complexity
to $\poly(dk/\e)$, but an analogous improvement of the time
complexity is unlikely (because finding $k$-long paths
in graphs is NP-Hard, when $k$ is part of the input).
We mention that, subsequent to our initial posting of this work,
Reznik~\cite{Reznik} presented a $\poly(dk/\e)$-time algorithm
for the special case that the input graph is cycle-free.
\medskip

\BPF
We call a vertex $v$ {\tsf bad} if there is a simple path
of length $k$ starting at $v$. Let $\rho$ denote the
density of bad vertices in $G$. Then, on the one hand,
we reject $G$ with probability at least $\rho/d^k$.
On the other hand, $\rho \geq \e/2$, because omitting all
bad vertices (or rather their incident edges) from $G$
we obtain a graph that has no simple $k$-length paths.
\EPF

\subsection{Testing that the graph contains no $k$-star as a minor}
\label{k-star.sec}
Here we consider the special case where $H=T_k$,
where $T_k$ denotes the $k$-star
(i.e., the $(k+1)$-vertex tree that has $k$ leaves).
%
The key observation here is  that a graph $G=([N],E)$
is $T_k$-minor free
if and only if for every set $S$ such that $G_S$ is connected
it holds that the set $S$ has less than $k$ neighbors
(in $[N]\setminus S$).
%
This implies that if for every connected set $S$ of size at most $k/\eps$,
the set $S$ has less than $k$ neighbors in $[N]\setminus S$,
then the graph is $\eps$-close to being $T_k$ minor free.
The reasoning (which is detailed in the proof of Claim~\ref{k-star.clm})
is that if the premise of the statement (the
small-cuts condition) holds, then
by removing less than $k \cdot d \cdot (N/(k/\eps))= \eps d N$ edges
we can partition the graph into connected components such that
each is $T_k$-minor free.
Another useful observation is that
searching for sets that violate the condition can be 
done efficiently by performing a BFS with a bounded depth (and width)
and running a polynomial-time procedure on the subgraph
induced by the BFS.

\BA{\em(the $k$-star-minor-freeness tester):}
\label{k-star.alg}
On input $G=([N],E)$ and proximity parameter $\e$, proceed as follows.
\BE
\item Select uniformly a start vertex $s\in[N]$.
\item Perform a BFS starting at $s$ and stopping
as soon as either $2k/\e$ layers were explored
or a layer with at least $k$ vertices was encountered.

Note that it may also be that the BFS terminates before
either of these conditions hold; this can only happen if $s$
resides in a connected component of size smaller than $2k^2/\e$.
\item
Accept if and only if the explored graph is $T_k$-minor free.
\EE
\EA
Clearly, Algorithm~\ref{k-star.alg} never rejects
a $T_k$-minor free graph.
The query complexity of Algorithm~\ref{k-star.alg} is
$q(k,\eps) = O(k^2/\e)$
(the maximum depth of the BFS times the maximum number of vertices
in each level, assuming the degree $d$ is a constant).
By Corollary 1.2 in~\cite{KaKoRe}, the time complexity is
of the form $f(k)\cdot O(q^2(k,\eps))$ for some function $f$
(which is not explicitly specified in~\cite{KaKoRe}).
Thus, all that is left is to prove the following claim.

\BCM
\label{k-star.clm}
If $G$ is $\e$-far from being a $T_k$-minor free graph,
then Algorithm~\ref{k-star.alg} rejects with probability at least $\e/2$.
\ECM
Thus, $T_k$-minor freeness can be tested by a one-sided error tester
of query complexity $O(k^2/\e^2)$
and time complexity $f(k)\cdot O(1/\eps^3)$ for some
function $f$. 
\medskip

\BPF
We call a vertex $v$ {\tsf bad} if
there exists a set $S$ containing $v$ such that
(i) $G_S$ is connected and has radius at most $2k/\e$ from $v$
(i.e., all vertices in $S$ are at distance at most $2k/\e$ from $v$),
and (ii) the set $S$ has at least $k$ neighbors in $G$
(i.e.,
$|\{u\!\in\![N]\setminus S:\exists w\!\in\!S\mst\{u,w\}\!\in\!E\}|\geq k$).
Note that if a bad vertex is chosen in Step~1,
then Algorithm~\ref{k-star.alg} rejects in Step~3
(because either a $2k/\e$-step BFS of $G$ starting at $v$
reaches a layer with at least $k$ vertices, or it reaches
all vertices in the witness set $S$).
Let $\rho$ denote the fraction of bad vertices in $G$.
By the above, Algorithm~\ref{k-star.alg} rejects
with probability at least $\rho$.
We next show that $G$ must be $(\rho+(\e/2))$-close to $T_k$-minor free,
and so $\rho\geq\e/2$ follows.

Let $G^{(0)}$ denote the graph obtained from $G$ by omitting
all the edges that are incident at bad vertices.
Indeed, $G^{(0)}$ is $\rho$-close to $G$.
The rest of our analysis proceed in iterations.
If the current graph $G^{(i-1)}$ is $T_k$-minor free,
then we are done. Otherwise, we pick an arbitrary
vertex $s^{(i)}$ that resides in some $T_k$-minor.
Since $s^{(i)}$ is not bad, it must reside in a connected component
of $G^{(i-1)}$ that has radius at least $2k/\e$ from $s^{(i)}$
(because otherwise the existence of a $T_k$-minor containing $s^{(i)}$
contradicts the hypothesis that $v$ is not bad).
Consider an arbitrary set $S^{(i)}\ni s^{(i)}$ of $2k/\e$ vertices
such that $G^{(i-1)}_{S^{(i)}}$ is connected.
Since $s^{(i)}$ is not bad,
it follows that $S^{(i)}$ has less than $k$ neighbors (in $G^{(i-1)}$).
We now obtain $G^{(i)}$ by omitting from $G^{(i-1)}$
 the (less than $k d$) edges
of the cut $(S^{(i)},[N]\setminus S^{(i)})$,
and observe that $G^{(i)}_{S^{(i)}}$ is $T_k$-minor free
(and that $S^{(i)}$ will not intersect with any future $S^{(j)}$).
When the process ends, we have a $T_k$-minor free graph.
In total, we omitted at most $ t k\cdot d$ edges (from $G^{(0)}$),
where $t \leq N/(2k/\e)$ denotes the number of iteration.
Noting that $tdk \leq (\e/2)d N$,
we conclude that $G^{(0)}$ is $\e/2$-close to $G^{(t)}$
and thus $G$ is $(\rho+(\e/2))$-close to $T_k$-minor free.
\EPF

\subsection{The general case: Testing $T$-minor freeness for any tree $T$}
\label{general-tree.sec}
Following is a presentation of the main result of this section:
a one-sided tester for $T$ minor-freeness,
where $T$ is an arbitrary rooted tree with $k$ vertices.
The algorithm is 
an extension of the algorithm for stars:
We perform a BFS from a random starting vertex (but for more levels)
and check if we find a $T$-minor.

The analysis of this algorithm, in the current (general) case,
is far more involved;
nonetheless, the basic intuition remains the same.
Suppose our procedure is typically unable to find a $T$-minor in $G$.
We shall show that we can split up the graph into many small pieces,
each being $T$-minor free and having few edges leaving it.
Removing the few edges going between these pieces,
we get a $T$-minor free graph, which proves that $G$ is close
to  being $T$-minor free.

The main challenge is to perform the foregoing decomposition.
For that, we will define an auxiliary procedure, called $\find$,
that attempts to find $T$-minors.
This procedure will not be used by our algorithm;
it will be used solely in the analysis.
\ifnum\submit=0
But, first, let us detail the alleged tester.
\else
But, first, let us recall the alleged tester.
\fi
In all that follows we assume that $\eps \leq \eps_0$ for
some sufficiently small constant $\eps_0$ (or else we can run
the algorithm with $\eps$ set to $\eps_0$).

\ifnum\submit=0
\BA{\em(the tree-minor-freeness tester):}
\label{general-tree.alg}
Given as input a proximity parameter $\e$ and given
query access to a graph $G=([N],E)$ with maximum degree at most $d$,
set $D=k\cdot(4d/\e)^{4k+2}$ and proceed as follows.
\BE
\item
Select uniformly, independently at random, $8/\eps$ start vertices in $[N]$.
\item For each selected start vertex $s$, perform a BFS starting at $s$
and stop as soon as $D$ layers are explored (or the BFS reaches all
the vertices of a connected component in $G$).
\item
Accept if and only if all explored subgraphs are $T$-minor free.
\EE
\EA
\else
\medskip\noindent{\bf Algorithm~\ref{general-tree.alg}.}~
{\it
Given as input a proximity parameter $\e$ and given
query access to a graph $G=([N],E)$ with maximum degree at most $d$,
set $D=k\cdot(8d/\e)^{4k+2}$ and proceed as follows.
\BE
\item
Select uniformly, independently at random, $4/\eps$ start vertices in $[N]$.
\item For each selected start vertex $s$, perform a BFS starting at $s$
and stop as soon as $D$ layers are explored (or the BFS reaches all
the vertices of a connected component in $G$).
\item
Accept if and only if all explored subgraphs are $T$-minor free.
\EE
}

\fi
\ifnum\verTR=0    
We note that better complexity bounds can be obtained in the case
that $T$
is of depth-two; specifically, $D$ can be set to $\poly(k/\e)$,
rather than being exponential in $k$.
The interested reader is referred to our TR~\cite{tr}.
\fi
Clearly, Algorithm~\ref{general-tree.alg} never rejects
a $T$-minor free graph. Its query complexity is exponential in $D$,
and its time complexity is polynomial in its query complexity
(for constant $k$, by Corollary 1.2 of~\cite{KaKoRe}).
\ifnum\submit=0
The correctness of the algorithm
thus follows from the next theorem.
\BT
\label{general-tree.thm}
If $G$ is $\e$-far from being a $T$-minor free graph,
then Algorithm~\ref{general-tree.alg} rejects with probability
at least $2/3$.
\ET
\else
The correctness of the algorithm thus follows from Lemma~\ref{general-tree.lem},
which is restated below.

\ifnum\submit=0
\medskip\noindent{\bf Lemma~\ref{general-tree.lem}}~
\else
\medskip\noindent{\bf Lemma~\ref{general-tree.lem}.}~
\fi
{\it
If $G$ is $\e$-far from being a $T$-minor free graph,
then Algorithm~\ref{general-tree.alg} rejects with probability
at least $2/3$.
}

\fi
As noted in the introduction,
one of the byproducts of our analysis is a combinatorial theorem claiming
that any graph with ``local" expansion must contain all tree minors of some
constant size.
\BD
\label{local-expander.def}
Let $G$ be a graph of maximum degree $d$ and $s$ be
a vertex of $G$. We say that the
{\dsf $R$-neighborhood of $s$ in $G$ is $\e$-expanding}, if
for every vertex set $S$ such that $\max_{u\in S}\{\dist(v,u)\}\leq R$,
it holds that the number of edges in the cut $(S,[N]\setminus S)$
is at least $\e|S|d$.
\ED

\BT
\label{local-expander.thm}
For any $k$ and $d$,
if the $k(4d/\e)^{4k+2}$-neighborhood of $s$ in $G$
is $\e$-expanding, then this neighborhood contains
a $T$-minor of any tree $T$ of at most $k$ vertices.
\ET
Both theorems are proved using
a procedure called $\find$, which tries to find small $T$-minors.
When invoked at a certain vertex and failing to find a small $T$-minor,
the procedure provides us with a sort of ``explanation for its failure''
in the form of a sparse cut, that is, a cut with relatively
few edges crossing it.
Thus,
if the graph $G$ is accepted by the tester with high probability,
then we can use this procedure to get the desired decomposition.
As may be expected, the procedure $\find$ is designed
by a (tedious, but not obvious) induction on the size of $T$.
Following is an overview of our approach.

Consider the tree $T$ and remove an edge so as
to obtain two trees $T_1$ and $T_2$.
Let the roots of these trees be the endpoints of the edge
removed. A $T$-minor can be broken up into a $T_1$-minor and
$T_2$-minor with a path connecting the two respective roots.
So, it seems that we should try to find ``rooted minors",
where we specify a vertex $v$ that must be present in the connected
component that is the root. Inductively, assume that we
have a procedure $\find$ for $T_1$ and $T_2$. We can use $\find$ to get these
minors and try to connect the roots by a path. The problem
is that we have to get \emph{disjoint} minors to get
a $T$-minor. Suppose we find a $T_1$-minor in the original graph.
Because we want to find a disjoint $T_2$-minor, we make the
vertices in this minor a \emph{forbidden} set $F$ (and effectively
remove them from $G$). This means
that $\find$ is not allowed to use the vertices of $F$ in
the $T_2$-minor. But now, $\find$ may return a sparse cut,
instead of a $T_2$-minor, in the modified graph. This cut,
between a set $S$ and the rest of the vertices,
is only sparse in the modified graph (without $F$),
but it is possible that it is not sparse in the graph $G$.
That is, there may be
many edges between $S$ and the rest of the vertices in $G$,
which include the vertices in $F$.
To get around this,
we somehow need to ensure that whenever a cut is found, the
number of vertices in the smaller side of the cut is much larger
than $|F|$. Then, a sparse cut in the modified graph
remains sparse in the original. We will give an indication
of how this is done when we describe the parameters of $\find$.

\subsubsection{Setting the stage}

We introduce some definitions and notation, after which we can formally
express
the claim about the procedure $\find$. Given that claim, we will prove
Theorem~\ref{general-tree.thm} and Theorem~\ref{local-expander.thm}.
In the next subsection, we will
prove the claim about $\find$.
For a graph $H= (V(H),E(H))$ and a subset of vertices
$S \subseteq V(H)$, we use the standard notation $H_S$
to denote the subgraph of $H$ that is induced by $S$.
\BD[Distances] \label{rad.def}
Let $H = (V(H),E(H))$ be a fixed graph.
For any pair of vertices $v,u \in V(H)$, let $\dist_{H}(v,u)$ be
the shortest-path distance between $u$ and $v$ in $H$.
Given a set of vertices $T\subset V(H)$ and a vertex $v\in V(H)$,
let $\Delta_H(v,T) \eqdef \max_{u \in T}\{\dist_H(v,u)\}$.
More generally,
for two sets of vertices $S,T \subseteq V(H)$, let
$\Delta_H(S,T) \eqdef \max_{u \in T} \min_{v \in S} \dist_H(v,u)$.
\ED

\BD[Sparse Cuts]
For a graph $H = (V(H),E(H))$ with degree bound $d$,
a cut $(S,V(H)\setminus S)$ is {\dsf $\f$-sparse\/} with respect to $H$,
if the number of edges in $E(H)$ that cross the
cut is at most $\f |S|d$.
We denote the cut $(S,V(H)\setminus S)$ by $cut_H(S)$.
\ED

\paragraph{The parameters of $\find$:}
The procedure $\find$ takes as input a vertex $v$ in a graph $G'=([N],E')$,
a set of vertices $U$ containing $v$, a rooted tree $T$ with $k$
nodes, and a set of \emph{forbidden} vertices $F$ (not containing $v$).
There is also a proximity parameter $\f$, but since it is fixed for our
discussion, we will not consider it as an input parameter. We will set
$\f = \Theta(\e)$ for the final analysis. For now, it will be convenient
to have it as a separate parameter.

Let $f =  \max\{|F|,k(4d/\f)^{4k+2}\}$,
and $G'' = G'_{[N]\setminus F}$.
The procedure works under the conditions that $U$ is disjoint from $F$,
$|U| \geq 4f/\f$, and $\Delta_{G''}(v,U) \leq (4/\f)\ln(f/\f)$.
The procedure $\find(v,U,T,F)$ outputs a pair $(\sigma,S)$
such that $\sigma\in\{\minor,\cut\}$
and $S\subseteq[N]\setminus F$, where there is a path in
$G''$ between $v$ and every vertex in $S$.
It will be convenient to express quantities in terms
of $\kh = 4k-2$.

\paragraph{The requirement from $\find$:}
The output $(\sigma,S)$ of $\find(v,U,T,F)$ should satisfy
the following conditions.
\BI
\item  $\Delta_{G''} (v,S) \leq (4d/\f)^{\kh}\ln(f/\e)$.
\item
\BDes
\item[If $\sigma=\minor$,]
then the graph $G'_{S}$ contains a $T$-minor not involving
$F$ that is rooted at $v$
(i.e., $v$ resides in the connected component that is contracted
to fit the root $r$ of $T$).

\item[If $\sigma=\cut$,]
Then the cut $cut_{G'}(S)$ is $\f$-sparse. 

\EDes
\EI
Intuitively,
the set $U$ acts as a kind of large buffer around $v$. This deals
with the issue that we raised earlier. When we try to find a $T_2$-minor
by making the vertices of the $T_1$-minor forbidden, we could
get a sparse cut in this modified graph. The buffer $U$ ensures
that this cut contains sufficiently many vertices.

\BCM \label{find-proc.clm}
There exists a procedure $\find$
that satisfies the foregoing requirements.
\ECM

We  next show how Theorem~\ref{general-tree.thm} and
Theorem~\ref{local-expander.thm} can be proved based on this claim.
In what follows, when we say we perform a BFS in a graph
$H = (V(H),E(H))$ {\tsf from a subset of vertices $M$},
we mean the following.
Consider the graph $H'(M)$
whose vertex set is $(V(H)\setminus M)\cup \{v(M)\}$ (so that
$M$ is replaced by a single vertex $v(M)$), and whose edge set
is $\{(u,w)\in E(H)\,:\,u,w \in V(H)\setminus M\}
     \cup \{(u,v(M))\,:\,u \notin M \mbox{ and }
          \exists w \in M \mbox{ s.t. } (u,w) \in E(H)\}$.
A BFS from $M$ in $H$ corresponds to a BFS in $H'(M)$ that
starts from $v(M)$.
We first state a fairly simple claim that gives one
of the basic ``dichotomies" that we will repeatedly use:
Either a BFS starting from some set $M$ finds
a $\f$-sparse cut, or it leads to a level set that is quite large.
We will introduce a forbidden set $F$, and make all our arguments
with respect to $G'_{[N] \setminus F}$.

\BCM \label{cut-or-exp.clm}
Let $F$ and $M$ be two disjoint subsets of vertices in $G'$
such that $|M| \geq (2/\f)|F|$.
Suppose we perform a BFS up to depth $t$
in $G'_{[N] \setminus F}$, starting from $M$, and
let $\ell$ be the size of the last level reached.
Then either there exists a subset of vertices $R$ that are reached
by the BFS and such that $cut_{G'}(R)$
is $\f$-sparse, or $\ell \geq |M|\cdot e^{(\f/3) t}$.
\ECM

\BPF
Consider some intermediate level in the BFS, and let $R$ be
the set of vertices reached up to that level (including it). Suppose
that the next level has at most $\f |R|/2$ vertices.
All edges in $\cut_{G'}(R)$
are either incident to vertices in the next level (which contains at most
$\f|R|/2$ vertices) or to $F$. Since $|R| \geq |M| \geq 2|F|/\f$, the size
of the cut is at most $\f |R|d$,  and hence it is $\f$-sparse.

Otherwise, the size of the levels keeps expanding by a factor
of at least $(1+\f/2)$.
Since the depth of the BFS is $t$, 
the size of the last level is at least
$|M|\cdot (1+\f/2)^{t} \geq |M| \cdot e^{(\f/3)t}$.
\EPF

\bigskip


\BPFOF{Theorem~\ref{general-tree.thm}}
Recall that $D=k\cdot(8d/\e)^{4k+2}$, and that
Algorithm~\ref{general-tree.alg} performs a BFS
from $4/\eps$ start vertices, up to depth $D$ for each,
and rejects if any of the subgraphs observed contains
a $T$ minor.
We call a vertex $v$ {\tsf bad} if its $D$-neighborhood
(i.e., the subgraph induced by all vertices at distance at most $D$ from $v$)
contains a $T$-minor, and denote the fraction of bad vertices
(in $G$) by $\rho$.
We shall show that $G$ is $(\rho+\e/2)$-close to
being $T$-minor free. The lemma follows since this implies
that if $G$ is $\eps$-far from being $T$-minor free, then
$\rho > \eps/2$. In such a case, the probability that
no bad vertex is selected as a start vertex by the algorithm
is at most $(1-\eps/2)^{4/\eps} < e^{-2} < 1/3$.

In order to prove that $G$ is $(\rho+\e/2)$-close to
being $T$-minor free, we will remove at most
$(\rho+\e/2)d N$ edges from $G$ to make it $T$-minor free.
We start by removing all edges incident to bad vertices,
so that the number of edges removed at this stage is
at most $\rho d N$.
Let the resulting graph be $G^{(0)}$.
The rest of our analysis proceed in iterations
in which we invoke the procedure $\find$
with proximity parameter $\f = \e/2$. Note that $D = k(4d/\f)^{4k+2}$.
At the start of each iteration
we have a current graph $G^{(i-1)}$ where some
connected components are marked ``minor free''.
These components are certified to have no
$T$-minor. If all the components are marked,
then we are done.
Otherwise, consider some unmarked component
$C$. Suppose there is $v \in C$, such that
$\Delta_{G^{(i-1)}}(v,C) \leq D$.
If $C$ contains a $T$-minor, then $v$ must be bad.
This contradicts that fact that $C$ is a connected
component containing $v$. Therefore $C$ has no $T$-minor,
and can be marked. We proceed in this fashion till
we get a component $C$ that cannot be marked.

For such an unmarked component $C$
we take an arbitrary vertex $s^{(i)} \in C$ and
observe that $\Delta_{G^{(i-1)}}(s^{(i)},C) > D$.
Let $F$ be initialized to $\emptyset$.
Set $f = \max(|F|,k(4d/\f)^{4k+2})$.
We perform a BFS from $s^{(i)}$ up to depth $D_0 = (3/\f)\ln(4f/\f)$
steps, and invoke  Claim~\ref{cut-or-exp.clm} with
$M = \{s^{(i)}\}$, $F = \emptyset$, and $t = D_0$.
We have $f = k(4d/\zeta)^{4k+2}$ (and $k \geq 1$), so
\begin{eqnarray}
D_0 = (3/\f)\ln(4f/\f) & \leq & (4/\f)\ln(4k(4d/\f)^{4k+2}/\f) \nonumber\\
& = & (4/\f)\ln(4k/\f) + ((16k+8)/\f)\ln(4d/\f) \leq k(4d/\f)^{4k+2} = D\;.
\label{D0-D.eq}
\end{eqnarray}
Suppose
we get a set $S^{(i)}$ such that $cut_{G^{(i-1)}}(S^{(i)})$
is $\f$-sparse.
By Claim~\ref{cut-or-exp.clm} and the bound in
Equation~(\ref{D0-D.eq}),
$\Delta_{G^{(i-1)}}(s^{(i)},S^{(i)}) = D_0 \leq D$.
Hence, the subgraph $G_{S^{(i)}}$ cannot contain
a $T$-minor. We remove
all edges in the cut $cut_{G^{(i-1)}}(S^{(i)})$
and mark the connected components in $G_{S^{(i)}}$ as minor free.
This gives us the graph $G^{(i)}$, and we continue with the next iteration.

Otherwise, the BFS gives a set $U$,
such that $|U| \geq e^{(\f/3) D_0} \geq 4f/\f$,
and $\Delta_{G^{(i-1)}}(s^{(i)},U) \leq D_0 \leq (4/\f)\ln(f/\f)$.
We therefore call $\find(s^{(i)},U,T,F)$ (on
the graph $G' = G^{(i-1)}$). If it outputs $(\minor,S^{(i)})$,
then $s^{(i)}$ must be bad. This is a contradiction, and hence
the output must be $(\cut,S^{(i)})$.
We have $\Delta_{G^{(i-1)}}(s^{(i)},S^{(i)}) \leq D$, where
$cut_{G^{(i-1)}}(S^{(i)})$ is $\f$-sparse. We proceed as before by
removing all edges in $cut_{G^{(i-1)}}(S^{(i)})$ to get $G^{(i)}$.
%

When the process ends, we have a $T$-minor free graph.
Since all the $S^{(i)}$'s considered are disjoint,
in total,
we omitted at most $\sum_i \f d|S^{(i)}| \leq \e d N/2$ edges
(from $G^{(0)}$),
and thus $G$ is $(\rho+\e/2)$-close to $T$-minor freeness.
\EPFOF

%

\medskip
\BPFOF{Theorem~\ref{local-expander.thm}}
The theorem follows from Claim~\ref{find-proc.clm},
where the key observation is that $\find$ works
for any $k$-vertex tree $T$ and that $\find$ may not return a
$\f$-sparse cut (because no such cut exists by the hypothesis).
Specifically, set $\f = \e$, $F = \emptyset$,
and $f = k(4d/\e)^{4k+2}$.
Let $U$ be a set
such that $|U| \geq 4f/\e$ and $\Delta_{G}(v,U) \leq (4/\e)\ln(f/\e)$
(which exists since the said neighborhood contains no $\f$-sparse cuts).
Now, for any $k$-vertex tree $T$,
we run $\find(v,U,T,F)$ and get the output $(\sigma,S)$,
where $\sigma \neq \cut$. Thus, we get the desired $T$-minor.
\EPFOF

\subsubsection{The procedure $\find$} \label{find.sec}
We first introduce the notion of a boundary.
\BD[Boundaries]
Given sets of vertices $S$ and $F$,
let $\bound{F}{S}$ denote the {\dsf boundary} of $S$
in $G'_{[N]\setminus F}$. That is,
$\bound{F}{S} \eqdef\{u\in S\,:\,\exists w \in[N]\setminus (S\cup F) \mbox{ s.t. }
                (u,w) \in E(G')\}$ .
We use  $\nbound{F}{S}$ to denote the set
$S \setminus \bound{F}{S}$.
\ED

We build a little on the basic dichotomy of Claim~\ref{cut-or-exp.clm}.
Given a starting set $M$ and forbidden set $F$, we try to find a
$\f$-sparse cut that is not
too far from the boundary $\bound{F}{M}$. If we fail, then we can find
a vertex $v \in \bound{F}{M}$ and a set $U_v \ni v$ disjoint from $F$,
such that $U_v$ is large, but vertices in $U_v$ are quite close to $v$.

\BCM \label{cut-or-good.clm}
Let $F$ and $M$ be two disjoint subsets of vertices
such that $|M| \geq (2/\f)|F|$, and
let $\widetilde{F} = \nbound{F}{M} \cup F$.  
There is a procedure that, given a graph $G'$ and an integer parameter $t$,
outputs one of the following:
\begin{itemize}
   \item A set $R$ such that the $cut_{G'}(R)$
is $\f$-sparse and $\Delta_{G'_{[N]\setminus F}}(\bound{F}{M},R) \leq t$.
  \item A vertex $v \in \bound{F}{M}$ and a set $U_v$ disjoint from $\widetilde{F}$
  such that $v \in U_v$,
$|U_v| \geq e^{(\f/3)t}$,
and $\Delta_{G'_{[N]\setminus \widetilde{F}}}(v,U_v) \leq t$
\end{itemize}
\ECM

\BPF
We start by performing a BFS from $M$ in $G'' = G'_{[N]\setminus F}$
up to depth $t$.
By the definition of the BFS, all the vertices reached
in levels $1,\ldots, t$
are disjoint from $M$ and $F$.
Applying Claim~\ref{cut-or-exp.clm}, in the process of this BFS
either we find a $\f$-sparse cut, thus satisfying the first condition,
or the size of the last level
is at least $|M|\cdot e^{(\f/3)t}$.
In the latter case,
for each vertex $v \in \bound{F}{M}$, perform a BFS in
$G'_{[N]\setminus \widetilde{F}}$
up to depth $t$, and let $U_v$ be the set of vertices
reached. Since the last level of the original BFS is
contained in $\bigcup_v U_v$,
we have that $\sum_{v\in \bound{F}{M}} |U_v| \geq |M|\cdot e^{(\f/3)t}$.
Therefore, there exists a vertex $v \in \bound{F}{M}$ such that
$|U_v| \geq |M|\cdot e^{(\f/3)t}/|\bound{F}{M}| \geq e^{(\f/3)t}$.
\EPF

\medskip\noindent
With these tools in hand, we are ready to describe the procedure
$\find$.\\

\BPFOF{Claim~\ref{find-proc.clm}}
We prove the claim by induction over the size
of the tree $T$. For the base case, let $T$ be a singleton
vertex. Then, the procedure $\find$ just outputs the pair
$(\minor,U)$. Now for the induction step.

Take an edge $e$ of $T$ that is incident to the root $r$.
Removing this edge gives us two trees $T_1$ and $T_2$
with roots $r_1$ and $r_2$ (these are the respective endpoints
of $e$). We let $T_1$ be the tree still rooted at $r$ (so that $r_1=r$).
Using subscripts to denote the respective size parameters
of these trees, we have $\kh = \kh_1 + \kh_2 + 2$ (recall
that  $\kh = 4k -2 $). We also
have that $\kh_1, \kh_2 \geq 2$.

We will describe the procedure $\find(v,U,T,F)$
using the respective procedures for $T_1$ and $T_2$.
We set $D_1 = (4d/\f)^{\kh_1}\ln(6f/\f^2)$
(recall that $f = \max\{|F|,k(4d/\f)^{4k+2}\} $).
We will be dealing mainly with the graph $G'' = G'_{[N]\setminus F}$
and hence all our boundaries are in this graph.
Recall that the procedure is required to work under the conditions that
$U$ is disjoint from $F$, $|U| \geq 4f/\f$, and
$\Delta_{G''}(v,U) \leq (4/\f)\ln(f/\f)$.
We may actually assume that $|U| = 4f/\f$. Suppose this is not the
case. Take the vertex in $U$ farthest from $v$ and remove it
from $U$. We keep repeating this until $|U| = 4f/\f$.
Note that the upper bound on $\Delta_{G''}(v,U)$ remains.
%
%

We now describe the steps of the procedure $\find$.
We will encapsulate each step through a different claim.
These claims will be stated here, and their proofs shall be given after
we describe $\find$ (and provide guarantees on its behavior
on the basis of these claims). These proofs will depends heavily
on the notation used here as well as the induction hypotheses. Nonetheless,
we defer their proofs so that the reader can more easily follow
the flow of our argument.
\ifnum\LatexVer=1
Refer to {\tt missing figure} to understand
the various claims.
\else
Refer to Figure~\ref{treefree.fig} to understand
the various claims.
\fi
For a set $R$, it will be convenient to refer to the following
as Condition (*): $R \ni v$, $\cut_{G'}(R)$ is $\f$-sparse, and
$\Delta_{G''}(v,R) \leq (4d/\f)^{\kh}\ln(f/\f)$.
If $R$ satisfies this, we will say ``$R$ satisfies Condition (*)".

\ifnum\LatexVer=1
\onote{I miss a figure that would latex on my LaTeX.}
\else
\begin{figure}[tb]
    \centering
  \psfrag{v}{$v$} \psfrag{v1}{$v_1$} \psfrag{v2}{$v_2$} \psfrag{dU}{$\partial U$} \psfrag{dA}{$\partial A$}
  \psfrag{U}{$U$} \psfrag{A}{$A$} \psfrag{d2}{$2D_1$} \psfrag{P}{$P$} \psfrag{S2}{$S_2$}
  \ \psfig{file=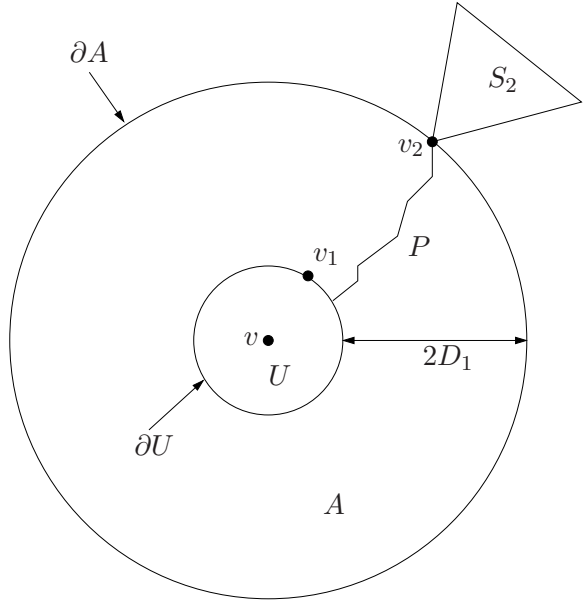,width=3in}
    \ \caption{The various sets in $\find$} \label{treefree.fig} \hfill\break
\end{figure}
\fi

\BCM \label{find-step1.clm} We can either find
a set $R$ satisfying Condition (*) or a set $A$ with the following
properties.
The set $A$ is disjoint from $F$ and exactly contains all vertices
(outside $F$) at distance at most $2D_1$ from $U$.
Also, $|U|\cdot e^{(\f/2)D_1} \leq |\bound{F}{A}|$ and
$|A| \leq |U|\cdot  e^{2D_1\ln d}$.
\ECM

\BCM \label{find-step2.clm} Suppose we have a set $A$
satisfying the conditions given in Claim~\ref{find-step1.clm}.
Then, we can either find a set $R$ satisfying Condition (*), or a vertex $v_2 \in \bound{F}{A}$ and set $S_2 \ni v_2$ with the following properties.
The set $S_2$ is disjoint from $F \cup \nbound{F}{A}$ and contains a $T_2$-minor such that $v_2$ belongs to the set whose contraction is the root.
Furthermore, $\Delta_{G''}(v,S_2) \leq (4d/\f)^{\kh}\ln(f/\f)$.
\ECM

\BCM \label{find-step3.clm} Let
$A$, $v_2$ and $S_2$ have the properties stated in the previous
claims. We can either find a set $R$ satisfying Condition (*), or
a path $P \subseteq A$, a vertex $v_1 \in \bound{F}{U}$,
and set $S_1 \ni v_1$ satisfying the following.
The sets $P$ and $S_1$ are disjoint from $F$ and disjoint from each other.
The path
$P$ connects $v_2$ to $U$ in $G''$. The set
$S_1$ contains a $T_1$-minor such that $v_1$ belongs to the set whose
contraction is the root, and
$\Delta_{G''}(v,S_1) < 2D_1$.
\ECM

With the claims, the proof becomes fairly direct. The procedure $\find$
invokes Claim~\ref{find-step1.clm}. If we find a $\f$-sparse cut,
we are done.
Otherwise, we have a set $A$ with the properties stated
in Claim~\ref{find-step1.clm}. Now, $\find$
invokes Claim~\ref{find-step2.clm}. Again, if we do not find
a $\f$-sparse cut, we have a vertex $v_2$ and set $S_2$.
Applying Claim~\ref{find-step3.clm},
we either find a $\f$-sparse cut, or get a path $P$,
vertex $v_1$ and set $S_1$.

We show how to construct a $T$-minor using $U$, $S_1$, $S_2$, and $P$.
Refer to Figure~\ref{treefree.fig} to see how these sets are laid out.
Note that all these sets are disjoint from $F$. Furthermore, since $\Delta_{G''}(v,S_1) < 2D_1$
and all vertices in $S_2$ have distance at least $2D_1$ from $U$,
the set $S_1$ is disjoint from $S_2$.
Also, $S_1$ is disjoint from $P$ (Claim~\ref{find-step3.clm}).
Our aim is to connect $v_1$ to $v_2$ (in $G''$) by a path that is disjoint
from $S_1 \cup S_2$.
If this path contains $v$, we will get a $T$-minor
rooted at $v$ that involves no vertex of $F$.
Take the path $P$ in $G''$ that connects $\bound{F}{U}$ to $v_2$.
This path is disjoint from $S_1 \cup S_2$. The vertex
$v_1$ is in $\bound{F}{U}$ and connected to all of $U$ in $G''$.
We take a path from $v$ to $P$ and a path from $v$ to $v_1$.
This connects $v_1$ to $v_2$ (via $v$) in $G''$ and completes
the construction of the $T$-minor.
\EPFOF

\medskip

In the proofs of Claims~\ref{find-step1.clm}, ~\ref{find-step2.clm}
and~\ref{find-step3.clm}, the following bounds will be repeatedly used.
By assumption, $\Delta_{G''}(v,U) \leq (4/\f) \ln(f/\f)$, $|U| = 4f/\f$
and $D_1 = (4d/\f)^{\kh_1}\ln(6f/\f^2)$. Since $\kh = \kh_1 + \kh_2 + 2$
and $\kh_1, \kh_2 \geq 2$, $\max(\kh_1,\kh_2) \leq \kh - 4$.
We first state a technical claim.

\BCM \label{tech-bound.clm}
$$
\Delta_{G''}(v,U) + (2+6(\ln d)/\f)D_1 +
   (4d/\f)^{\kh_2}\ln(|U|/\f) + (3/\f)\ln(4|U|/\f)
     \;\;\leq\;\; (1/2) (4d/\f)^{\kh}\ln(f/\f)\;.
$$
\ECM

\BPF As argued earlier, $|U| = 4f/\f$, so $(3/\f)\ln(4|U|/\f) \leq (4/\f)\ln(16f/\f^2)$.
Assuming $\f$ is at most a sufficiently small constant,
we can bound:
\begin{eqnarray*}
\Delta_{G''}(v,U) + (3/\f)\ln(4|U|/\f)
  &\leq& (4/\f)(\ln(f/\f) + \ln(16f/\f^2)) \\
&=&(4/\f)\ln(16f^2/\f^3)\\
&\leq& (4/\f)^4\ln(f/\f) \;.
\end{eqnarray*}
We trivially bound $2+6(\ln d)/\f$ by $8d/\f$. We now bound
the summation in the claim as follows. We use the fact that $\kh_1 + 2 \geq 4$,
and that $\kh - 2 \geq \max(\kh_1 + 2, \kh_2 + 2)$.
\begin{eqnarray*}
\lefteqn{(4/\f)^4\ln(f/\f) + 8dD_1/\f + (4d/\f)^{\kh_2}\ln(|U|/\f)}
\;\;\;\;\;\;\;\;\;\;\; \\
& = & (4/\f)^4\ln(f/\f) + 2(4d/\f)^{\kh_1+1}\ln(6f/\f^2) + (4d/\f)^{\kh_2}\ln(4f/\f^2) \\
& \leq & (4d/\f)^{\kh_1+2}\ln(f/\f) + (4d/\f)^{\kh_1+2}\ln(f/\f) + (4d/\f)^{\kh_2+2}\ln(f/\f) \\
& \leq & (1/2) (4d/\f)^{\kh}\ln(f/\f)\;,
\end{eqnarray*}
and the proof is completed.
\EPF

\medskip

\BPFOF{Claim~\ref{find-step1.clm}}
Initiate a BFS in the residual graph
$G'' = G'_{[N]\setminus F}$ starting from $U$ for $2D_1$ steps.
Let $A$ denote the set of all vertices reached (including $U$).
We now invoke Claim~\ref{cut-or-exp.clm} with $F$, $M := U$, and $t := 2D_1$.
Suppose we find a $\f$-sparse cut $cut_{G'}(R)$. We have
$\Delta_{G''}(v,R) \leq \Delta_{G''}(v,U) + 2D_1$.
By Claim~\ref{tech-bound.clm}, this is at most $(4d/\f)^{\kh}\ln(f/\f)$.
Hence, $R$ satisfies Condition (*). Otherwise, the BFS reaches
$2D_1$ levels, and (by Claim~\ref{cut-or-exp.clm}), $|U|\cdot e^{(\f/2)D_1} \leq |\bound{F}{A}|$.
Since the degree bound is $d$, the size (after $2D_1$ steps of a BFS)
can only blow up by $d^{2D_1}$. Hence, $|A| \leq |U|\cdot  e^{2D_1\ln d}$.
\EPFOF

\medskip

\BPFOF{Claim~\ref{find-step2.clm}}
Note that $|\bound{F}{A}| \geq |U| = 4f/\f \geq |F|$. Set
$F_2 = \nbound{F}{A} \cup F$, so that
\begin{eqnarray*}
|F_2| = |\nbound{F}{A}\cup F|
= |\nbound{F}{A}| + |F| \leq |\nbound{F}{A}| + |\bound{F}{A}|
= |A| \leq |U|\cdot e^{2D_1\ln d}\;.
\end{eqnarray*}
We invoke Claim~\ref{cut-or-good.clm} with $F$, $M := A$,
and $t := (3/\f)\ln(4|F_2|/\f)$.
%
Suppose we get a $\f$-sparse cut $cut_{G'}(R)$.
\begin{eqnarray*}
    \Delta_{G''}(v,R) & \leq & \Delta_{G''}(v,U) + \Delta_{G''}(U,\bound{}{A}) + t \\
  & \leq & \Delta_{G''}(v,U) + \Delta_{G''}(U,\bound{}{A}) + (3/\f)\ln(4|F_2|/\f) \\
    & \leq & (4/\f)\ln(f/\f) + 2D_1 + (3/\f)\ln(4|U|/\f) + 6D_1(\ln d)/\f
\end{eqnarray*}
An application of Claim~\ref{tech-bound.clm} proves that this is at most $(4d/\f)^{\kh}\ln(f/\f)$.
So $R$ satisfies Condition (*).
By Claim~\ref{cut-or-good.clm}, if we do not get a set $R$, we get a vertex $v_2 \in \bound{}{A}$ and
a set 
$U_2$ disjoint from $F_2$ such that $v_2 \in U_2$,
$|U_2| \geq e^{(\f/3)t}$, and
$\Delta_{G'_{[N]\setminus F_2}}(v_2,U_2) \leq (3/\f)\ln(4|F_2|/\f)$.
By choice of $t$, $|U_2| \geq 4|F_2|/\f$.
Let $f_2 = \max\{|F_2|,k_2(4d/\f)^{4k_2+2}\}$.
To call the procedure $\find(v_2,U_2,F_2,T_2)$,
we need to argue that $|U_2| \geq 4f_2/\f$ and $\Delta_{G'_{[N]\setminus F_2}}(v,U_2) \leq (4/\f)\ln(f_2/\f)$.
We have chosen $f = \max\{|F|,k(4d/\f)^{4k+2}\}$,
and $|F_2| \geq f$. Hence, $f_2 = |F_2|$.
So $|U_2| \geq 4|F_2|/\f = 4f_2/\f$
and $\Delta_{G'_{[N]\setminus F_2}}(v_2,U_2)
   \leq (3/\f)\ln(4|F_2|/\f) \leq (4/\f)\ln(4f_2/\f)$.
%
%
%

Let $(\sigma,S^*)$ be the output of this call to $\find$.
We have,
$$ \Delta_{G''}(v,S^*) \leq \Delta_{G''}(v,U) +
  \Delta_{G''}(U,\partial A) + \Delta_{G''}(v_2,S^*) \;.$$
By the induction hypothesis $\Delta_{G''}(v_2,S^*) \leq (4d/\f)^{\kh_2}\ln(f_2/\f)$.
Since $f_2 \leq |U|e^{2D_1\ln d}$, this is at most $(4d/\f)^{\kh_2}\ln(|U|/\f)
+ (2\ln d)(4d/\f)^{\kh_2}D_1$. We now get
$$ \Delta_{G''}(v,S^*) \leq \Delta_{G''}(v,U) + \Delta_{G''}(U,\partial A) +
(4d/\f)^{\kh_2}\ln(|U|/\f)+ (2\ln d)(4d/\f)^{\kh_2}D_1$$
Barring the last term, we have that $\Delta_{G''}(v,U) + 2D_1 +
(4d/\f)^{\kh_2}\ln(|U|/\f)$ is at most
$(1/2)(4d/\f)^{\kh}\ln(f/\f)$ by Claim~\ref{tech-bound.clm}.
Turning to the last term,
\begin{eqnarray*}
(2\ln d)(4d/\f)^{\kh_2}D_1 & = & (2\ln d)(4d/\f)^{\kh_2}
   \cdot  2(4d/\f)^{\kh_1}\ln(6f/\f^2) \\
& = & (4\ln d)(4d/\f)^{\kh_1+\kh_2}\ln(6f/\f^2) \\
& = &(4\ln d)(4d/\f)^{\kh-2}\ln(6f/\f^2) \\
& \leq & (1/2)(4d/\f)^{\kh}\ln(f/\f)\;.
\end{eqnarray*}
Putting it all together, $\Delta_{G''}(v,S^*) \leq (4d/\f)^{\kh}\ln(f/\f)$.
If the output is a cut, then $S^*$ is the desired $R$ (satisfying Condition (*)).
Otherwise (by the induction hypothesis), the set $S^*$ (disjoint from $F_2$) 
contains a $T_2$-minor 
such that $v_2$ belongs to the subset whose contraction corresponds
to the root $r_2$ of $T_2$. We have $F_2 = \nbound{F}{A} \cup F$,
where the set $A$ contains all vertices
whose distance (in $G''$) from $U$ is at most $2D_1$. Since $S^*$ is disjoint from $F_2$,
the distance in $G''$
of any vertex in $S^*$ from $U$ is at least $2D_1$.
So $S^*$ is the desired set $S_2$.
\EPFOF

\medskip

\BPFOF{Claim~\ref{find-step3.clm}}
Consider the shortest path $P$ from $U$ to $v_2$ (in $G''$).
Since $v_2 \in \bound{F}{A}$, the path $P\setminus\{v_2\}$ is entirely contained in $\nbound{F}{A}$.
Hence, $S_2$ is disjoint from $P\setminus\{v_2\}$.
By construction, $|P| \leq 2D_1$.

We have $f = \max\{|F|,k(4d/\f)^{4k+2}\}$.
By some elementary calculations (given by Claim~\ref{func.clm}), $f \geq (4d/\f)^{\kh}\ln(6f/\f^2) \geq 2D_1 \geq |P|$.
Hence, $|F \cup P| \leq 2f$.
Let  $F' := F \cup P$, $F_1 := \nbound{}{U} \cup F'$
and $f_1 = \max\{|F_1|,k_1(4d/\f)^{4k_1+2}\}$.
We invoke Claim~\ref{cut-or-good.clm} with $F'$ as the forbidden set,
$M := U$, and $t :=  (3/\f)\ln(4f_1/\f)$.
Since $|F_1| \leq |U| + 2f \leq 4f/\f + 2f < 6f/\f$, we have that
$t = (3/\f)\ln(4f_1/\f) < (4d/\f)^{\kh_1}\ln(6f/\f^2) = D_1$.
If we get a $\f$-sparse cut $cut_{G'}(R)$, we end with the desired $R$.

\sloppy
Otherwise, we get a vertex $v_1 \in \bound{}{U}$ and a set $U_1 \ni v_1$
disjoint from $F_1$ such that
$|U_1| \geq e^{(3/\f)t} = 4|f_1|/\f$ and
$\Delta_{G'_{[N]\setminus F_1}}(v_1,U_1) \leq t
    \leq  4\ln(f_1/\f)/\f$.
We thus have the necessary conditions
to call $\find(v_1,U_1,F_1,T_1)$. By the induction hypothesis,
for the set $S^*$ returned,
$\Delta_{G'_{[N]\setminus F_1}}(v_1,S^*) \leq (4d/\f)^{\kh_1}\ln(f_1/\f)
\leq (4d/\f)^{\kh_1}\ln(6f/\f^2) \leq D_1$ (using the bound $f_1 < 6f/\f$).
Hence $\Delta_{G''}(v,S^*) \leq \Delta_{G''}(v,U) + D_1 < 2D_1 \leq (4d/\f)^{\kh} \ln(f/\f)$.
Regardless of whether $S^*$ is output as a cut or a minor,
we complete the proof.
\EPFOF

\BCM \label{func.clm} For
$x \geq k(4d/\f)^{4k+2}$, we have that $x \geq (4d/\f)^{\kh}\ln(6x/\f^2)$.
\ECM

\BPF Consider the function $x/\ln(\alpha x)$. The derivative
is given by $ \frac{d}{dx}\Big(\frac{x}{\ln(\alpha x)}\Big)
= \frac{1}{\ln(\alpha x)} - \frac{1}{\ln^2(\alpha x)}$.
Hence, this function is increasing when $x \geq e/\alpha$. We set $\alpha = 6/\f^2$,
and note that for small enough $\f$, $k(4d/\f)^{4k+2} > e\f^2/6$.
For $x \geq k(4d/\f)^{4k+2}$,
$$ \frac{x}{\ln(6x/\f^2)} \geq \frac{k(4d/\f)^{4k+2}}{\ln(6k/\f^2) + (4k^2+2k)\ln(4d/\f)}
\geq (4d/\f)^{4k-2}\;,$$
and the proof is completed.
\EPF


\ifnum\verTR=1    
\subsection{Testing $T$-minor freeness for any depth-two tree $T$}
\label{tree-depth2.sec}
Let $T$ be an arbitrary depth-two tree with $k$ vertices;
that is, $T$ consists of a root, denoted $r$,
and $m$ stars, denoted $T_1,\dots,T_m$, that are
rooted at neighbors of $r$, where here we consider
also the singleton vertex as a star (with 0~leaves).
Denote the $m$ corresponding roots by $r_1,\dots,r_m$,
and denote the number of leaves in these stars by $k_1,\dots,k_m$
(i.e., $k=1+m+\sum_{i\in[m]}k_i$).
The following algorithm is tailored for this tree $T$.

\BA{\em(tailored for the foregoing $T$):}
\label{depth2-tree.alg}
On input $G=([N],E)$ and proximity parameter $\e$,
set $D=(5d^2k/\e)^2$ and proceed as follows.
\BE
\item Select uniformly a start vertex $s\in[N]$.
\item Perform a BFS starting at $s$
and stopping as soon as $D$ layers are explored
(or the BFS reaches all
the vertices of a connected component in $G$).
\item
Accept if and only if the explored graph is $T$-minor free.
\EE
\EA
Clearly, Algorithm~\ref{depth2-tree.alg} never rejects
a $T$-minor free graph. Its query complexity is exponential
in $D$, and its time complexity is polynomial in its query
complexity (by Corollary 1.2 of~\cite{KaKoRe}).
\BL
\label{depth2-tree.lem}
If $G$ is $\e$-far from being a $T$-minor free graph,
then Algorithm~\ref{depth2-tree.alg} rejects with probability
at least $\e/4$.
\EL

\BPF
We call a vertex $v$ {\tsf bad} if its $D$-neighborhood
(i.e., the vertices that are distance at most $D$ from $v$)
contains a $T$-minor, and denote the fraction of bad vertices
(in $G$) by $\rho$.
As in the proof of Claim~\ref{k-star.clm},
it suffices to show that $G$ is $(\rho+(\e/2))$-close to
being $T$-minor free, and we again start by omitting all edges
incident at bad vertices and considering the resulting graph,
denoted $G^{(0)}$. Indeed, $G^{(0)}$ is $\rho$-close to $G$.

The rest of our analysis proceed in iterations.
If the current graph $G^{(i-1)}$ is $T$-minor free,
then we are done. Otherwise, we pick an arbitrary
vertex $s^{(i)}$ that resides in (the root of) some $T$-minor.
Since $s^{(i)}$ is not bad, it must reside in a connected component
of $G^{(i-1)}$ that has radius at least $D$ from $s^{(i)}$.
We shall show how to identify a set $S^{(i)}$
such that $G^{(i-1)}_{S^{(i)}}$ has radius at most $D$
and the cut $(S^{(i)},[N]\setminus S^{(i)})$
has less that $\e d|S^{(i)}|/2$ edges.
Omitting these cut edges yields a graph $G^{(i)}$
such that $G^{(i)}_{S^{(i)}}$ is $T$-minor free
(and $S^{(i)}$ will not intersect with any future $S^{(j)}$).
When the process ends, we have a $T$-minor free graph. In total,
we omitted at most $\sum_i \e d|S^{(i)}|/4\leq\e d N/2$ edges
(from $G^{(0)}$),
and thus $G$ is $(\rho+(\e/2))$-close to $T$-minor free.

The crux of the proof is indeed the process of identifying
a suitable set $S'=S^{(i)}$ in $G'\eqdef G^{(i-1)}$.
The identification procedure is initiated at $s'=s^{(i)}$
and proceeds in two stages.
In the first stage, the procedure tries to find
either a set $S_0$ of size at least $4m/\e$ such that
the cut $(S_0,[N]\setminus S_0)$ has less than $m$ edges
or a set $S_0$ of size at most $2dm/\e$ such that $G'_{S_0}$
contains an $m$-star as a minor rooted at $s'$.
Clearly, in the first case we are done.
In the second case, we get to the second stage of the procedure,
which explores $G'$ (somewhat) beyond $S_0$
in an attempt to extend the $m$-star minor into a $T$-minor,
but this attempt is bound to fail,
and this failure will allow finding the desired cut.
Loosely speaking, this second stage proceeds by trying to find
disjoint $T_j$-minors, for $j=1,\dots,m$.
This is done by invoking a ``$k'$-star-minor finding'' procedure,
denoted $\FS_{k'}$, which generalizes the procedure
that is described in the proof of Claim~\ref{k-star.clm}.
The procedure $\FS_{k'}$ is invoked on a vertex, $v$,
and a set of forbidden vertices, denoted $F$,
and tries to either find a $k'$-star rooted at $v$
in $G'_{[N]\setminus F}$ or find a good cut.
Indeed, $F$ will contain the set $S_0$ as well as adequate
sets that will prevent the current search from entering
any of the previously found star minors.
We first provide a specification of $\FS$,
and then turn to its actual implementation.

\medskip\noindent{\sf Specification of the procedure $\FS$.}
On input a vertex $v$ and a forbidden set $F$,
the procedure $\FS_{k'}$ outputs a triplet $(\sigma,\pS,F')$
such that $\sigma\in\{\minor,\cut,\free\}$
and $F'\subseteq\pS\subseteq[N]\setminus F$
such that $|F'|<dk'$ and $|\pS|<(4dk'/\e)\cdot(|F|+1)$.
In addition, it always holds that all vertices of $G'_{\pS}$
are connected to $v$,
and one of the following cases holds.
\BDes
\item[$\sigma=\minor$.]
The graph $G'_{\pS}$ contains a $k'$-star as a minor that is rooted at $v$
(i.e., $v$ resides in the connected component that is contracted
to fit the root of the $k'$-star).
Furthermore, all edges of
the cut $(\pS\setminus F',[N]\setminus(\pS\setminus F'))$
are incident at $F\cup F'$.
\item[$\sigma=\cut$.]
The cut $(\pS,[N]\setminus\pS)$ contains less that $\e d|\pS|/2$ edges.
\item[$\sigma=\free$.]
All edges of the cut $(\pS,[N]\setminus\pS)$ are incident at $F$.
\EDes
Let $T'$ denote a generic $k'$-star,
where we may assume that $k'\geq1$.

\medskip
\noindent{\sf Implementing the procedure $\FS$.}
Our aim is to either find a (relatively small) $T'$-minor
or find a set with a relatively small cut from
the rest of the graph. This is done by initiating a BFS
in the residual graph $G'_{[N]\setminus F}$ starting at $v$,
and stopping as soon as one of the following three cases occurs.
\BDes
\item[{\sf Case 1:}]
{\em A layer containing at least $k'$ vertices is found
before $4(|F|+k')/\e$ vertices are encountered}.
In this case the procedure returns $(\minor,\pS,F')$,
where $\pS$ is the set of encountered vertices
and $F'$ is the set of vertices in the last BFS layer.

Note that in this case $G'_{\pS}$ contains a $T'$-minor rooted at $v$,
and that $|F'|<dk'$ (as otherwise the BFS would have terminated
in a previous layer). Furthermore, by structure of the BFS, all edges
of the cut $(\pS\setminus F',[N]\setminus(\pS\setminus F'))$
are incident at $F\cup F'$.

\item[{\sf Case 2:}]
{\em The search encountered at least $4(|F|+k')/\e$ vertices,
while Case~1 does not hold}.
In this case the procedure returns $(\cut,\pS,\emptyset)$,
where $\pS$ is the set of encountered vertices.

Note that in this case the cut $(\pS,[N]\setminus\pS)$
contains less than $(|F|+k')\cdot d \leq \e d|\pS|/4$ edges.

\item[{\sf Case 3:}]
{\em The search cannot be extended any further,
while Cases~1 and~2 do not hold}.
In this case the procedure returns $(\free,\pS,\emptyset)$,
where $\pS$ is the set of encountered vertices.

Note that in this case the cut $(\pS,[N]\setminus\pS)$
contains only edges that are incident at $F$.
\EDes
Observe that in the first case
 $|\pS| < 4(|F|+k')/\eps$, in the second case
$|\pS| < 4d(|F|+k')/\eps $ (since otherwise the BFS
could have stopped in the previous level), and in the third case
$|\pS| < 4(|F|+k')/\eps$.
Hence, in all cases
\BEQ\label{bound-ps.eq}
|\pS|\;<\; 4dk'(|F|+1)/\e\;.
\EEQ
Thus, this implementation satisfies the specification.
We note that the above description applies also
in case $k'\in\{0,1\}$, where $k'=0$ is trivial%
\footnote{Actually, this case never occurs;
that is, we never invoke $\FS_{0}$.
The case $k'=1$ may occur, but we could have avoided it too,
by a direct treatment.}
(i.e., always return $(\minor,\{v\},\{v\})$)
and $k'=1$ is almost trivial
(i.e., return $(\minor,\{v,w\},\Gamma_{G'}(v)\setminus F)$
if $v$ has a neighbor $w$ in $G'_{[N]\setminus F}$
and $(\free,\{v\},\emptyset)$ otherwise).

Using the star finding procedure $\FS$, we now turn to the main
identification procedure, which is invoked on input vertex $s'=s^{(i)}$
and aims at finding an adequate set $S'=S^{(i)}$.
Recall that $r$ denotes the root of $T$,
and $r_1,\dots,r_m$ denote the roots of
the subtrees $T_1,\dots,T_m$, where $T_j$ is a $k_j$-star.
The main procedure operates as follows.

\BE
\item
It initiates a BFS in the graph $G'$ starting at $s'$,
stopping as soon as at least $B=4dk/\e$ vertices are encountered.
Let $S_0$ denote the set of encountered vertices.
Note that $|S_0| \geq B$ must hold, because $s'=s^{(i)}$
resides in a set of vertices that can be contracted to
the root of some $T$-minor having radius greater than $D$.
%


Note that necessarily $|S_0| < dB$
(because otherwise we would have stopped at the previous BFS-layer).
\item 
Let $F_0$ denote the last layer in the BFS performed
in the previous step.
If $|F_0|< m$, then we just use $S_0$ as the desired set
(i.e., let $S^{(i)}=S_0$).

Note that, in this case, the cut $(S_0,[N]\setminus S_0)$
contains less than $m\cdot d$ edges,
whereas by the case hypothesis $|S_0|\geq B>4m/\e$
(as $k >m$).
So the conditions regarding this set are satisfied.

We continue to the next step only if $|F_0| \geq m$.
\item
(The purpose of the current step is to generate calls to $\FS$
that will eventually lead to returning a set as in the second output
case (i.e., $\cut$), which can serve as $S^{(i)}$ (see above).
The presentation, however, pretends that we attempt to find
a $T$-minor as in the first output case (i.e., $\minor$).
Observing that $S_0\setminus F_0$ can serve as a contraction
of the root of $T$, we attempt to find disjoint sets $S_j$
that contain $T_j$-minors rooted at some $v_j\in F_0$.)

For $j=1,\dots,m$, we try to find $S_j$ as follows.
Let $F'=\bigcup_{a\in[j-1]}F_a$ and $V'=\{v_1,\dots,v_{j-1}\}$.
For every $v\in F_0\setminus V'$, we proceed as follows.
\BDes
\item[{~~~~}] 
We invoke $\FS_{k_j}$, letting
$(\sigma,X,Y)\gets \FS_{k_j}((F_0\setminus\{v\})\cup F',v)$.

We note that by
the specification of $\FS_{k_j}$ and Equation~(\ref{bound-ps.eq}),
 we have that  $|X| \leq (4dk_j/\e)\cdot(|F_0|+|F'|+1)$
and $|Y| \leq dk_j$.
Recall that $|F_0| < |S_0| < dB = 4d^2k/\e$
and $|F'|=\sum_{a\in[j-1]}|F_a| < d\sum_{a\in[j-1]} k_a<d(k-m)$,
where $k=1+m+\sum_{a\in[m]} k_a$.
Thus, $|X| < (5d^2k/\e)^2$.

We consider the following three cases regarding $\sigma$.
\BDes
\item[$\sigma={\minor}$.]
In this case we set $v_j\gets v$ and $(S_j,F_j)\gets(X,Y)$,
and proceed to the next value of $j$ (i.e., $j\gets j+1$);
see comment below.

Note that $|S_j| < (5dk/\e)^2$.
In fact, the same upper bound can be proved
for $\sum_{a=0}^{j}|S_a|$.

Note that this case cannot occur when $j=m$,
because this would yield a small $T$-minor rooted in $s'$
in contradiction to the hypothesis that $s'=s^{(i)}$ is not bad.
\item[$\sigma={\cut}$.]
In this case we just use $X$ as the desired set
(i.e., let $S^{(i)}=X$).

Note that, by the specification of $\FS$,
the cut $(S^{(i)},[N]\setminus S^{(i)})$
contains relatively few edges.
\item[$\sigma={\free}$.]
In this case we do nothing,
and continue to the next candidate $v$.
\EDes
\EDes
Note that we halted with a desired cut if
either Step~2 found such a cut
or any of the invocations of $\FS$ returned a {\cut}-value.
Furthermore, as noted in the above discussion
concerning the case $\sigma={\minor}$, it cannot be the case that in Step~3
we obtained a {\minor}-value for each $j\in[m]$.
Thus, we remain with the case that, for some $j\in[m]$,
all invocations of $\FS$ returned a {\free}-value.
In this case, we let $X'$ be the union of all sets $X$
that were returned in the corresponding $|F_0|-(j-1)$ invocations,
and use $S_0\cup X'$ as the desired set (i.e., let $S^{(i)}=S_0\cup X'$).
In this case,
the size of the cut $(S^{(i)},[N]\setminus S^{(i)})$
is at most $d\cdot|F'|<d^2k$,
because for each $X$ all edges of the cut $(X,[N]\setminus X)$
are incident at $F_0\cup F' \subseteq S_0\cup F'$.
Thus, the cut is sufficiently small,
because $|S^{(i)}|\geq|S_0|\geq B=4dk/\e$.
On the other hand, the size of $S_0\cup X'$
is at most $|F_0|\cdot (4dk/\e)\cdot|F'| < (4dk/\e)^2$.
\EE
This completes the description of the operation of the procedure $I$
as well as the showing that it satisfies its specification.
It follows that for any $s^{(i)}$ that reside in the root
of some $T$-minor in $G^{(i-1)}$,
we obtain a set $S^{(i)}$ such that
the cut $(S^{(i)},[N]\setminus S^{(i)})$
has less than $4d|S^{(i)}|/\e$ edges.
Using the fact $|S^{(i)}|<D$,
it follows that $G^{(i-1)}_{S^{(i)}}$ is $T$-minor free,
and the lemma follows.
\EPF
\fi

\section{The unbounded-degree graph model}
\label{general-model}
In this section we consider (one-sided error) testing cycle-freeness
(and tree-minor freeness) in what we shall refer to as
the {\tsf unbounded-degree incidence-lists model}.
In this model, introduced in~\cite{PR},
the maximum degree $d$ may be as large as $N-1$,
so there is effectively no degree-bound, and a graph $G$ is
represented by a function $g : [N] \times [N-1] \to \{0,\ldots,N\}$.
Similarly to the bounded-degree model, the algorithm may ask for the identity
of the $i^\xth$ neighbor of a vertex $v$, for any $v\in [N]$ and $i\in [N-1]$
of its choice, by querying the function $g$.
(If $v$ has less than $i$ neighbors, then the answer returned is `$0$').
For the sake of simplicity, we assume that the algorithm
can also query the degree of any vertex of its choice
(where such a query can, of course,
be replaced by $O(\log N)$ neighbor queries).

The main and crucial difference between the unbounded-degree model
and the bounded-degree model is
in the {\em distance measure between graphs}.
Rather than measuring distance between graphs in terms of
the size of the domain of $g$, as done in the bounded-degree model,
we measure it with respect to the number of edges $|E|$ in $G = ([N],E)$.
That is, we shall say that a graph $G$ is $\eps$-far from
being cycle-free (in the unbounded-degree model),
if the number of edges that must be removed in order to make it cycle-free
is greater than $2\eps |E|$ (see Footnote~\ref{edge-count:fn}).
Letting $\davg$ denote the average degree in $G$
(and assuming that $G$ is connected),
this is equivalent to saying that the number of edges in $G$
is greater than $(N-1) + \eps\davg N$.

We note that while the bounded-degree model is appropriate
for testing graphs in which the maximum degree is of the same order
as the average degree (and in particular for constant-degree graphs),
the unbounded-degree model is appropriate for testing graphs
in which the maximum degree may be much larger than the average degree.
We mention that the model considered in~\cite{KKR}
(see also \secref{general-model:kkr})
also allows adjacency queries (as in~\cite{GGR}),
but such queries are useless for us when the degree is
smaller than $\sqrt N$.

\paragraph{A necessary assumption.}
Throughout this section, we assume that the number of edges
in the graph is at least linear in the number of vertices.
As noted in~\cite{KKR}, without this assumption the model
becomes intractable for sublinear algorithms,
because one can always hide a tiny graph
(which may be far from having the property)
inside a huge graph that contains mostly isolated vertices.
Also, unless explicitly stated otherwise, we also assume that
the graphs are simple (i.e., have no self-loops or parallel edges).

\subsection{Testing cycle-freeness}
\label{general-model:cycle}
In this subsection, we show that the result of \thmref{intro:c3.thm}
(and thus also \thmref{intro:find-cycle.thm})
extends to the unbounded-degree (incidence lists) model.
Furthermore, the extended result is actually stronger than stated
in Theorem~\ref{intro:c3.thm}, since we eliminate the dependence
of the complexities on the average (let alone maximum)
degree of vertices in the input graph.
This is done by viewing the randomized reduction that underlies
Algorithm~\ref{cycle-free.alg} in a slightly different manner,
which actually yields an alternative tester
(which is closely related to
but different from Algorithm~\ref{cycle-free.alg}).
We then show that this algorithm extends easily
to the unbounded-degree model.

The pivot of our exposition is the following generalization
of 2-colorability in which edges of the graph are labeled
by either $\tt eq$ or $\tt neq$. That is, an instance
of this problem is a graph $G=([N],E)$ along with
a labeling $\lab:E\to\{{\tt eq},{\tt neq}\}$.
We say that $\chi:[N]\to\bitset$ is a legal 2-coloring
of this instance if for every $\{u,v\}\in E$
it holds that $\chi(u)=\chi(v)$ if and only if $\lab(\{u,v\})={\tt eq}$.
That is, a legal 2-coloring (of the vertices) is one in which
every two vertices that are connected by an edge labeled $\tt eq$
(resp. $\tt neq$) are assigned the same color (resp., opposite colors).
Note that the standard notion of 2-colorability corresponds
to the case in which all edges are labeled $\tt neq$.

We first observe that the (one-sided error)
Bipartite testers of~\cite{GR2} and~\cite{KKR}
can be extended to test this generalization of 2-colorability.%
\footnote{A similar observation refers to the $k$-colorability
testers of~\cite{GGR}, which operate in the dense graph model.
Thus, for every $k\geq2$,
the foregoing generalization of $k$-colorability can be tested
(with one-sided error)
in the dense graph model by using $\poly(1/\e)$ queries.}
Next, we observe that the randomized reduction that
underlies Algorithm~\ref{cycle-free.alg} can be viewed
as a randomized reduction of cycle-freeness to
generalized 2-coloring, while keeping the graph intact.
Combining these two observations, we obtain:

\BT{\em(Theorems~\ref{intro:c3.thm} and~\ref{intro:find-cycle.thm},
generalized to the unbounded-degree model):}
Cycle-freeness in $N$-vertex graphs can be tested with one-sided error
within time complexity $\tildeO(\poly(1/\e)\cdot\sqrt{N})$,
where $\e$ denotes the proximity parameter that refers to
the number of {\em(omitted)} edges as a fraction of
the total number of edges in the input graph.
Furthermore, whenever the tester rejects, it outputs
a simple cycle of length $\poly(\e^{-1}\log N)$.
\ET

\BPF
We first present a (local) randomized reduction of testing cycle-freeness
to testing the foregoing generalization of 2-colorability.
Unlike the reduction that underlies Algorithm~\ref{cycle-free.alg},
the current reduction does not modify the input graph $G=([N],E)$;
it merely introduces a random labeling $\lab:E\to\{{\tt eq},{\tt neq}\}$.
Specifically, the graph $G=([N],E)$ is mapped to
a random instance of the generalized 2-coloring problem
such that the graph equals $G$ itself
and the labeling is selected uniformly
among all possible $\lab:E\to\{{\tt eq},{\tt neq}\}$.
Thus, invoking any generalized 2-coloring (one-sided error) tester
on the resulting instance, we are done.
(Indeed, unlike in the case of Algorithm~\ref{cycle-free.alg},
here the emulation of the generalized 2-coloring tester
is straightforward. Also note that here we obtained a true
local reduction, whereas in Section~\ref{C3:section} we only
showed how to emulate bipartite testers of a special type
(i.e., such that only select uniformly distributed vertices
and make random-neighbor queries).)

The analysis of the forgoing randomized reduction
is analogous to the proof of Lemma~\ref{cycle-free.ana}.
For sake of clarity, we spell out what this means.
\BE
\item
If $G$ is cycle-free, then,
for every choice of $\lab:E\to\{{\tt eq},{\tt neq}\}$,
the graph $G$ is 2-colorable with respect to the labeling $\lab$
(i.e., the exists a legal 2-coloring of the instance $(G,\lab)$).
\item
If $G$ is not cycle-free, then, with probability at least $1/2$
(over the random choice of $\lab:E\to\{{\tt eq},{\tt neq}\}$),
the graph $G$ is not 2-colorable with respect to the labeling $\lab$.
\item
There exist universal constants $c_1 > 1$ and $c_2,c_3>0$
such that, for every $\eps \geq c_1/|E|$,
if $G$ is $\e$-far from being cycle free, then, with probability
at least $1-\exp(-c_2\e dN)$
(over the random choice of $\lab:E\to\{{\tt eq},{\tt neq}\}$),
at least $c_3\e\cdot|E|$ edges must be omitted from $G$
in order to obtain a graph that is 2-colorable
with respect to the labeling $\lab$.
\EE
Each of the above three claims may be proved by mimicking
the proof of the corresponding item of Lemma~\ref{cycle-free.ana}.
In particular, note that in the proof of the third item,
we actually established that (w.v.h.p.) $c_3\cdot(|E|-N+1)$ edges
must be omitted from $G_\tau$ in order to obtain a bipartite graph,%
\footnote{The stated distance of $c_3\e/2d$ in the bounded degree
model, was obtained by using $|E|-N+1=\e dN$ (which holds there),
and dividing the result by $d\cdot (d+1)N$ (which represents
the product of the degree of $G_\tau$ by an upper bound on
the number of its vertices).}
which means that these many edges must be omitted from $G$
in order to obtain a graph that is 2-colorable with respect to
the labeling $\lab$.

We turn to presenting a generalized 2-coloring
(one-sided error) tester for the unbounded-degree model.
One way of doing so it to observe that the 2-coloring tester of~\cite{KKR}
(which builds upon~\cite{GR2}) extends to the generalized version.
We shall {\em not}\/ follow this way,
but rather present an alternative one,
but before doing so we comment that all that is needed
in order to support this observation is
to (fictitiously) {\em define}\/ edges labeled $\tt eq$ as
having even length (say, length zero or two), whereas edges
labeled $\tt neq$ are defined as having odd length (say, length one).
Modulo this definition, the entire analysis of~\cite{GR2}
remains intact. Specifically, all references in~\cite{GR2}
to the length of paths and cycles are re-interpreted as
referring to the foregoing definition. In particular,
an odd length cycles (under this label-dependent definition of length)
indicates that the graph cannot be 2-colored
(under the corresponding labeling of edges),
whereas the non-existence of odd length cycles
enables such a 2-coloring.
(The same holds for~\cite{KKR}, which operates by a (local)
reduction to~\cite{GR2}.)

Seeking a self-contained presentation,
we present an alternative way of deriving a generalized 2-coloring
(one-sided error) tester for the unbounded-degree model.
Specifically, we shall reduce generalized 2-coloring testing
to standard 2-coloring testing as follows.
Given an instance $(G,\lab)$ of the generalized 2-coloring problem,
where $G=([N],E)$ and $\lab:E\to\{\tteq,\ttneq\}$,
we consider the following multi-graph $G'=([2N],E')$,
where a {\sf multi-graph} is a graph with parallel edges:
Each vertex $v \in [N]$ of $G$ is replaced by two vertices, $v$ and $N+v$,
which are connected by $2|\Gamma_G(v)|$ parallel edges,
and the edges of $G$ are replaced by edges among the corresponding copies
such that edges labeled $\ttneq$ are replaced by edges
between ``matching copies'' and edges labeled $\tteq$ are replaced
by edges between opposite copies.
Specifically, if the edge $\{u,v\}\in E$ is labeled $\ttneq$
(i.e., $\lab(\{u,v\})=\ttneq$),
then $E'$ contains the edges $\{u,v\}$ and $\{N+u,N+v\}$,
otherwise (i.e., (i.e., $\lab(\{u,v\})=\tteq$)
the edges $\{u,v+N\}$ and $\{N+u,N\}$ are placed in $E'$.

Note that the degree of each vertex $v$ in $G'$ is thrice
the degree of the original vertex $v'\eqdef (v-1)\bmod N+1$ in $G$,
where the first $2|\Gamma_G(v)|$ edges go to the other copy of $v'$
and the rest are connected to the adequate copies of neighbors of $v'$.
Clearly, this reduction is local;
specifically, given oracle access to $G$ and $\lab$,
one can implement oracle access to $G'$ by issuing
a single query to $G$ per each query to $G'$.
The reader can easily verify that the number of edges
that must be omitted from $G$ in order to obtain a graph
that is 2-colorable with respect to $\lab$
equals half the number of edges that must be omitted from $G'$
in order to obtain a bipartite graph.
(The key observation is that when 2-coloring $G'$,
the number of edges that are violated is minimized
by assigning the two copies of each vertex of $G$
different colors).

It seems that we are done, except that the multi-graph
obtained in the reduction has parallel edges,
whereas the algorithm of~\cite{KKR} only refers to simple graphs
(i.e., no parallel edges). This assumption is used in~\cite{KKR}
for a single purpose~-- for the construction of a randomized
subroutine that samples vertices with probability that
is (approximately) proportional to their degree.
Other than that, both the algorithm and the analysis of~\cite{KKR}
are oblivious to whether the input is a simple graph or a multi-graph.
The algorithm and its analysis refer to the behavior of random walks
taken on an auxiliary graph, and at that level it does not matter
whether the original graph (or the auxiliary graph) have parallel
edges or not. Thus, we merely need to address the problem of
sampling vertices in the graph $G'$ obtained via our reduction
such that vertices are sampled with probability that
is (approximately) proportional to their degree.

The latter problem is easily solved by noticing that if
we drop all parallel edges from $G'$, we obtain a graph $G''$
such that the degree of each vertex in $G''$ is one third
of its degree in $G'$. Hence, we may just as well invoke
the vertex-sampling procedure of~\cite{KKR} on $G''$,
which is a simple graph. The theorem follows.
\EPF

\paragraph{Digest.} 
Note that combining the two reductions presented
in the foregoing proof, we obtain a randomized reduction
of (one-sided error) testing cycle-freeness to
(one-sided error) testing bipartiteness
(of graphs having parallel edges).
The advantage of this reduction over the one presented
in Section~\ref{C3:section} is that the number of vertices
in the resulting graph is only a constant factor larger
than the number of vertices in the original graph.
(The same holds for the number of edges,
but this was true also for the reduction in Section~\ref{C3:section}.)
The disadvantage of the current reduction (in comparison to the one
presented in Section~\ref{C3:section}) is that it yields graphs
with parallel edges. Fortunately, the number of parallel edges
incident at each vertex is a fixed fraction of the degree of
this vertex, which in turn allows for an easy adaptation of the
Bipartiteness tester presented in~\cite{KKR} so that it can be
applied to these graphs.

\paragraph{Finding longer cycles.}
We were not able to extend our results regarding finding $C_k$-minor
for $k>3$ to the unbounded-degree model. Our reductions to finding
simple cycles (i.e., to the case of $k=3$) increase the size of
the graph by a factor of at least $d^{k-3}$, where $d$ is the
initial degree bound, putting aside the difficulty of handing
the case that the average degree is significantly smaller than $d$.
Indeed, this topic is left for further study.

\subsection{Testing tree-minor-freeness}
\label{general-model:tree}
In contrast to \SAref{general-model:cycle},
we show that the result of \thmref{intro:find-tree.thm}
cannot be extended to the unbounded-degree model.
This follows by considering an $N$-vertex graph $G$
that consists of a cycle of length $N-\sqrt{N}$
and a clique of size $\sqrt{N}$
(i.e., $G=C_{N-\sqrt{N}}+K_{\sqrt{N}}$).
Denoting the 3-star by $T_3$,
note that $G$ is $\Omega(1)$-far from being $T_3$-minor-free
(since we must omit $\sqrt{N}-3$ edges from each vertex of
the $\sqrt{N}$-clique in order to eliminate all copies of $T_3$ itself).
On the other hand, no $o(\sqrt{N})$-query algorithm can find
a $T_3$-minor in a random isomorphic copy of $G$,
except with probability $o(1)$.
Furthermore, any algorithm of query complexity $o(\sqrt{N})$
cannot distinguish a random copy of $G$ from a random copy
of a $N$-vertex graph that consists of
a cycle of length $N-\sqrt{N}$ and $\sqrt{N}$ isolated vertices.
Thus, in this model, even two-sided error testing of $T_3$-minor
freeness requires $\Omega(\sqrt{N})$ queries.

We mention that an $O(\sqrt{N})$-query one-sided error tester
for $T_k$-minor-freeness does exist for any $k$,
where $T_k$ denotes the $k$-star.
This tester may be obtained by combining
the tester for the bounded-degree model
(for $d=k-1$, as presented in \secref{k-star.sec})
with an $O(\sqrt{N})$-query procedure for
finding a vertex of degree at least $k$.
The argument is detailed in the proof of the following result.

\BT{\em(Testing $T_k$-minor freeness in the unbounded-degree model):}
For every $k\geq3$,
testing $T_k$-minor freeness of $N$-vertex graphs,
in the unbounded-degree model,
has query complexity $\tildeT(\sqrt{N})$,
where the upper bound hides factors that are polynomial in $k/\e$.
Furthermore, the lower bound holds also for two-sided error testers,
whereas the upper bound holds with respect to a one-sided error tester
that outputs a $T_k$-minor of size $O(k^2/\e)$ whenever it rejects.
\ET

\BPF
The lower bound follows from the foregoing discussion.
Recall that we consider an $N$-vertex graph $G$
that consists of a cycle of length $N-\sqrt{N}$
and a clique of size $\sqrt{N}$, which is $\Omega(1)$-far
from being $T_k$-free.
We claim that any algorithm of query complexity $o(\sqrt{N})$
cannot distinguish a random copy of $G$ from a random copy
of a $N$-vertex graph that consists of
a cycle of length $N-\sqrt{N}$ and $\sqrt{N}$ isolated vertices.
This claim is proved by noting that, conditioned on
the case that all the prior queries refer to vertices
that are on the long cycle, the probability that the next query
refers to a vertex not on this cycle is at most ${\sqrt N}/N$,
where equality holds if the next query refers to a vertex that
was not seen before (as either part of a query or an answer).

We now turn to the upper bound.
The suggested tester invokes
an arbitrary (one-sided error) $T_k$-minor free tester
for the bounded-degree model (e.g., Algorithm~\ref{k-star.alg}),
where the degree bound, $d$, is set to $k-1$.
Indeed, this algorithm is invoked on a graph that may
not satisfy the degree bound, and so while emulating
this algorithm we check whether each encountered vertex
has degree at most $k-1$.
If (during the emulation) we ever encounter a vertex
of degree at least $k$, then we reject
(while outputting this vertex and $k$ of its neighbors).
If the emulation terminates outputting a $T_k$-minor,
then we just output this minor.
Otherwise, using the edge sampling algorithm of~\cite[Fig.~5]{KKR}
(which has query complexity $\tildeO(\sqrt{N/\e})$ per sample),
we take an almost uniform sample of $O(1/\e)$ edges
and check the degrees of the endpoints of all these edges.
Needless to say, if any of these vertices
is found to have degree at least $k$, then we reject
(while outputting this vertex and $k$ of its neighbors),
otherwise we accept.

Suppose that $G=([N],E)$ is $\e$-far from being $T_k$-minor free,
which means that at least $2\e|E|$ edges have to be removed from $G$
in order to obtain a $T_k$-minor free graph.
Let $H$ denote the set of vertices in $G$ that have degree at least $k$,
let $G'=(V',E')$ be the subgraph of $G$
that is induced by $V'\eqdef[N]\setminus H$,
and $G''=([N],E')$ be $G'$ augmented by $|H|$ isolated vertices.
If at least $\e |E|$ edges have to be removed from $G'$
in order to obtain a $T_k$-minor free graph,
then $G''$ is $\e/4k$-far from being a $T_k$-minor free
(with respect to the bounded-degree model with $d=k-1$,
while assuming $|E|>N/2$).
Thus, Algorithm~\ref{k-star.alg} invoked on $G''$
(with proximity parameter $\e/4k$),
will output a $T_k$-minor with probability at least $2/3$.
The same would happen if Algorithm~\ref{k-star.alg}
is invoked on $G$, because as long as vertices in $H$
are not visited both graphs are indistinguishable,
whereas visiting any graph in $H$ will cause the
algorithm to output a $T_k$-minor.

So we are left with the complimentary case
in which $G$ can be made $T_k$-minor free by
omitting at least $\e|E|$ edges that have at
least one endpoint in $H$.
In this case, the edge sampling algorithm of~\cite{KKR}
(invoked with proximity parameter $\e/2$),
will hit such an edge with probability at least $\Omega(\e)$
per each invocation (see~\cite[Lem.~6]{KKR}).
(Needless to say, hitting any edges with an endpoint in $H$
will cause the algorithm to output a $T_k$-minor.)
The theorem follows by noting that we invoked
this edge sampling algorithm $O(1/\e)$ times,
whereas its query complexity is $\tildeO(\sqrt{N/\e})$
(and the query complexity of Algorithm~\ref{k-star.alg}
is $O(k^2/\e)$).
\EPF

\paragraph{Finding other tree-minors.}
Our star-minor free (one-sided error) tester to the unbounded-degree model,
begs the question of whether similar results can be obtained
with respect to other tree-minors.
Indeed, this question is left for further study.

\subsection{Testing with adjacency queries}
\label{general-model:kkr}
Here we consider an augmentation of the model with adjacency queries.
This augmentation was first considered in~\cite{KKR}, and it was
shown to be useful (for testing bipartiteness) when the average
degree, $\davg$, exceeds $\sqrt N$. We observe that the same holds
with respect to testing cycle-freeness (see details below).
(In contrast, recall that in the bare model (i.e., without adjacency queries)
the upper bound presented in \secref{general-model:cycle} are optimal.)

We note that the reduction
presented in \secref{general-model:cycle} remains valid,
except that in this case the generalized 2-coloring tester
(derived from~\cite{KKR}) may use adjacency queries.
In this case, the resulting cycle-freeness tester will have
complexity $\min(\tildeO(\sqrt{N}),\tildeO(N)/\davg)\cdot\poly(1/\e)$
(just as the 2-coloring tester of~\cite{KKR}).
%

We also note that the results regarding testing tree-minors
(see Section~\ref{general-model:tree}) extend similarly.
Specifically, for any $k\geq3$,
the complexity of testing $T_k$-minor freeness
is $\min(\tildeT(\sqrt{N}),\tildeT(N)/\davg))$.

%
%
%

\section{Open Problems}\label{open:section}
The current paper leaves open many questions regarding the complexity
of one-sided error testing of $H$-minor-freeness,
for arbitrary (fixed) graphs $H$.
While some significant progress has been done
regarding this general question, much is left to be desired
(even if we restrict ourselves to the bounded-degree model).

Let us denote by $\QT_H:\N\times[0,1]\to\N$ the query complexity of
one-sided error testing of $H$-minor-freeness in the bounded-degree model.
(For simplicity, we fix the degree bound and consider the number
of queries made as a function of the number of vertices and
the proximity parameter.)
The most begging open problem is whether or not,
for any fixed $\e_0>0$ and graph $H$,
it holds that $\QT_H(N,\e_0)=o(N)$.
We showed that $\QT_H(N,\e_0)=\tildeT(\sqrt{N})$
for any $H$ that is a cycle,
and $\QT_H(N,\e_0)=O(1)$ for any cycle-free $H$,
but for other graphs $H$ that have cycles
we only know that $\QT_H(N,\e_0)=\Omega(\sqrt{N})$.
The only step beyond this state of knowledge
is represented by the following result,
which we discovered recently.

\BP
\label{new:prop}
Let $H$ be he 4-vertex graph consisting of a triangle
and an additional edge. Then, for any fixed $\e_0>0$,
it holds that $\QT_H(N,\e_0)=\tildeO(\sqrt{N})$.
\EP
There seems to be hope to obtain a similar result
for any graph $H$ that contains exactly one cycle
(along with some additional edges),
but we do not see how to do this at the moment.%
\footnote{Even the two cases of 5-vertex graphs that
contain a single cycle and a total of five edges
cannot be handled by the idea that underlies the
proof of Proposition~\ref{new:prop}.}
Another natural case to consider is the one in which $H$
is the 4-vertex clique
(or the 4-vertex clique with one edge omitted).
\medskip

\BPF
The lower bound follows as a special case of Theorem~\ref{lower-bound.thm},
and so our focus is on the upper bound.
The idea is to invoke a one-sided tester for cycle-freeness
(e.g., Algorithm~\ref{cycle-free.alg}),
and scan the simple cycles in the subgraph
that this algorithm has explored.
Our goal is to find a cycle that contains a vertex of degree
greater than two, since any such cycle yields a desired $H$-minor.
(If this vertex has a neighbor outside the cycle,
then the cycle combined with the corresponding edge
yields an $H$-minor. Otherwise the non-cycle edge
is a chord, and we can use this chord together with
half of the cycle and some other edge to form an $H$-minor.)

Note that if the current cycle that we scan does not yield
an $H$-minor, then it resides on an isolated cycle in the
input graph. In this case, we omit it from the subgraph
explored by the algorithm and continue looking for other
cycles (which may still exist) in this subgraph.

To analyze this algorithm assume that $G$ is $\e$-far from
being $H$-minor free, and consider the isolated cycles
(of vertices having degree $G$) that may exist in $G$.
Let $G'$ be a graph obtained from $G$ by omitting a single
edge from each such isolated cycle.
Then, $G'$ is also $\e$-far from being $H$-minor free
(and so also $\e$-far from being cycle-free),
whereas invoking our algorithm on $G'$ will yield
an $H$-minor with probability at least $2/3$
(since any cycle in $G'$ contains a vertex of degree exceeding 2).
But then the same must happen also when we invoke our algorithm on $G$,
because the manner in which Algorithm~\ref{cycle-free.alg}
explores one connected component is independent
of its exploration of a different connected component.
\EPF

\paragraph{Finer bounds.}
Focusing on the case that $H$ is cycle-free,
we know that $\QT_H(N,\e)=F_H(\e)$ for some function $F_H$.
A natural question refers to the exact dependence
of $F_H(\e)$ on the graph $H$.
Specific questions include:
\BE
\item We know that if $H$ is a $k$-path,
then $F_H(\e)$ is at most exponential in $k$.
{\em Can $F_H(\e)$ be polynomial in $k$}?

(Partial progress is reported in~\cite{Reznik},
which deals with the special case that the input graph is a tree.)

\item We know that if $H$ is a $k$-vertex tree,
then $F_H(\e)$ is at most double-exponential in $k$.
{\em Can $F_H(\e)$ be at most exponential in $k$}?

(In case $H$ has at most depth two, the answer is
positive, as reported in~\cite[Sec.~7.5]{tr}.
Also recall that if $H$ has depth one (i.e., is a star),
then $F_H(\e)$ is polynomial in $k$.)
\EE
Also recall that when $H$ is a $k$-cycle,
we have $\QT_H(N,\e)=\tildeO(\sqrt N)\cdot\poly(2^k/\e)$,
and one may ask
whether $\QT_H(N,\e)=\tildeO(\sqrt N)\cdot\poly(k/\e)$ is possible.

\paragraph{The unbounded degree model.}
Turning to the unbounded degree model,
which was considered in Section~\ref{general-model},
we let $\QT'_H:\N\times\N\times[0,1]\to\N$ denote the corresponding
query complexity (assuming only neighbor queries),
which is now a function of the number of vertices,
the (approximate) average degree, and the proximity parameter.
We showed that, for every fixed $\e_0>0$ and $H$ that is either
a 3-cycle or a star, it holds that $\QT'_H(N,d,\e_0)=\tildeT(\sqrt N)$.
The open questions here refer even to other cycles and trees
(for which the complexity in the bounded-degree model is known).

\section*{Acknowledgments}
We are grateful to Aviv Reznik for useful comments on a prior version,
and to the anonymous reviewers for their comments and suggestions.


\addcontentsline{toc}{section}{Bibliography}


\ifnum\LatexVer=2
\ifnum\submit=0
\bibliographystyle{alpha}
\else
\bibliographystyle{plain}
\fi
\bibliography{arxiv-version}
\else

\fi


\end{document}

%% file: C4c.pstex_t
\begin{picture}(0,0)%
\includegraphics{C4c.pstex}%
\end{picture}%
\setlength{\unitlength}{3947sp}%
\begingroup\makeatletter\ifx\SetFigFont\undefined%
\gdef\SetFigFont#1#2#3#4#5{%
  \reset@font\fontsize{#1}{#2pt}%
  \fontfamily{#3}\fontseries{#4}\fontshape{#5}%
  \selectfont}%
\fi\endgroup%
\begin{picture}(6908,1212)(366,-549)
\put(995,547){\makebox(0,0)[lb]{\smash{{\SetFigFont{11}{13.2}{\rmdefault}{\mddefault}{\updefault}{\color[rgb]{0,0,0}$G$}%
}}}}
\put(6247,555){\makebox(0,0)[lb]{\smash{{\SetFigFont{11}{13.2}{\rmdefault}{\mddefault}{\updefault}{\color[rgb]{0,0,0}$G'$}%
}}}}
\put(517,-459){\makebox(0,0)[lb]{\smash{{\SetFigFont{11}{13.2}{\rmdefault}{\mddefault}{\updefault}{\color[rgb]{0,0,0}$c$}%
}}}}
\put(366,181){\makebox(0,0)[lb]{\smash{{\SetFigFont{11}{13.2}{\rmdefault}{\mddefault}{\updefault}{\color[rgb]{0,0,0}$a$}%
}}}}
\put(1548,167){\makebox(0,0)[lb]{\smash{{\SetFigFont{11}{13.2}{\rmdefault}{\mddefault}{\updefault}{\color[rgb]{0,0,0}$d$}%
}}}}
\put(2024,176){\makebox(0,0)[lb]{\smash{{\SetFigFont{11}{13.2}{\rmdefault}{\mddefault}{\updefault}{\color[rgb]{0,0,0}$a$}%
}}}}
\put(3258,157){\makebox(0,0)[lb]{\smash{{\SetFigFont{11}{13.2}{\rmdefault}{\mddefault}{\updefault}{\color[rgb]{0,0,0}$d$}%
}}}}
\put(2167,-479){\makebox(0,0)[lb]{\smash{{\SetFigFont{11}{13.2}{\rmdefault}{\mddefault}{\updefault}{\color[rgb]{0,0,0}$c$}%
}}}}
\put(3128,-475){\makebox(0,0)[lb]{\smash{{\SetFigFont{11}{13.2}{\rmdefault}{\mddefault}{\updefault}{\color[rgb]{0,0,0}$e$}%
}}}}
\put(4157,-489){\makebox(0,0)[lb]{\smash{{\SetFigFont{11}{13.2}{\rmdefault}{\mddefault}{\updefault}{\color[rgb]{0,0,0}$c$}%
}}}}
\put(3994,166){\makebox(0,0)[lb]{\smash{{\SetFigFont{11}{13.2}{\rmdefault}{\mddefault}{\updefault}{\color[rgb]{0,0,0}$a$}%
}}}}
\put(5168,-455){\makebox(0,0)[lb]{\smash{{\SetFigFont{11}{13.2}{\rmdefault}{\mddefault}{\updefault}{\color[rgb]{0,0,0}$e$}%
}}}}
\put(5827,-479){\makebox(0,0)[lb]{\smash{{\SetFigFont{11}{13.2}{\rmdefault}{\mddefault}{\updefault}{\color[rgb]{0,0,0}$c$}%
}}}}
\put(6253,196){\makebox(0,0)[lb]{\smash{{\SetFigFont{11}{13.2}{\rmdefault}{\mddefault}{\updefault}{\color[rgb]{0,0,0}$b$}%
}}}}
\put(5624,186){\makebox(0,0)[lb]{\smash{{\SetFigFont{11}{13.2}{\rmdefault}{\mddefault}{\updefault}{\color[rgb]{0,0,0}$a$}%
}}}}
\put(6788,-465){\makebox(0,0)[lb]{\smash{{\SetFigFont{11}{13.2}{\rmdefault}{\mddefault}{\updefault}{\color[rgb]{0,0,0}$e$}%
}}}}
\put(6998,167){\makebox(0,0)[lb]{\smash{{\SetFigFont{11}{13.2}{\rmdefault}{\mddefault}{\updefault}{\color[rgb]{0,0,0}$d$}%
}}}}
\put(5328,167){\makebox(0,0)[lb]{\smash{{\SetFigFont{11}{13.2}{\rmdefault}{\mddefault}{\updefault}{\color[rgb]{0,0,0}$d$}%
}}}}
\put(4663,166){\makebox(0,0)[lb]{\smash{{\SetFigFont{11}{13.2}{\rmdefault}{\mddefault}{\updefault}{\color[rgb]{0,0,0}$b$}%
}}}}
\put(2613,166){\makebox(0,0)[lb]{\smash{{\SetFigFont{11}{13.2}{\rmdefault}{\mddefault}{\updefault}{\color[rgb]{0,0,0}$b$}%
}}}}
\put(2607,535){\makebox(0,0)[lb]{\smash{{\SetFigFont{11}{13.2}{\rmdefault}{\mddefault}{\updefault}{\color[rgb]{0,0,0}$G'$}%
}}}}
\put(985,181){\makebox(0,0)[lb]{\smash{{\SetFigFont{11}{13.2}{\rmdefault}{\mddefault}{\updefault}{\color[rgb]{0,0,0}$b$}%
}}}}
\put(1468,-465){\makebox(0,0)[lb]{\smash{{\SetFigFont{11}{13.2}{\rmdefault}{\mddefault}{\updefault}{\color[rgb]{0,0,0}$e$}%
}}}}
\put(4645,540){\makebox(0,0)[lb]{\smash{{\SetFigFont{11}{13.2}{\rmdefault}{\mddefault}{\updefault}{\color[rgb]{0,0,0}$G$}%
}}}}
\end{picture}%

%% file: spot-paths.pstex_t
\begin{picture}(0,0)%
\includegraphics{spot-paths.pstex}%
\end{picture}%
\setlength{\unitlength}{3947sp}%
\begingroup\makeatletter\ifx\SetFigFont\undefined%
\gdef\SetFigFont#1#2#3#4#5{%
  \reset@font\fontsize{#1}{#2pt}%
  \fontfamily{#3}\fontseries{#4}\fontshape{#5}%
  \selectfont}%
\fi\endgroup%
\begin{picture}(2716,2014)(1493,-1568)
\put(2551,314){\makebox(0,0)[lb]{\smash{{\SetFigFont{12}{14.4}{\rmdefault}{\mddefault}{\updefault}{\color[rgb]{0,0,0}$S$}%
}}}}
\put(1854,-226){\makebox(0,0)[lb]{\smash{{\SetFigFont{11}{13.2}{\rmdefault}{\mddefault}{\updefault}{\color[rgb]{0,0,0}$u$}%
}}}}
\put(1868,-1261){\makebox(0,0)[lb]{\smash{{\SetFigFont{11}{13.2}{\rmdefault}{\mddefault}{\updefault}{\color[rgb]{0,0,0}$v$}%
}}}}
\put(3646,-795){\makebox(0,0)[lb]{\smash{{\SetFigFont{11}{13.2}{\rmdefault}{\mddefault}{\updefault}{\color[rgb]{0,0,0}$w$}%
}}}}
\put(2416,-525){\makebox(0,0)[lb]{\smash{{\SetFigFont{11}{13.2}{\rmdefault}{\mddefault}{\updefault}{\color[rgb]{0,0,0}$x$}%
}}}}
\end{picture}%

%% file: 2spot.pstex_t
\begin{picture}(0,0)%
\includegraphics{2spot.pstex}%
\end{picture}%
\setlength{\unitlength}{3947sp}%
\begingroup\makeatletter\ifx\SetFigFont\undefined%
\gdef\SetFigFont#1#2#3#4#5{%
  \reset@font\fontsize{#1}{#2pt}%
  \fontfamily{#3}\fontseries{#4}\fontshape{#5}%
  \selectfont}%
\fi\endgroup%
\begin{picture}(2565,1551)(968,-1118)
\put(3158,-488){\makebox(0,0)[lb]{\smash{{\SetFigFont{11}{13.2}{\rmdefault}{\mddefault}{\updefault}{\color[rgb]{0,0,0}$w$}%
}}}}
\put(2663,-203){\makebox(0,0)[lb]{\smash{{\SetFigFont{11}{13.2}{\rmdefault}{\mddefault}{\updefault}{\color[rgb]{0,0,0}$u'$}%
}}}}
\put(2626,-751){\makebox(0,0)[lb]{\smash{{\SetFigFont{11}{13.2}{\rmdefault}{\mddefault}{\updefault}{\color[rgb]{0,0,0}$v'$}%
}}}}
\put(3008,208){\makebox(0,0)[lb]{\smash{{\SetFigFont{12}{14.4}{\rmdefault}{\mddefault}{\updefault}{\color[rgb]{0,0,0}$S_2$}%
}}}}
\put(1943,-83){\makebox(0,0)[lb]{\smash{{\SetFigFont{11}{13.2}{\rmdefault}{\mddefault}{\updefault}{\color[rgb]{0,0,0}$u$}%
}}}}
\put(1906,-676){\makebox(0,0)[lb]{\smash{{\SetFigFont{11}{13.2}{\rmdefault}{\mddefault}{\updefault}{\color[rgb]{0,0,0}$v$}%
}}}}
\put(1315,277){\makebox(0,0)[lb]{\smash{{\SetFigFont{12}{14.4}{\rmdefault}{\mddefault}{\updefault}{\color[rgb]{0,0,0}$S_1$}%
}}}}
\end{picture}%

%% file: cycle-spot.pstex_t
\begin{picture}(0,0)%
\includegraphics{cycle-spot.pstex}%
\end{picture}%
\setlength{\unitlength}{3947sp}%
\begingroup\makeatletter\ifx\SetFigFont\undefined%
\gdef\SetFigFont#1#2#3#4#5{%
  \reset@font\fontsize{#1}{#2pt}%
  \fontfamily{#3}\fontseries{#4}\fontshape{#5}%
  \selectfont}%
\fi\endgroup%
\begin{picture}(4934,1400)(207,-983)
\put(1290,-23){\makebox(0,0)[lb]{\smash{{\SetFigFont{11}{13.2}{\rmdefault}{\mddefault}{\updefault}{\color[rgb]{0,0,0}$u$}%
}}}}
\put(1314,-683){\makebox(0,0)[lb]{\smash{{\SetFigFont{11}{13.2}{\rmdefault}{\mddefault}{\updefault}{\color[rgb]{0,0,0}$v$}%
}}}}
\put(3915,-694){\makebox(0,0)[lb]{\smash{{\SetFigFont{11}{13.2}{\rmdefault}{\mddefault}{\updefault}{\color[rgb]{0,0,0}$v$}%
}}}}
\put(2034, 51){\makebox(0,0)[lb]{\smash{{\SetFigFont{11}{13.2}{\rmdefault}{\mddefault}{\updefault}{\color[rgb]{0,0,0}$w_1$}%
}}}}
\put(2100,-743){\makebox(0,0)[lb]{\smash{{\SetFigFont{11}{13.2}{\rmdefault}{\mddefault}{\updefault}{\color[rgb]{0,0,0}$w_2$}%
}}}}
\put(4711,-39){\makebox(0,0)[lb]{\smash{{\SetFigFont{11}{13.2}{\rmdefault}{\mddefault}{\updefault}{\color[rgb]{0,0,0}$w_1$}%
}}}}
\put(3210,-503){\makebox(0,0)[lb]{\smash{{\SetFigFont{11}{13.2}{\rmdefault}{\mddefault}{\updefault}{\color[rgb]{0,0,0}$w_2$}%
}}}}
\put(3914, -4){\makebox(0,0)[lb]{\smash{{\SetFigFont{11}{13.2}{\rmdefault}{\mddefault}{\updefault}{\color[rgb]{0,0,0}$u$}%
}}}}
\put(3481,267){\makebox(0,0)[lb]{\smash{{\SetFigFont{12}{14.4}{\rmdefault}{\mddefault}{\updefault}{\color[rgb]{0,0,0}$S$}%
}}}}
\put(775,285){\makebox(0,0)[lb]{\smash{{\SetFigFont{12}{14.4}{\rmdefault}{\mddefault}{\updefault}{\color[rgb]{0,0,0}$S$}%
}}}}
\end{picture}%